\documentclass[]{aa}
\input psfig.tex
\usepackage{graphicx}
\usepackage{natbib}
\usepackage{array}
\usepackage{graphics}
\usepackage{latexsym}
\usepackage{amssymb}
\usepackage{amsmath}
\usepackage{fancyhdr}
\usepackage{morefloats}
\bibpunct{(}{)}{;}{a}{}{,}
\include{hyphe}

\begin{document}

\title{Formation and Evolution of the Dust in Galaxies. II.\\ The Solar Neighbourhood}

\author{L. Piovan \inst {1,2}, C. Chiosi \inst{1}, E. Merlin \inst{1}, T. Grassi \inst{1},
R. Tantalo \inst{1}, U. Buonomo \inst{1} and L. P. Cassar\`{a}
\inst{1}}

\institute{    $^1$ Department of Astronomy, Padova University,
Vicolo dell'Osservatorio 3, I-35122, Padova, Italy\\
$^2$Max-Planck-Institut f\"ur Astrophysik, Karl-Schwarzschild-Str.
1, Garching bei M\"unchen, Germany\\
\email{  {lorenzo.piovan\char64unipd.it} }   }

\date{Received: July 2011; Revised: *** ****;  Accepted: *** ***}

\abstract   {Over the past decade a new generation of chemical
models, in addition to the gas, have included the dust in the
treatment of the interstellar medium (ISM). This major
accomplishment has been spurred by the growing amounts of data on
the highly obscured high-z Universe and the intriguing local
properties of the Solar Neighbourhood (SoNe) of the Milky Way (MW)
Disk.} {We present here a new model able to simulate the formation
and evolution of dust in the ISM of the MW. The model follows the
evolution of 16 elemental species, with particular attention to
those  that are simultaneously present in form of gas and dust, e.g.
C, N, O, Mg, Si, S, Ca and Fe. In this study we focus on the SoNe
and the MW Disk as a whole which are considered as laboratories to
test the physical ingredients governing the dust evolution.} {The MW
is described as a set of concentric rings of  which we follow the
time evolution of  gas and dust. Infall of primordial gas, birth and
death of stars, radial flows of matter between contiguous shells,
presence of a central bar, star-dust emission by SN{\ae} and AGB
stars, dust destruction and accretion are taken into account. The
model  reproduces the local depletion of the elements in the gas,
and simultaneously satisfies other constraints obtained from the
observations.} {The evolution of the element abundances  in the gas
and dust has been well reproduced for plausible choices of the
parameters. The Mg/Si ratio, in particular, drives the formation of
silicates. We show that for most of the evolution of the MW, the
main process for dust  enrichment is the accretion in the cold
regions of the ISM. SN{\ae} dominate in the early phases of the
evolution. We have also examined the main factors controlling the
temporal window in which SN{\ae} govern the dust budget both in low
and high star forming environments. The role played by AGB stars is
also discussed. We find that IMFs with regular slope in the range of
massive stars better reproduce the observed depletions.} {The
classical chemical models nicely reproduce the abundances, depletion
factors and dust properties of the SoNe and the main ingredients of
the models are tested against observational data. The results
obtained for the SoNe lead us to safely extend the model to the
whole Galactic Disk or galaxies of different morphological types.}
\keywords{Galaxies - Dust; Galaxies -- Spirals; Galaxies -- Milky
Way }

\titlerunning{Formation and evolution of the dust in galaxies}

\authorrunning{L. Piovan  et al.}
\maketitle



\section{Introduction}\label{intro}

In the fascinating subject of the origin and evolution of galaxies, the
interstellar dust is acquiring a primary role because of its growing
importance in the observations of the high-z Universe
\citep{Omont01,Shapley01,Bertoldi03,Robson04,Wang08a,Wang08b,Gallerani10,Michalowski10a,Michalowski10b}
and the theoretical spectro-photometric,
dynamical, and chemical modeling of galaxies
\citep{Schurer09,Narayanan10,Jonsson10,Grassi10,Pipino11,Popescu11}. \\
\indent Indeed, the evidence of highly obscured QSOs and galaxies
already in place at high-z leads necessarily to a new generation of
theoretical models where dust is a key ingredient that cannot be
neglected, if we want to obtain precious clues on the fundamental
question about when and how galaxies formed and evolved. First of
all, dust absorbs the stellar radiation and re-emits it in the
infrared deeply changing the shape of the observed spectral energy
distributions (SEDs) of obscured galaxies
\citep{Silva98,Piovan06b,Popescu11}; second, it strongly affects the
production of molecular hydrogen and the local amount of UV
radiation in galaxies thus playing a strong role in the star
formation process via the cooling mechanisms \citep{Yamasawa11}. The
inclusion of dust in the models leads to a growing complexity and
typically to a much larger set of parameters influencing the results
of the simulations to be then compared with the observations. Indeed
several question must be addressed, each one easily expanding the
model: who are the main stardust injectors in the interstellar
medium (ISM) \citep{Gail09,Valiante09,Gall11a,Piovan11a}? How much
dust do they produce and on which timescales
\citep{Draine09,Dwek09}? What is the contribution and the role
played by the molecular clouds (MCs)-grown dust that form in the
cold dense regions of the ISM \citep{Zhukovska08,Pipino11}? How much
dust is destroyed by SN{\ae} shocks
\citep{Nozawa06,Nozawa07,Bianchi07,Jones11}? What is the typical
minimal set of dust grains whose evolution should be followed and
what could be a minimal set of dust grains to be used for a
satisfactory description
of the chemical or spectrophotometric properties of the galaxy? \\
\indent To answer all these questions the theoretical models must
include (i) a set of grains with suitable composition and properties
and/or an ISM made of gas and dust in which the abundances of the
elements are followed, (ii) a recipe for their formation/accretion
and destruction in the ISM and (iii) a prescription for the yields
of dust by the stellar sources. The duty cycle of the dust can be
schematically summarized as follows \citep{Jones04}. Stars, mainly
AGBs and SN{\ae}, inject material in the ISM, mainly in form of gas,
but with a variable amount that condenses into the so-called
star-dust. Once injected into the ISM, star-dust grains are
subjected to destruction processes that restitute the material to
the gaseous phase. However in cold and dense regions dust can
accrete on the so-called seeds: the competition between the
accretion and destruction processes, mainly via shocks, determines
the total budget of dust in the ISM and the observed depletion of
the elements that are involved in the formation of dust grains
\citep{Dwek98}. Dust accretion mainly occurs in the very cold
molecular clouds (MCs), where it induces strong cooling thus leading
to the formation of new stars. The stellar winds from AGB stars
SN{a} explosions more and more enrich the ISM with new metals and
star-dust grains that are able to survive to the local shocks
caused by  SNa explosions. \\
\indent The Milky Way (MW) is the ideal laboratory to our disposal
to study the dust cycle  \citep{Zhukovska08} and its impact on the
wider subject of galaxy formation. For obvious reasons, the MW
provides plenty of observational data to which we can compare
theoretical predictions, thus setting useful constraints on
theoretical simulations and highlighting the role of the most
important physical quantities leading the whole problem. Once this
important step is accomplished, our modelling of the role played by
dust can be extended to other galaxies such as local disk and
spheroidal galaxies, high-z galaxies
and QSOs. \\
\indent Starting from these considerations, in this paper we
simulate the formation, evolution and composition of dust in the MW,
both locally in the Solar Neighbourhood (SoNe) and radially along
the Galactic Disk. We build up a detailed chemical model (the
theoretical simulation is still the  main tool to investigate the
formation and evolution of dust in galaxies) starting from the
pioneering study  by \citet{Dwek98} and taking into account the more
recent ones by \citet{Zhukovska08}, \citet{Calura08},
\citet{Valiante09}, \citet{Gall11a}, \citet{Gall11b},
\citet{Mattsson11}, \citet{Valiante11}, \citet{Kemper11} and
\citet{Dwek11}. The theoretical model we are building up stems from
the basic one with infall by \citet{Chiosi80}, however updated to
the more recent version with radial flows of matter and presence of
a central bar  developed by \citet{Portinari00}. The stellar yields
of chemical elements in form of gas  are those calculated by
\citet{Portinari98}. The model follows the evolution of the
abundances of  a number of elements composing the ISM gas, includes
the formation/destruction and evolution of dust and, finally,
follows in detail also the abundances of those elements that are
embedded in the dust grains. To this aims, the model makes use of
the best prescriptions available in literature concerning dust
accretion, destruction and condensation in the AGB  winds and
SN{\ae} explosions.  These prescriptions have been already presented
by \citet{Piovan11a} to whom the reader should refer
and will also be discussed in some detail here.  \\
\indent The main test for any  model of dust formation is given by
the data on element depletion provided by SoNe of the MW to which we
will compare our results. In a forthcoming paper \citep{Piovan11c}
we will investigate the radial dependence of chemical abundances and
dust depletion across the Disk of the MW. \\
\indent The plan of the paper is as follows. In Sect.
\ref{Che_Evo_Mod} we introduce the formalism and basic equations
governing the temporal evolution of the gas, stars, and dust in an
open model with radial flows of gas and dust for the galactic Disk
of the MW. In Sect. \ref{SFRandIMF} we summarize the current
prescriptions for the star formation rate and initial mass function.
In Sect. \ref{graingrowth} we introduce and describe in some detail
the various processes responsible for the formation and growth of
dust grains in the ISM, whereas in  Sect. \ref{grainevolution} we
present the accretion rates into dust grains for a number of
important elements we have considered. The yields of dust from AGB
stars and SNa explosions are adopted according to \citet{Piovan11a}
to whom the reader should refer for all the details. In Sect.
\ref{SNEdestruction} the problem of dust destruction by SN{a} shocks
is faced. Then, in Sect. \ref{AbundancesSS} we present the
observational data for the elemental abundances in the Solar
Neighborhood and define the reference set of abundances we have
adopted. In Sect. \ref{Dustymodels} after summarizing the main ones
between the many available parameters, we discuss and compare the
effect of them on the formation and evolution of dust, at varying
them between the many possible choices. In particular we examine the
influence on dust of a different CO fraction in the ISM (Sect.
\ref{COFraction}), the effect of the IMF (Sect. \ref{IMFeffect}) and
of the SF law (Sect. \ref{SFeffect}), the choice between different
models for the accretion of dust in cold regions (Sect.
\ref{Accreffect}) and, finally we discuss some interesting
parameters however not discussed in our model. In Sect.
\ref{Models_Observ} we present the results for our models of the
Solar Vicinity in presence of dust and as a function of three
important ingredients, namely, the initial mass function, the
efficiency of star formation, the accretion time scale of primordial
gas onto the system mimicking the evolution of the MW Disk. The
effect of radial flows and central bar are always included according
to the prescription developed in previous studies
\citep{Portinari00} and also adopted in \citet{Piovan11c}. The
simulations are compared with the depletion of the elements in the
SoNe under the constraints that we want that the local chemical
properties are satisfied, such as the time evolution of the
elemental abundances, the metallicity and iron enrichment. In Sect.
\ref{Discus_Concl} we discuss the results we have obtained and draw
some general conclusions.

\section{Chemical Evolution Model}\label{Che_Evo_Mod}

\noindent In classical models of chemical evolution,  the Disk of
the MW  is subdivided in $N$ concentric circular rings of a certain
thickness $\Delta r$, where $r$ is the galacto-centric distance, in
the case of plane geometry or $N$ concentric cylindrical shells if
the third dimension is considered. Each ring or shell is identified
by the mid radius $r_k$ with $k=1,.....,N$.  In most cases, radial
flows of interstellar gas and dust are neglected, so that each ring
/ shell evolves independently from the others. The physical quantity
used to describe the Disk is the surface mass density as a function
of the radial coordinate $r$ and time $t$: $\sigma \left(r_{k},t
\right)$ is the mass surface density at radius $r_{k}$ and time $t$.
Depending on the model,  $\sigma$ can refer to the ISM
($\sigma^{\mathcal{M}}$), in turn split into   dust or gas
($\sigma^{D}$ or $\sigma^{G}$ respectively), to the stars
($\sigma^{*}$) or to the total mass (simply $\sigma$). At every
radius $r_{k}$,  the surface mass density is supposed to slowly grow
by infall of either primordial or already enriched gas and to fetch
at the present age $t_G$ the mass density profile across the
Galactic Disk for which an exponential profile is best suited to
represent  the surface mass density distribution: $\sigma
\left(r_{k}, t_{G}\right) \varpropto
\exp{\left(-r_{k}/r_{d}\right)}$, where $r_{d}$ is the scale radius
of the Galactic Disk, that is typically estimated of the order of
$4-5$ kpc. Since the final density profile is \textit{a priori}
known, one may   normalize to it the current total surface mass
density of the ISM "$\mathcal{M}$" (sum of gas and dust),

\vspace{-7pt}
\begin{equation}
\mathcal{M}\left(r_{k},t\right)=\frac{\sigma^{\mathcal{M}}\left(r_{k},t\right)}
{\sigma\left(r_{k},t_{G}\right)}.
\end{equation}
\vspace{-7pt}

\noindent Introducing the fractionary mass of the generic $i$-th
element we have:

\vspace{-7pt}
\begin{equation}
\mathcal{M}_{i}\left(r_{k},t\right)=\frac{\sigma_{i}^{\mathcal{M}}\left(r_{k},t\right)}{\sigma\left(r_{k},t_{G}\right)}=
\chi_{i}\left(r_{k},t\right)\mathcal{M}\left(r_{k},t\right)
\end{equation}
\vspace{-7pt}

\noindent and therefore the fractional mass abundance
\begin{small}$\displaystyle
\chi_{i}\left(r_{k},t\right)=\displaystyle
\mathcal{M}_{i}\left(r_{k},t\right)/\displaystyle
\mathcal{M}\left(r_{k},t\right)$\end{small}, with
$\sum_{i}\chi_{i}\left(r_{k},t\right)=1$.

Similar expressions can be derived for the dust (indicated by $"D$)
and the gas (indicated by $"G"$)

\vspace{-7pt}
\begin{equation}
D\left(r_{k},t\right)=\frac{\sigma^{D}\left(r_{k},t\right)}{\sigma\left(r_{k},t_{G}\right)}
\end{equation}
\vspace{-6pt}

\vspace{-6pt}
\begin{equation}
G\left(r_{k},t\right)=\frac{\sigma^{G}\left(r_{k},t\right)}{\sigma\left(r_{k},t_{G}\right)}
\end{equation}
\vspace{-7pt}

\noindent with
$\sigma^{\mathcal{M}}\left(r_{k},t\right)=\sigma^{G}\left(r_{k},t\right)+\sigma^{D}\left(r_{k},t\right)$
and
$\sigma\left(r_{k},t\right)=\sigma^{\mathcal{M}}\left(r_{k},t\right)+\sigma^{*}\left(r_{k},t\right)$,
where $\sigma^{*}\left(r_{k},t\right)$ is the surface mass density
of stars. For single chemical elements we may write:

\vspace{-7pt}
\begin{equation}
D_{i}\left(r_{k},t\right)=
\frac{\chi_{i}^{D}\left(r_{k},t\right)\sigma_{D}\left(r_{k},t\right)}
{\sigma\left(r_{k},t_{G}\right)}=
\chi_{i}^{D}\left(r_{k},t\right)D\left(r_{k},t\right)
\end{equation}
\vspace{-7pt}

\vspace{-7pt}
\begin{equation}
G_{i}\left(r_{k},t\right)=
\frac{\chi_{i}^{G}\left(r_{k},t\right)\sigma_{G}\left(r_{k},t\right)}
{\sigma\left(r_{k},t_{G}\right)}=
\chi_{i}^{G}\left(r_{k},t\right)G\left(r_{k},t\right)
\end{equation}
\vspace{-7pt}

\noindent with
$\sum_{i}\left[\chi_{i}^{D}\left(r_{k},t\right)+\chi_{i}^{G}\left(r_{k},t\right)\right]=1$,
from which it follows that
$\sum_{i}\chi_{i}^{D}\left(r_{k},t\right)\neq 1$ and
$\sum_{i}\chi_{i}^{G}\left(r_{k},t\right)\neq 1$.

The fundamental equation describing the evolution of the
ISM  in absence of  radial flows of matter between
contiguous shells \citep{Portinari00} and   processes of dust
accretion/destruction  \citep{Dwek98} is:

\vspace{-7pt}
\begin{eqnarray}
\frac{d}{dt}\mathcal{M}_{i}\left(r_{k},t \right) &=&
-\chi_{i}\left(r_{k},t\right)\psi\left(r_{k},t\right)
\nonumber \\
&+& \int_{M_{l}}^{M_{u}}
\psi\left(r_{k},t-\tau_{M}\right)R_{i}\left(M\right)\phi\left(M\right)dM
\nonumber \\
&+& \left[\frac{d}{dt}\mathcal{M}_{i}\left(r_{K},t \right)
\right]_{inf}
\end{eqnarray}
\vspace{-7pt}

\noindent where $\phi\left(M\right)$ is the IMF and $M_{l}$ and
$M_{u}$ are the lower and upper limits for the stellar masses,
$\psi\left(r_{k},t-\tau_{M}\right)$ is the star formation rate (SFR)
at the radius $r_{k}$  and at the time $t^{\prime}=t-\tau_{M}$,
$R_{i}\left(M\right)=E_{iM}/M$ \citep{Portinari98} is the fraction
of a star of initial mass $M$ ejected back in form of the chemical
species $i$-th. The three terms at the r.h.s. represent the
depletion of the ISM due to star formation, its increase by stellar
ejecta,  and the increase by infall of external gas (either
primordial or already enriched).

\textsf{Adding supernovae and radial flows}. Type Ia supernovae
originate in binary systems and have a fundamental role, in
particular concerning the iron enrichment. The supernovae rate and
the adopted formalism are the ones of \citet{Greggio83} and the
formulation of equation describing chemical evolution of the ISM is
modified following \citet{Matteucci86}. The contribution of single
stars, corresponding to a fraction $\left(1-A\right)$ of the total,
is separated from the contribution of binary system, a fraction $A$
of the total. Inserting the contribution of type Ia supernovae and
integrating in time, instead that in mass, the equation for the
evolution of the $i$-th component of the ISM is,

\vspace{-7pt}
\begin{small}
\begin{flalign} \label{GISM}
\frac{d}{dt}\mathcal{M}_{i}&\left(r_{k},t \right) = -\chi_{i}\psi+ \nonumber \\
&+ \int_{0}^{t-\tau_{M_{B,l}}}\psi \left[\phi\left( M \right)
R_{i}\cdot\left(-\frac{dM}{d\tau_{M}}\right)
\right]_{M\left(\tau\right)}dt^{\prime}+\nonumber\\
&+ \left(1-A \right)\int_{t-\tau_{M_{B,l}}}^{t-\tau_{M_{B,u}}}\psi
\left[\phi R_{i}\cdot\left(-\frac{dM}{d\tau_{M}}\right)
\right]_{M\left(\tau\right)}dt^{\prime}+\nonumber\\
&+ \int_{t-\tau_{M_{B,u}}}^{t-\tau_{M_{u}}}\psi \left[\phi
R_{i}\cdot\left(-\frac{dM}{d\tau_{M}}\right)
\right]_{M\left(\tau\right)}dt^{\prime}+ \nonumber\\
&+ A\int_{t-\tau_{M_{1,min}}}^{t-\tau_{M_{1,max}}}\psi
\left[\textit{f}\left( M_{1} \right)
R_{i,1}\cdot\left(-\frac{dM_{1}}{d\tau_{M_{1}}}\right)
\right]_{M_{1}\left(\tau\right)}dt^{\prime}+ \nonumber\\
&+ R_{SNI}\cdot E_{SNI,i}+ \nonumber\\
&+\left[\frac{d}{dt}\mathcal{M}_{i}\left(r_{k},t \right)
\right]_{inf}-\left[\frac{d}{dt}\mathcal{M}_{i}\left(r_{k},t \right)
\right]_{out} \nonumber \\
&+\left[\frac{d}{dt}\mathcal{M}_{i}\left(r_{k},t \right)
\right]_{rf}
\end{flalign}
\end{small}
\vspace{-2pt}

\noindent where $\psi=\psi\left(r_{k},t^{\prime} \right)$,
$\phi=\phi\left( M \right)$, $\chi=\chi_{i}\left(r_{k},t \right)$,
$R_{i}=R_{i}\left(M\right)$, $R_{i,1}=R_{i}\left(M_{1}\right)$ and
$M\left(t-t^{\prime}\right)=M\left(\tau\right)$. The first term at
the r.h.s. is as usual the one describing the depletion of
interstellar material because of the process of star formation and
it depends from the star formation rate and from the abundance of
the $i$-th element considered. The next three terms  represent the
contribution of single stars to the enrichment of the $i$-th
element. The fifth term  is the contribution of the primary star in
a binary system (assumed to be independent from the secondary star
as far as it concerns the chemical yields). The sixth term is the
contribution of type Ia supernovae. Finally, the last three terms
are the infall rate of external gas, the outflow rate of matter due
for example to the onset of galactic winds powered by supernov{\ae}
explosions, and the radial flows of gas that determine the ISM
exchange between contiguous shells \citep{Portinari00},
respectively. Furthermore, $\textit{f}\left( M_{1} \right)$ is the
distribution function of the mass of the primary star $M_{1}$ in a
binary system, between $M_{1,min}=M_{B,l}/2$ and
$M_{1,max}=M_{B,u}$, where $M_{B,l}$ and $M_{B,u}$ are the lower and
upper limit of the binary systems assumed respectively $3M_{\odot}$
and $12M_{\odot}$. $R_{SNI}$ is the rate of type Ia SN{\ae} and
$E_{SNI,i}$ their ejecta of the $i$-th chemical species.
$M\left(\tau\right)=M\left(t-t^{\prime}\right)$ is the mass of a
star of lifetime $\tau$, born at $t^{\prime}$. It is worth noticing
that various quantities depend on the metallicity $Z\left( t\right)$
as well as on $M$:
$M\left(\tau\right)=M\left(t-t^{\prime}\right)=M\left(t-t^{\prime},
Z\left(t-t^{\prime}\right)\right)$ and
$R_{i}\left(M\right)=R_{i}\left(M,Z\left(t-t^{\prime}\right)\right)$
as stellar lifetimes and ejecta depend on metallicity. $R_i(M)$ are
calculated on the base of the detailed stellar yields from
\citet{Portinari98} and keep track of finite stellar lifetimes (no
instantaneous recycling approximation). Eqns. (\ref{GISM}) govern
the evolution of the ISM.

\textsf{Separating gas from dust}.  For our purposes we need to
formulate the equations governing the evolution of the dust in the
ISM. Separating the ISM in gas and dust, the equations governing the
evolution of the generic elemental species $i$-th in the  dust are

\vspace{-7pt}
\begin{small}
\begin{flalign} \label{DUST}
\frac{d}{dt}&D_{i}\left(r_{k},t \right)  = -\chi_{i}^{D}\psi+ \nonumber \\
&+ \int_{0}^{t-\tau_{M_{B,l}}}\psi\left[\phi
\delta^{w}_{c,i}R_{i}\cdot\left(-\frac{dM}{d\tau_{M}}\right)\right]
_{M\left(\tau\right)}dt^{\prime} +  \nonumber  \\
&+ \left(1-A
\right)\int_{t-\tau_{M_{B,l}}}^{t-\tau_{M_{SN{\ae}}}}\psi\left[\phi
\delta^{w}_{c,i}R_{i}\cdot\left(-\frac{dM}{d\tau_{M}}\right)\right]
_{M\left(\tau\right)}dt^{\prime} + \nonumber \\
&+ \left(1-A
\right)\int_{t-\tau_{M_{SN{\ae}}}}^{t-\tau_{M_{B,u}}}\psi\left[\phi
\delta^{II}_{c,i}R_{i}\cdot\left(-\frac{dM}{d\tau_{M}}\right)\right]
_{M\left(\tau\right)}dt^{\prime} +  \nonumber \\
&+ \int_{t-\tau_{M_{B,u}}}^{t-\tau_{M_{u}}}\psi\left[\phi
\delta^{II}_{c,i}R_{i}\cdot\left(-\frac{dM}{d\tau_{M}}\right)\right]
_{M\left(\tau\right)}dt^{\prime} +  \nonumber  \\
&+
A\int_{t-\tau_{M_{SN{\ae}}}}^{t-\tau_{M_{1},max}}\psi\left[\textit{f}
\left(M_{1}\right)\delta^{II}_{c,i}R_{i,1}\cdot\left(-\frac{dM_{1}}{d\tau_{M_{1}}}\right)\right]
_{M\left(\tau\right)}dt^{\prime} +  \nonumber   \\
&+
A\int_{t-\tau_{M_{1},min}}^{t-\tau_{M_{SN{\ae}}}}\psi\left[\textit{f}
\left(M_{1}\right)\delta^{w}_{c,i}R_{i,1}\cdot\left(-\frac{dM_{1}}{d\tau_{M_{1}}}\right)\right]
_{M\left(\tau\right)}dt^{\prime} +  \nonumber   \\
&+ R_{SNI}E_{SNI,i}\delta^{I}_{c,i}+ \nonumber  \\
&-\left[\frac{d}{dt}D_{i}\left(r_{k},t \right) \right]_{out}
+\left[\frac{d}{dt}D_{i}\left(r_{k},t \right) \right]_{rf}+
\nonumber \\
&+\left[\frac{d}{dt}D_{i}\left(r_{k},t \right) \right]_{accr}
-\left[\frac{d}{dt}D_{i}\left(r_{k},t \right) \right]_{SN}
\end{flalign}
\end{small}

\noindent where $\chi_{i}^{D}=\chi_{i}^{D}\left(r_{k},t\right)$. The
first term at the r.h.s. of Eqn. (\ref{DUST})  is the depletion of
dust because of star formation that consumes both gas and dust
(uniformly mixed in the ISM). The second term is the contribution by
stellar winds from low mass  stars to the enrichment of the $i$-th
component of the dust. Following \citet{Dwek98}, we introduce the
so-called condensation coefficients $\delta^{w}_{c,i}$ that
determines the fraction of material in stellar winds that goes into
dust with respect to that in gas (local condensation). The third
term is the contribution by stars not belonging to binary systems
and not going into type II SN{\ae} (the same coefficients
$\delta^{w}_{c,i}$ are used). The fourth term is the contribution by
stars not belonging to binary systems, but going into type II
SN{\ae}. For the condensation efficiency in the ejecta of type II
SN{\ae} we introduce the coefficients $\delta^{II}_{c,i}$, the
analog of $\delta^{w}_{c,i}$. The possible choices for these
coefficients are discussed in detail in \citet{Piovan11a}. The fifth
term is the contribution of massive stars going into type II
SN{\ae}. The sixth and seventh term represent the contribution by
the primary star of a binary system, distinguishing between those
becoming  type II SN{\ae} from those failing this stage and using in
each situation the correct coefficients. The eighth term is the
contribution of type Ia SN{\ae}, where again we introduced the
condensation coefficients $\delta^{I}_{c,i}$ to describe the mass
fraction of the ejecta going into dust. The last four terms
describe: (1) the  outflow of dust due to galactic winds (in the
case of disk galaxies this term can be set to zero); (2) the radial
flows of matter between contiguous shells; (3) the accretion term
describing the accretion of grain onto bigger particles in cold
clouds; (4) the destruction term taking into account the effect of
the shocks of SN{\ae} on grains, obviously giving a negative
contribution. The infall term in the case of dust can be neglected
because we can safely assume that the  material entering the galaxy
is made by gas only without a solid dust component mixed to it.\\
\indent Finally, from the equation for $D_{i}\left(r_{k},t \right)$
we can get the equation describing the evolution of the gaseous
component $G_{i}\left(r_{k},t \right)$, where $G_{i}\left(r_{k},t
\right)=\mathcal{M}_{i}\left(r_{k},t \right)-D_{i}\left(r_{k},t
\right)$:

\vspace{-7pt}
\begin{footnotesize}
\begin{flalign} \label{GAS}
\frac{\displaystyle d}{\displaystyle dt}&G_{i}\left(r_{k},t\right) =
-\chi_{G,i}\psi+ \nonumber \\
&+ \left(1-A
\right)\int_{t-\tau_{M_{B,l}}}^{t-\tau_{M_{SN{\ae}}}}\psi
\left[\phi\left(1-\delta^{w}_{c,i}\right)R_{i}\cdot\left(-\frac{dM}{d\tau_{M}}\right)\right]
_{M\left(\tau\right)}dt^{\prime} + \nonumber \\
&+ \left(1-A \right)\int_{t-\tau_{M_{SN{\ae}}}}^{t-\tau_{M_{B,u}}}
\psi\left[\phi\left(1-\delta^{II}_{c,i}\right)R_{i}\cdot\left(-\frac{dM}{d\tau_{M}}\right)\right]
_{M\left(\tau\right)}dt^{\prime} +  \nonumber \\
&+
\int_{0}^{t-\tau_{M_{B,l}}}\psi\left[\phi\left(1-\delta^{w}_{c,i}\right)R_{i}\cdot\left(-\frac{dM}{d\tau_{M}}\right)\right]
_{M\left(\tau\right)}dt^{\prime} +  \nonumber  \\
&+ \int_{t-\tau_{M_{B,u}}}^{t-\tau_{M_{u}}}\psi\left[\phi
\left(M\right)\left(1-\delta^{II}_{c,i}\right)R_{i}\cdot\left(-\frac{dM}{d\tau_{M}}\right)\right]
_{M\left(\tau\right)}dt^{\prime} +  \nonumber  \\
&+
A\int_{t-\tau_{M_{SN{\ae}}}}^{t-\tau_{M_{1},max}}\psi\left[\textit{f}
\left(M_{1}\right)\left(1-\delta^{II}_{c,i}\right)R_{i,1}\cdot\left(-\frac{dM_{1}}{d\tau_{M_{1}}}\right)\right]
_{M\left(\tau\right)}dt^{\prime} +  \nonumber   \\
&+
A\int_{t-\tau_{M_{1},min}}^{t-\tau_{M_{SN{\ae}}}}\psi\left[\textit{f}
\left(M_{1}\right)\left(1-\delta^{w}_{c,i}\right)R_{i,1}\cdot\left(-\frac{dM_{1}}{d\tau_{M_{1}}}\right)\right]
_{M\left(\tau\right)}dt^{\prime} +  \nonumber   \\
&+R_{SNI}E_{SNI,i}\left(1-\delta^{I}_{c,i}\right)
+\left[\frac{d}{dt}G_{i}\left(r_{k},t \right) \right]_{inf}+\nonumber  \\
&-\left[\frac{d}{dt}G_{i}\left(r_{k},t \right) \right]_{out}
+\left[\frac{d}{dt}G_{i}\left(r_{k},t \right) \right]_{rf}+
\nonumber \\
&-\left[\frac{d}{dt}D_{i}\left(r_{k},t \right) \right]_{accr}
+\left[\frac{d}{dt}D_{i}\left(r_{k},t \right) \right]_{SN}
\end{flalign}
\end{footnotesize}
\vspace{-3pt}

\noindent where again the outflow term
$-\left[\frac{d}{dt}G_{i}\left(r_{k},t \right) \right]_{out}$ will
be fixed to zero because we do not have galactic wind for spirals
with continuous star formation. Since the primordial material is
likely dust-free we have $\left[\frac{d}{dt}\mathcal{M}_{i}\left(r,t
\right) \right]_{in}=\left[\frac{d}{dt}G_{i}\left(r,t \right)
\right]_{in}$.\\
\indent It is worth noticing the following point: the stellar
models, upon which are based our yields, predict that stars with
mass higher than $6 M_{\odot}$ go into SN{\ae}, whereas those with
mass lower than $6 M_{\odot}$ first become   AGB stars and later
White Dwarfs. We must therefore  split the third and the fifth
member of Eqn. (\ref{GISM}) in two parts, both in Eqns. (\ref{GAS})
and (\ref{DUST}), because the minimum mass dividing the intervals of
AGB and/or SN{\ae} belongs to the mass interval $\left(3-12
M_{\odot}\right)$ describing binary systems going into type Ia
SN{\ae}\footnotemark[1].

\footnotetext[1]{For example in Eqn. (\ref{DUST}), using
$\delta_{c,i}$ to indicate the generic condensation coefficient we
have the following split:
\begin{equation}
\left(1-A \right)\int_{t-\tau_{M_{B,l}}}^{t-\tau_{M_{B,u}}}\psi
\left[\phi
\delta_{c,i}R_{i}\left(M\right)\left(-\frac{dM}{d\tau_{M}}\right)
\right]_{M\left(\tau\right)}dt^{\prime} \nonumber
\end{equation}
\noindent is divided into:
\begin{flalign}
&\left(1-A
\right)\int_{t-\tau_{M_{B,l}}}^{t-\tau_{M_{SN{\ae}}}}\psi\left[\phi
\delta^{w}_{c,i}R_{i}\left(-\frac{dM}{d\tau_{M}}\right)\right]
_{M\left(\tau\right)}dt^{\prime} \nonumber \\
&+ \left(1-A
\right)\int_{t-\tau_{M_{SN{\ae}}}}^{t-\tau_{M_{B,u}}}\psi\left[\phi
\delta^{II}_{c,i}R_{i}\left(-\frac{dM}{d\tau_{M}}\right)\right]
_{M\left(\tau\right)}dt^{\prime} \nonumber
\end{flalign}
\noindent where $M_{SN{\ae}}$ is the separation mass that tell us if
we must use the condensation coefficients of stellar winds
(condensation of dust in the envelopes of AGB stars) in the mass
interval between $M_{B,l}$ and $M_{SN{\ae}}$ or the condensation
coefficients of supernov{\ae} between $M_{SN{\ae}}$ e $M_{B,u}$. In
the same way:
\begin{equation}
A\int_{t-\tau_{M_{1,min}}}^{t-\tau_{M_{1,max}}}\psi
\left[\textit{f}\left( M_{1} \right)\delta_{c,i}
R_{i,1}\left(-\frac{dM_{1}}{d\tau_{M_{1}}}\right)
\right]_{M_{1}\left(\tau\right)}dt^{\prime} \nonumber
\end{equation}
\noindent splits itself into:
\begin{flalign}
&A\int_{t-\tau_{M_{SN{\ae}}}}^{t-\tau_{M_{1},max}}\psi\left[\textit{f}
\left(M_{1}\right)\delta^{II}_{c,i}R_{i,1}
\left(-\frac{dM_{1}}{d\tau_{M_{1}}}\right)\right]
_{M\left(\tau\right)}dt^{\prime}  \nonumber   \\
&+
A\int_{t-\tau_{M_{1},min}}^{t-\tau_{M_{SN{\ae}}}}\psi\left[\textit{f}
\left(M_{1}\right)\delta^{w}_{c,i}R_{i,1}
\left(-\frac{dM_{1}}{d\tau_{M_{1}}}\right)\right]
_{M\left(\tau\right)}dt^{\prime}. \nonumber
\end{flalign}}

\textsf{To summarize}. Indicating the contribution to the yields by
stellar winds and type Ia and II SN{\ae}  with the symbols
$W_{i,D}\left(r_{k},t\right)$, $W_{i,G}\left(r_{k},t\right)$ and
$W_{i,\mathcal{M}}\left(r_{k},t\right)$ (they can easily be
reconstructed by comparison) and  neglecting the outflow term, Eqns.
(\ref{GISM}), (\ref{DUST}) and (\ref{GAS}) become:

\vspace{-7pt}
\begin{flalign} \label{GISM_B}
\frac{d}{dt}&\mathcal{M}_{i}\left(r_{k},t \right)  =
-\chi_{i}^{\mathcal{M}}\left(r_{k},t \right)\psi\left(r_{k},t
\right)+ \nonumber+W_{i,\mathcal{M}}\left(r_{k},t\right) \nonumber \\
&+\left[\frac{d}{dt}\mathcal{M}_{i}\left(r_{k},t \right)
\right]_{rf}
\end{flalign}
\vspace{-7pt}

\vspace{-7pt}
\begin{flalign} \label{DUST_B}
\frac{d}{dt}&D_{i}\left(r_{k},t \right)  =
-\chi_{i}^{D}\left(r_{k},t \right)\psi\left(r_{k},t \right)+
\nonumber+W_{i,G}\left(r_{k},t\right) \nonumber \\
&+\left[\frac{d}{dt}D_{i}\left(r_{k},t \right) \right]_{accr}
-\left[\frac{d}{dt}D_{i}\left(r_{k},t \right) \right]_{SN} \nonumber
\\ &+\left[\frac{d}{dt}D_{i}\left(r_{k},t \right) \right]_{rf}
\end{flalign}
\vspace{-7pt}

\vspace{-7pt}
\begin{flalign} \label{GAS_B}
\frac{\displaystyle d}{\displaystyle dt}&G_{i}\left(r_{k},t\right)
= -\chi_{G,i}\left(r_{k},t \right)\psi\left(r_{k},t \right)+W_{i,G}\left(r_{k},t\right) \nonumber \\
&-\left[\frac{d}{dt}D_{i}\left(r_{k},t \right) \right]_{accr}
+\left[\frac{d}{dt}D_{i}\left(r_{k},t \right) \right]_{SN} \nonumber
\\ &+\left[\frac{d}{dt}G_{i}\left(r_{k},t \right) \right]_{inf}
+\left[\frac{d}{dt}G_{i}\left(r_{k},t \right) \right]_{rf}.
\end{flalign}
\vspace{-2pt}

\noindent It is soon evident that  the  dust creation/destruction
and the radial flows  make the system of differential equations more
complicated than the original one by \citet{Talbot75} for a one-zone
closed-box model. As the  ISM is given by the sum of gas and dust,
only two of these equations are required, furthermore Eqn.
(\ref{GISM_B}) can be used only if gas and dust flow with the same
velocity. To  proceed further we must now specify the law of star
formation, the IMF, the stellar ejecta and the various rates
describing gas infall, dust accretion/destruction, and radial
flows/bar effect.  No details will be given about these ones. They
are included into the model and they are mainly useful in order to
reproduce  the radial gradients of abundance in the MW. The reader
should refer to \citet{Portinari00} and \citet{Piovan11c}.

\section{The Star Formation Laws and Initial Mass Functions}
\label{SFRandIMF}

\textsf{Star Formation}. The law of Star Formation (SF) is a key
ingredient of any model of galaxy formation and evolution.
Unfortunately it is poorly known, so that many prescriptions for the
SF rate can be found in literature. In this study we have considered
several well known SF laws adopted for the MW \citep[see][\, for
details]{Portinari99}.

A very popular prescription is the  \citet{Schmidt59} law. In our formalism
it becomes:

\begin{footnotesize}
\begin{equation}
\Psi(r_{k},t) = - \left[ \frac{d G(r_{k},t)}{dt} \right]_*  = \nu \,
\left[ \frac{\sigma(r_{k},t_G)}{\sigma(r_{\odot},t_G)}
\right]^{\kappa-1} \, G^{\kappa}(r_{k},t) \label{SFSchmidt}
\end{equation}
\end{footnotesize}

\noindent where the  normalization factor is
$\sigma(r_{\odot},t_G)^{-(\kappa-1)}$ and $\nu$ is in $[t^{-1}]$.
Following \citet[][]{Portinari99} we adopt $\kappa=1.5$.

This simple dependence of the SFR can be complicated by including
other physical effects. For instance, the SF suited to spiral
galaxies such as the MW, may include the effect of  gas compression
by density waves \citep{Roberts69,Shu72,Wyse89,Prantzos98} or
gravitational instabilities \citep{Wang94}. We have:

\begin{footnotesize}
\begin{equation}
\label{SFWyse} \Psi(r_{k},t) = \nu \left[\frac{r}{r_{\odot}}
\right]^{-1} \, \left[
\frac{\sigma(r_{k},t_G)}{\sigma(r_{\odot},t_G)} \right]^{\kappa-1}
G^{\kappa}(r_{k},t)
\end{equation}
\end{footnotesize}

\noindent  where $\nu$ is always in $[t^{-1}]$ and  $k=1$
\citep{Kennicutt98,Portinari99}.

Another possibility is to describe the SF as a balance between
cooling and heating processes, that is the gravitational settling of
the gas onto the Disk and the energy injection from massive stars
\citep{Talbot75,Dopita85,Dopita94}. In our formalism, we have

\begin{footnotesize}
\begin{equation}
\label{SFChiappini} \Psi(r_{k},t) = \nu \left[ \frac
{\sigma^n(r_{k},t) \, \sigma^{m-1}(r_{k},t_G)}
{\sigma(r_{\odot},t_G)^{n+m-1}} \right] \, G^m(r_{k},t)
\end{equation}
\end{footnotesize}

\noindent where $n=1/3$, $m=5/3$ \citep{Portinari99} and $\nu$ in
$[t^{-1}]$. This formulation is similar to the original one by
\citet{Talbot75} thus leading to similar results
\citep{Portinari98}.


\textsf{Initial Mass Function}.  The initial mass function (IMF) is
perhaps the most important ingredient of chemical models of any kind
\citep[see][for a recent review of the subject]{Kroupa02a}. Brown
dwarfs and very low mass stars whose lifetimes are longer than the
age of the Universe, in practice lock up forever the chemical
elements present in the ISM at the age of their birth, whereas
intermediate and high mass stars of short life continuously enrich
the environment with the products of thermonuclear reactions, thus
driving the chemical evolution of the host system.  They are also
the factories of star-dust to be injected into the ISM by  SN{\ae}
explosions and strong stellar winds. The adoption of an IMF has two
effects worth being mentioned here. First of all, a different slope
of the IMF in the intermediate-high mass range would imply a
different relative population  of the stars  contributing  to the
dust yields. Second, the net yield of metals and dust per stellar
generation varies. According to its definition \citep[see for
instance][]{Tinsley80b,Pagel97,Portinari04b} the net yield is the
the amount of metals globally produced by a stellar generation over
the the fraction of mass locked up in living stars and remnants.
Therefore, efficiency of metal and dust enrichment depends not just
on the amount of metals produced per unit mass involved in star
formation but on the ratio between this and the mass that remains
locked in remnants or ever-lived low mass stars. The locked-up
fraction, is therefore as crucial to the metal and dust enrichment
as is the absolute number of the high-mass stars directly
responsible for the production of dust itself. In a given model of
fixed total mass, it is clear that an IMF  bending down steeply at
low masses  will lead to a different locked up fraction with respect
to a power-law, low-mass oriented IMF. The metal and dust production
is accordingly affected.\\
\indent In this section we shortly present the IMFs we have included
in our model of the Galactic Disk and SoNe. For the purposes of our
study the IMF is assumed to be constant in time and space. All the
IMFs are normalized assuming that the total mass encompassed by the
IMF from the lower, M$_L$, to the upper, M$_U$, mass limit of stars
is equal to 1 M$_\odot$. To this aim, following \citet{Talbot75} and
\citet{Bressan94}, we define the parameter $\zeta$, which describes
the fraction of total mass in form of stars stored in the IMF above
a given mass M$_{*}$. In other words, M$_{*}$ is  the minimum mass
contributing to the nucleo-synthetic enrichment of the ISM over a
timescale of the order of the galaxy life

\begin{equation}\label{zita}
\zeta = \frac{\int^{M_{U}}_{M_{*}}\phi \left( M \right)
dM}{\int^{M_{U}}_{M_{L}}\phi \left( M \right) dM}.
\end{equation}

This equation, at varying $\zeta$ and for fixed M$_{U}$ and M$_{*}$,
can be reversed numerically to determine the lower limit of the
distribution M$_{L}$.  The IMFs included in our model are:

- \textsf{The Salpeter IMF}.  Salpeter-like IMFs are very popular.
These are an extension over the desired mass range of the original
Salpeter IMF \citep{Salpeter55}. This IMF is $\phi
\left(M\right)=C_{S}M^{-1.35}$ with $C_{S}$ depending on the value
of $\zeta$. For a mass range $\left[0.1-100 \right]M_{\odot}$
\citep{Portinari04b} we have $C_{s}=0.1716$ and a $\zeta=0.3925$.

- \textsf{The Kroupa IMF}. In a series of papers Kroupa revised and
updated the power-law IMF with a set of continuous multi-slope
power-laws
\citep[e.g.]{Kroupa93,Kroupa01,Kroupa02a,Kroupa02b,Kroupa07}. In the
following we consider two cases. First, the  IMF derived by
\citet{Kroupa98} for field stars in the SoNe and used by
\citet{Portinari04a}. \noindent This IMF is typical of models of
chemical evolution of disk galaxies
\citep{Boissier99,Boissier00,Prantzos00,Hou08} \footnotemark[2]
\footnotetext[2]{ Along this line it worth recalling that IMF with
slope $\left(1.6\sim 1.7 \right)$ in the high mass range, i.e.
steeper than the Salpeter value, is from \citet{Scalo86} and it is
widely used in literature in chemical models of the MW even with
dust
\citep{Matteucci89,Chiappini97a,Dwek98,Romano00,Francois04,Calura08}}.
For $M_{L}=0.1M_{\odot}$ and $M_{U}=100M_{\odot}$, we obtain
$\zeta=0.405$. Second, the \citet{Kroupa07} IMF, where taking
$M_{L}=0.01M_{\odot}$ and $M_{U}=100M_{\odot}$ we get  $\zeta =
0.38$, slightly lower than in the above \citet{Kroupa98} because of
the lower limit extended to brown dwarf regime.

- \textsf{The Larson IMF}. \citet{Larson98} proposed an IMF in which
the relative percentage of very low mass stars and sub-stellar
objects is decreased due to the presence of  an exponential cut-off.
As a consequence of this there is a negligible contribution to the
locked-up mass, and in contrast a very high net yield per stellar
generation, and a high production of metals and dust. The
\citet{Larson86} IMF is

\begin{equation}\label{LarsonOriginalIMF}
\phi \left(M\right)=C_{L}M^{-1.35}\exp
\left(-\frac{M_{L}}{M}\right).
\end{equation}

\noindent In practice this IMF recovers the Salpeter IMF at high
masses, whereas at low masses the exponential cut-off determines the
steep downfall after the peak mass $M_{P}=M_{L}/1.35=0.25
M_{\odot}$. For a typical mass range $[0.01-100]M_{\odot}$ we have
$\zeta=0.653$ thus allowing for a high number of intermediate-high
mass stars. For the present aims, we will simply keep $M_{L}$
constant with time or metallicity. We also consider the possibility
for a modified Larson IMF adapted to the SoNe, in which the slope in
the power-law factor for the high mass range has the value
$M^{-1.7}$, according to \citet{Scalo86} and in agreement with
recent IMFs proposed by Kroupa (see above). In this last case with
the same mass range $[0.01-100]$, we have $\zeta=0.5$, lower than
the other case.

- \textsf{The Chabrier IMF}. Along the line of thought of
\citet{Larson98}, \citet{Chabrier01} proposes:

\begin{equation}\label{ChabrierIMF}
\phi \left(M\right)=C_{C}M^{-2.3}\exp
\left[\left(-\frac{M_{C}}{M}\right)^{1/4}\right].
\end{equation}

\noindent The two parameters in equation (\ref{ChabrierIMF}) are
tuned on local field low mass stars and the functional form is
proved to be  o be valid down to the brown dwarfs regime
\citep{Chabrier02a}. With a mass range $[0.01-100]M_{\odot}$ and
$M_{C}=716.4$ we get $C_{C}=40.33$ and  $\zeta=0.545$, not as high
as in the Larson IMF, but still leading  to high net yield and low
locked up mass fraction.

- \textsf{The Kennicutt IMF}.
The \citet{Kennicutt83} IMF is used in literature to describe
the global properties of spiral galaxies and it is inspired by the
observations of $H\alpha$ luminosities and equivalent width in
external galaxies \citep{Kennicutt94,Portinari04a,SommerLarsen96}.
\noindent With a mass range $[0.1-100]M_{\odot}$, we get
$\zeta=0.59$.

- \textsf{The Arimoto IMF}. This top-heavy IMF has been suggested by
\citet{Arimoto87} to simulate elliptical galaxies and is introduced
just for the sake of comparison as an extreme case: $\phi
\left(M\right)=C_{Ari}M^{-1.0}$. With a mass range
$[0.1-100]M_{\odot}$, we get $\zeta=0.5$.

\section{Growth dust grains in the ISM}
\label{graingrowth}

Dust grains form by a number of physical processes (see below) and
once in the ISM they may grow in mass by accreting a number of atoms
or molecules. In the following we examine in some detail
the accretion mechanisms for which two different models are
proposed.

\subsection{The Dwek (1998) model or case A} \label{modelA}

\citet{Dwek98} view of grain growth can be summarized as follows.
Let us consider an ISM whose dust component is formed by grains made
of a single element  $i$-th  with mass  $m_{gr,i}$. Let $m_{i}$ be
the mass of an atom (or molecule) of the element $i$-th in the
gaseous phase. $N_{i}$ is the number of atoms (molecules)  of the
element $i$-th locked up into the mono-composition grains that we
are considering and $n_{gr,i}$ the density of grains of $i$-th type
in the ISM, considered in this model as \textit{a single phase},
without distinction between diffuse ISM and cold molecular clouds.
Let now $\alpha_{i}$ be the sticking coefficient, telling us the
probability that an atom (molecule)  of the element $i$ in the
gaseous phase binds to the grains increasing the number of
atoms/molecules on it. Finally $n_{gas,i}$ is the number  density
the gas made of atoms (molecules)  of type $i$-th and $\displaystyle
\overline{v}=\left(\frac{\displaystyle 8K_{B}T}{\displaystyle \pi
m_{i}}\right)^{\frac{1}{2}}$ is the mean thermal velocity of the
particles of the element $i$ in the gaseous phase with respect to
the dust grains. We get:

\vspace{-7pt}
\begin{equation}\label{Rate_Numerica}
\frac{dN_{i}}{dt}=\alpha_{i} \pi a^{2} n_{gr,i}  n_{gas,i}
\overline{v} \, .
\end{equation}
\vspace{-7pt}

\noindent Multiplying both members of the equation by the constant
mass $m_{i}$, multiplying and dividing the second member by the
mass of one grain $m_{gr,i}$, introducing the mean thermal
velocity $\overline{v}$ and using the following relation
\footnotemark[3] \citep{Dwek98}:

\vspace{-7pt}
\begin{equation}\label{MC_density}
m_{i} n_{gas,i} = 2 m_{H}  n_{H_{2}} \left(1 - \frac{\sigma_{i}^{D}
}{\sigma_{i}^{\mathcal{M}} }\right)
\end{equation}
\vspace{-7pt}

\noindent where $\sigma_{i}^{D}\left(r,t\right)$ and
$\sigma_{i}\left(r,t\right)$ are functions of position and time, we obtain:
\footnotetext[3]{The
above relation  says that the mass of the element $i$-th in the
gaseous phase can be expressed by means of the density of the
molecular clouds, i.e. the site in which the grains grow. $2 m_{H}
n_{H_{2}}$ is the mass of the molecular cloud, $\left(1 -
\displaystyle \sigma_{i}^{D} /\displaystyle \sigma_{i}^{\mathcal{M}}
\right)$ is the fraction of the element $i$-th in the gaseous phase
as a function of the abundance of the same element in the dust and
ISM.}

\vspace{-7pt}
\begin{equation}
\frac{d \sigma_{i}^{D} }{dt}= \frac{\sigma_{i}^{D} }{\tau_{i,0}}
\left(1 - \frac{\sigma_{i}^{D} }{\sigma_{i}^{\mathcal{M}}}\right).
\end{equation}
\vspace{-7pt}

\noindent For the dimensional consistency of the equation, the
multiplying quantity has the
dimension of the inverse of a time indicated by $\tau_{i,0}$

\vspace{-7pt}
\begin{equation}
\frac{1}{\tau_{i,0}}= 2 n_{H_{2}}\alpha_{i} \frac{\pi a^{2}}{A_{i}}
\left( \frac{8K_{B}T}{\pi m_{i}}\right)^{\frac{1}{2}} \, .
\end{equation}
\vspace{-7pt}

It may be worth of interest to give an estimate of the involved
timescale: by means of typical values of the involved quantities,
\citet{Dwek98} obtains $\thicksim 3 \times 10^{4}$ yrs. It is
possible now to define the accretion time scale of our $i$-th
element onto the grains as:

\vspace{-7pt}
\begin{equation}\label{tau_accrescimento}
\frac{1}{\tau_{i,accr}}=\frac{1}{\tau_{0,i}} \cdot \left(1 -
\frac{\sigma_{i}^{D} }{\sigma_{i}^{\mathcal{M}}} \right)
\end{equation}
\vspace{-4pt}

\noindent that is the inverse of $\tau_{0,i}$ multiplied by the
fraction of the $i$-th element in the gaseous phase. Again, we drop
the dependence from $r$ and $t$.  Dwek (1998) adopting the ratio
$\sigma_{i}^{D}/ \sigma_{i}^{\mathcal{M}} \thickapprox 0.7$  gets
$\tau_{i,accr} \thickapprox 10^{5}$ yrs  which turns out to be
significantly shorter than  the lifetime of molecular clouds where
accretion takes place. This lifetime is estimated to be  of the
order of $t_{MC} \thicksim 2-3 \times 10^6 - 10^{7}$ yrs. The
lifetime of a molecular cloud is comparable to the lifetime of the
most massive stars  born in it. These stars indeed injecting great
amounts of energy by  stellar winds and SN{\ae} explosions
eventually disrupt the cloud. Similar estimates have been made in
studies on the temporal evolution of dust in the ISM by
\citet{Calura08},  $\thicksim 5 \cdot 10^{7}$ yrs, and
\citet{Zhukovska08},  $\thicksim 10^{7}$ yrs, and are consistent
with  observational and theoretical estimates of the lifetimes of
molecular clouds  \citep[see e.g.]{Matzner02,Krumholz06,Blitz07}.
One may argue that we are  comparing two  different timescales, i.e.
with $t_{i,accr}\ll t_{MC}$. However,  inside molecular clouds there
are many  physical processes  continuously stripping atoms and  and
molecules from the grains, like UV radiation and cosmic rays. The
net effect of it  could be that   $\tau_{i,accr}$ is considerably
lengthened  \citep{Dwek98}, likely up to  $6 \cdot 10^{7}$ yrs, so
that  $t_{i,accr}\simeq t_{MC}$. Furthermore, all the dust grains
are in cold molecular clouds where the growing occurs. The simplest
way to take this into account is to divide  the above estimate for
$t_{i,accr}$ by the  fraction of dust in molecular clouds. The
resulting   accretion  timescale would be $t_{i,accr} \thicksim 1-2
\times 10^{8}$ yrs.\\
\indent There is another important point to consider on which indeed
several studies of the ISM evolution are based: both the destruction
timescale (driven by  SN{\ae} explosions)  and the accretion
timescale  $t_{i,accr}$ have the same dependence on the star
formation rate $\psi $ and surface mass density $\sigma_{i} $.
SN{\ae}  rate  can be expressed as $\tau_{i,snr} \propto \sigma
/\psi $. The lower the star formation rate, the lower is  the number
of supernovae explosions and finally the longer the destruction time
of grains due to  shocks. We clearly expect  $\tau_{i,accr}$ to vary
over the evolutionary history  of the Galaxy: \citet{Dwek98} divides
the accretion timescale  by $\sigma^{MC} / \sigma $ and reasonably
assumes that $\psi \propto \sigma^{MC} $. In this way,
$\tau_{i,accr} \propto \displaystyle \sigma \displaystyle \psi $.
Since both $\tau_{i,accr}$ and $\tau_{i,snr}$ have the same
dependence from $\displaystyle \sigma^{ISM} /\displaystyle \psi  $
\citet{Dwek98} suggests that the ratio between the two timescales is
constant: $\tau_{i,accr}/\tau_{i,snr} =1/2$. \noindent If now we
consider the accretion/destruction terms in eqn. (\ref{DUST_B}) we
have:

\vspace{-7pt}
\begin{eqnarray}\label{Eq_chimica}
\left( \frac{d \sigma_{i}^{D} }{dt}\right)_{accr}&+&\left( \frac{d
\sigma_{i}^{D} }{dt}\right)_{snr} = \frac{\sigma_{i}^{D}
}{\tau_{i,accr} }
-\frac{\sigma_{i}^{D} }{\tau_{i,snr}}= \nonumber \\
&=& 2\cdot\frac{\sigma_{i}^{D} }{\tau_{i,snr} }-\frac{\sigma_{i}^{D}
}{\tau_{i,snr}} = \frac{\sigma_{i}^{D} }{\tau_{i,snr} }\, .
\end{eqnarray}
\vspace{-3pt}

\noindent The contribution of the  creation/destruction terms is
greatly simplified and the numerical solution of the differential
equations describing the dust evolution becomes very simple. More
refined is the solution adopted by \citet{Calura08}. In brief, (i)
only $\tau_{0,i}$ is fixed and  $\tau_{i,accr}$ remains an explicit
function of the ratio $\displaystyle \sigma_{i}^{D}/\displaystyle
\sigma_{i}$; (ii) no \textit{a priori} correlation between
$\tau_{i,accr}$ and $\tau_{i,snr}$ is supposed to exist, the two
timescales are defined independently, and both plays a role in the
$\sigma_{i}^{D}$ evolution.  In our model we adopt the same
strategy.

\subsection{The Zhukovska et al. (2008) model or case B} \label{modelB}

A step forward in modelling grain accretion has been made by
\citet{Zhukovska08}. In brief,  \citet{Zhukovska08}'s picture  can
be summarized as follows: (i) grain accretion is tightly related to
molecular clouds and a multi-phase description of the ISM would be
required; (ii) the evolution of chemical elements is calculated in
presence of  some typical dust compounds that are  representative of
the grain families growing inside the cold regions of the ISM or
ejected by the stars, namely  silicates, carbonaceous grains, iron
grains, and silicon carbide. This model has two important
advantages. First,  it allows for different physical situations each
of which implying  different timescales for each elements and
different relationships  between $\tau_{i,accr}$ and $\tau_{i,snr}$
which are independent; second, accretion in cold regions is
described in a realistic way; third, the model can be easily
incorporated in a multi-phase description of the ISM to estimate the
amounts of gas in the warm and cold phases.

Let us now examine some of the key points and basic equations of the
model and how they are  adapted  to our description, where contrary
to what made in \citet{Zhukovska08} we track the evolution
\textit{abundances of single elements in dust} instead of
\textit{ad-hoc} families of dust grains.

The growth of  the generic type of grains $j$-th  is driven by the
least abundant element indicated as element $i$-th (otherwise called
the \textit{key element}) among those forming the  grain.   All the
other elements concurring to form the $j$-th compound adapt their
abundances to that of the \textit{key element}.  For a mixture of
accreting grains made of silicates (pyroxenes and olivines, i.e.
magnesium-iron sulfates), carbonaceous and iron grains, the key
elements are Mg, Si or Fe for silicates, depending on which has the
lowest abundance, C for carbon grains, and Fe for iron grains. Most
of these elements accrete some specific atomic or molecular species
called \textit{growth species}. The growth in mass of a single
$j$-th type grain is:

\vspace{-7pt}
\begin{equation}\label{crescita_jesimoA}
\frac{dm_{j}}{dt}= S_{j} \alpha_{j}  v_{j}^{gw} A_{j}^{D}m_{H}
\frac{\nu_{i,j}^{gw}}{\nu_{i,j}^{D}} n_{j}^{gw}
\end{equation}
\vspace{-7pt}

\noindent where $m_{j}$ is the mass of one grain of the $j$-th type,
$S_{j}$ is the surface area, $\alpha_{j}$ is the sticking
coefficient, $v_{j}^{gw}$ is the relative velocity of the accreting
species with respect to the grain, $A_{j}^{D}m_{\mathrm{H}}$ is the
mass of one atom (or molecule) that is accreting on the grain of
type of type $j$-th, $n_{j}^{gw}$ is the number density of the
growth species $j$, and finally $\nu_{i,j}^{gw}/\nu_{i,j}^{D}$ is
the ratio between the number of atoms of the key element $i$-th in
the growth species for  the dust grain of type $j$-th  and the
number of atoms of $i$-th element  we need to build one  grain of
the $j$-th type.

After some substitutions  and simplifications for which all the
details can be found in \citet{Zhukovska08}, it comes out:





\vspace{-7pt}
\begin{equation}\label{crescita_jesimoH}
\frac{df_{i,j}}{dt}= \frac{1}{\tau_{j}^{gr}}f_{i,j}\left(1-f_{i,j}
\right)
\end{equation}
\vspace{-7pt}

where $f_{i,j} \varpropto \rho_{j}^{D}$ is the fraction of the key element $i$-th
that is already condensed onto the dust
type $j$-th, called degree of condensation. $1-f_{i,j}$ is therefore
the fraction of the key element $i$-th still in the gaseous phase
and available to form growth species for the $j$-th dust component.
\noindent An explicit formula for $\tau_{j}^{gr}$ as a function of
all the relevant quantities  \citep[see Eqn. 31]{Zhukovska08} is:

\vspace{-5pt}
\begin{small}
\begin{equation}\label{tau_ZhukovskaB}
\tau_{j}^{gr}=46  \frac{(A^{gw}_{j})^{1/2} \nu_{i,j}^{D}}{A_{j}^{D}}
\left(\frac{\rho_{j}^{C}}{3}\right) \left(
\frac{10^{3}}{n_{H}}\right) \left(\frac{3.5 \cdot
10^{-5}}{\epsilon_{i,j}}\right)\mathrm{Myr}\, .
\end{equation}
\end{small}
\vspace{-5pt}

Integrating eqn. (\ref{crescita_jesimoH}) with the initial
conditions $f_{i,j}=f_{0,i,j}$ for $t=0$ we get the final equation
describing the evolution of the condensed fraction:

\vspace{-7pt}
\begin{equation}
f_{i,j}\left(t\right)=\frac{f_{0,i,j}e^{t/\tau_{j}^{gr}}}{1-f_{0,i,j}+
f_{0,i,j}e^{t/\tau_{j}^{gr}}}\, .
\end{equation} \label{fij-evol}
\vspace{-7pt}

\noindent As in real conditions the grain accretion takes place in
cold MCs with a finite lifetime $\tau_{MC}$, the above accretion
time scale must be compared with $\tau_{MC}$. As already discussed
for model A of accretion, the molecular lifetime constrains the real
time interval during  which the exchange of matter between cold
clouds and surrounding medium can happen. It is the interplay
between the accretion/disruption timescales to determine if and how
much material can be condensed inside the cloud before it dissolves
into the surrounding ISM. If $f_{0,i,j}$ is the initial degree of
condensation of the key element $i$-th of the grains $j$-th and the
molecular cloud will be destroyed after a lifetime $t$ we have that
the effective amount of material that is given back to the ISM after
the cloud dissolution is:

\vspace{-7pt}
\begin{align} \label{MjMC}
M_{j}^{MC}\left(t\right)=\left(f_{i,j} \left(t\right)-f_{0,i,j}
\right) \chi_{j,max}^{D} M_{MC}
\end{align}
\vspace{-7pt}

\noindent where $f_{i,j} \left(t\right)-f_{0,i,j}$ is the fraction
of key element condensed with respect to the initial condensation
fraction, $\chi_{j,max}^{D}$ is the maximum possible fraction of
dust of the kind $j$-th that can be formed in the molecular cloud
and that would exist  if all the material is condensed in   dust of
type $j$-th. Multiplying by  the mass of the cloud we get the mass
of newly formed material $M_{j}^{MC}$ returned to the ISM. The
fraction $\chi_{j,max}^{D}$ is given by

\vspace{-7pt}
\begin{equation}\label{chi}
\chi_{j,max}^{D}=\frac{A_{j}^{D}\epsilon_{i,j}}{\left(1+4\epsilon_{He}\right)\nu_{i,j}^{D}}
\end{equation}
\vspace{-7pt}

\noindent as shown in the footnote below
\footnotemark[4].

\footnotetext[4]{The number density  $n_{\mathrm{H}}$  of hydrogen
atoms is given by $n_{\mathrm{H}}=\displaystyle
\rho_{MC}/\left[\displaystyle
\left(1+4\epsilon_{He}\right)m_{\mathrm{H}}\right]$ which in turn
follows from the series  of equalities:
$\rho_{MC}=m_{\mathrm{H}}n_{\mathrm{H}}+4m_{\mathrm{H}}n_{\mathrm{He}}=
m_{\mathrm{H}}n_{\mathrm{H}}+4m_{\mathrm{H}}n_{\mathrm{H}}\epsilon_{\mathrm{He}}=
m_{\mathrm{H}}n_{\mathrm{H}}(1+4\epsilon_{He})$, where
$\epsilon_{He}=n_{\mathrm{He}}/n_{\mathrm{H}}$. The fraction
$\chi_{j,d,max}$ is given by $\chi_{j,max}^{D}=\frac{\displaystyle
\rho_{j,max}^{D}}{\displaystyle \rho_{MC}}=\frac{\displaystyle
A_{j}^{D}m_{\mathrm{H}}(\epsilon_{i,j}n_{\mathrm{H}})}
{\displaystyle \nu_{i,j}^{D} \cdot
m_{\mathrm{H}}n_{\mathrm{H}}(1+4\epsilon_{He})}=\frac{\displaystyle
A_{j}^{D}\epsilon_{i,j}} {\displaystyle
(1+4\epsilon_{He})\nu_{i,j}^{D}}$. The explanation is quite simple:
dividing the number of atoms $\epsilon_{i,j}n_{\mathrm{H}}$ of the
key element available in the molecular cloud by the number of atoms
$\nu_{i,j}^{D}$ of the same key element that are tied up when we
form one dust grain, we get the maximum number of dust grains that
we can form per unit volume.  Knowing the mass of a single dust
grain, $A_{j}^{D}m_{\mathrm{H}}$, the the mass density $
\rho_{j,max}^{D}$ immediately follows. }

Not all the MCs have the same lifetime: therefore there is a certain
probability  $P\left( t \right)$  that a molecular cloud is
destroyed in the  time interval  $t$ -- $t+dt$. According to
\citet{Zhukovska08} the probability is well represented by an
exponential law with time scale $\tau_{MC}$. The instantaneous
condensation fraction at the time $t$ of the key element $i$-th
relative to the dust compound $j$-th is derived from integrating
over time and considering for each cloud of lifetime $x$, born at
$t-x$ the  condensation fraction holding at the time of  birth
$f_{0,i,j}\left(t-x\right)$. Since the lifetime distribution
function quickly decays over time, only the MCs born some
$\tau_{MC}$ before $t$ will contribute in practice. The
instantaneous condensation fraction is

\vspace{-7pt}
\begin{equation} \label{fijintegralinst}
f_{i,j}(t)=\frac{1}{\tau_{MC}}\int_{0}^{t}
\frac{f_{0,i,j}(t-x)e^{x/\tau_{j}^{gr}}e^{-x/\tau_{MC}}}{1-f_{0,i,j}(t-x)(1-
e^{x/\tau_{j}^{gr}})}dx\, .
\end{equation}
\vspace{-3pt}

This procedure is however time consuming when applied in practice. A
much simpler approach is  provided by  using the average restitution
mass $\overline{R}_{j}^{MC}$ and the average condensation fraction
$\overline{f}_{i,j}$ assuming $f_{0,i,j}$ as a constant. Using  Eqn.
(\ref{MjMC}) and integrating between $0$ and $t=\infty$ we obtain
the  mean value of $f_{i,j}$:

\vspace{-7pt}
\begin{equation} \label{fijintegral}
\overline{f}_{i,j}=\frac{1}{\tau_{MC}}\int_{0}^{\infty}
\frac{f_{0,i,j}e^{t/\tau_{j}^{gr}}}{1-f_{0,i,j}+
f_{0,i,j}e^{t/\tau_{j}^{gr}}}e^{-t/\tau_{MC}}dt\, .
\end{equation}
\vspace{-3pt}

\noindent Inserting  Eqn. (\ref{fijintegral}) into Eqn. (\ref{MjMC})
we get $\overline{M}_{j,MC}$, that is the time averaged mass of dust
of type $j$-th returned to the ISM. However, this is only part of
the story, because the molecular clouds have also different mass, so
we need to evaluate their mean mass $\overline{M}_{MC}$ and use this
instead of $M_{MC}$.

\textsf{Fractionary mass  of dust grains of $j$-type}. With aid of
Eqn. (\ref{fijintegral}) we can derive the fractionary mass of dust
of type $j$-th produced by the accretion of grains in cold
clouds per unit area and unit time:

\vspace{-7pt}
\begin{equation} \label{Gjd}
\left(\frac{d\sigma_{i}^{D}}{dt}\right)_{accr}=\frac{1}{\tau_{MC}}\left(\overline{f}_{i,j}
-f_{0,i,j} \right) \cdot \chi_{j,max}^{D} \sigma_{MC}\, .
\end{equation}
\vspace{-7pt}

\noindent Looking at Eqn. (\ref{Gjd}), it requires to know the
amount of molecular gas of the galaxy we are going to model. To do
this, a  multi-phase description of the ISM taking  into account the
exchange of matter between cold and warm phases would be required.
This is not possible with the present model because  Eqns.
(\ref{GISM}), (\ref{DUST}), and (\ref{GAS}) have been formulated
for the single phase description and without a description of the
gas exchange between cold and warm phases. To overcome this
limitation some adjustment of the model are needed. Introducing
$\chi_{MC}=
 \displaystyle \sigma^{MC}/\sigma^{\mathcal{M}}$ as the
fraction of molecular gas and with some passages \citep{Zhukovska08} the
final equation for grain growth in MCs in the frame of a
single-phase ISM.

\vspace{-5pt}
\begin{equation} \label{Gjdequation}
\left(\frac{d\sigma_{i}^{D}}{dt}\right)_{accr}=\frac{\chi_{MC}}{\left(1-\chi_{MC}\right)\tau_{MC}}\left(
\overline{f}_{i,j} \sigma_{j,max}^{D} -\sigma_{j}^{D}\right)\, .
\end{equation}
\vspace{-5pt}

\noindent We need however to fix $\chi_{MC}$: in \citet{Zhukovska08}
is fixed to 0.2 in agreement to the observations for the SoNe at the
current time. However this approximation could not probably hold for
other phases of the evolution of the SoNe or other regions of the
MW. In \citet{Piovan11c}, by means of the data available for the MW
disk about molecular hydrogen, neutral hydrogen, total amount of gas
and star formation we suitably extended $\chi_{MC}$ by means of
Artificial Neural Networks (ANNs) in such a way to have the amount
of MCs as a function of the current local variables describing the
system. This recipe will be adopted in this work. For more details
see \citet{Piovan11c}.

\textsf{Degree of condensation}. Another critical parameter to be
examined with accuracy is the condensation degree
$\overline{f}_{i,j}$. It depends on the comparison between
$\tau_{MC}$, the average lifetime of the MC, and the typical
accretion time of grains $\tau_{j}^{gr}$ of Eqn.
(\ref{tau_ZhukovskaB}). If $\tau_{j}^{gr} \gg \tau_{MC}$ then we
expect that only small amounts of dust are produced because the cold
clouds will be destroyed before that dust has enough time to grow.
On the contrary, if $\tau_{j}^{gr} \ll \tau_{MC}$, we expect  the
condensation of dust grains to occur almost completely and
consequently dust to increase. \citet{Zhukovska08} propose an
analytical expression for the condensation degree as a function of
the ratio $\displaystyle \tau_{j}^{gr} /\displaystyle \tau_{MC}$
(assuming a constant $f_{0,i,j}$):

\vspace{-5pt}
\begin{equation} \label{fijanalytical}
\overline{f}_{i,j}= \left[\frac{1}{f_{0,i,j}^{2}\left(1+\frac{
\displaystyle \tau_{MC}}{\displaystyle \tau_{j}^{gr}}\right)^{2}}+1
\right]^{-\frac{1}{2}}\, .
\end{equation}
\vspace{-3pt}

\noindent According to \citet{Zhukovska08}, this expression is no
longer valid only for very small values of $f_{0,i,j}$. The
situation for our models is more complicated because we are  dealing
with large variations of $f_{0,i,j}$ and $\tau_{j}^{gr}$. To clarify
the issue, we checked in advance whether or nor  the above relation
can be still used.  To this aim we have calculated the  differences
between $\overline{f}_{i,j}$ from t Eqn. (\ref{fijanalytical}) and
the  relation of  Eqn. (\ref{fijintegral}). This comparison is made
for $\tau_{j}^{gr} < 1000 \tau_{MC}$ and $\tau_{j}^{gr} >
\tau_{MC}/50$. If $\tau_{j}^{gr}
> 1000\tau_{MC}$ or $\tau_{j}^{gr} < \tau_{MC}/50$ we use the Taylor
expansion suggested by \citet{Zhukovska08} deal with the  extreme
cases of very slow or very fast accretion time.

\begin{figure} \label{FIJTheoryAnalytical}
\centerline{\hspace{-15pt}
\includegraphics[height=7.5cm,width=8.5cm]{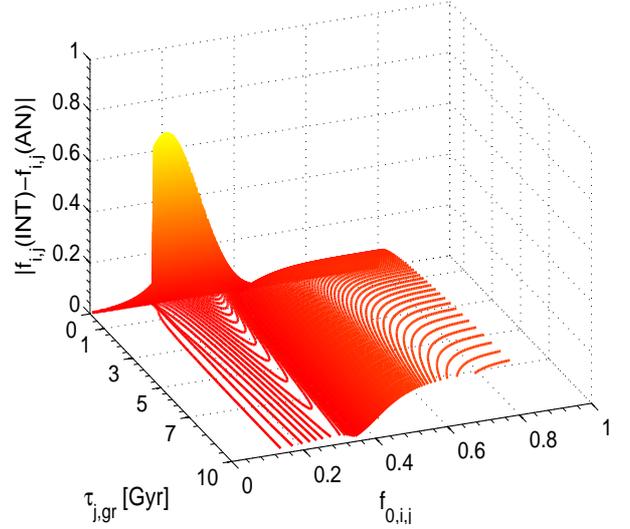}}
\caption{Absolute value of the differences between the
$\overline{f}_{i,j}$ calculated integrating and averaging on the MCs
lifetimes as in Eqn. (\ref{fijintegral}) and the same quantity
derived from  the analytical formula of Eqn. (\ref{fijanalytical})
suggested by \citet{Zhukovska08}. A grid of 200$\times$200 suitably
discrete values of $f_{0,i,j}$ and $\tau_{j}^{gr}$ has been
considered.}
\end{figure}

The results are presented in Fig. \ref{FIJTheoryAnalytical} for a
200$\times$200 grid of models with uniform spacing of $f_{0,i,j}$
and $\tau_{j}^{gr}$ growing linearly  from low to high values. The
error can be significant, in particular for very low $f_{0,i,j}$ and
very short $\tau_{j}^{gr}$ or high $\tau_{j}^{gr}$ and average
$f_{0,i,j}$.

\noindent As we discussed in Sect. \ref{modelA}, the same accretion term
in \citet{Dwek98} and \citet{Calura08} is

\vspace{-5pt}
\begin{equation}
\left( \frac{d \sigma_{i}^{D}}{dt}\right)_{accr} =
\frac{\sigma_{i}^{D}}{\tau_{i,accr}}
=\frac{\sigma_{i}^{D}}{\tau_{0,i}} \cdot \left(1 -
\frac{\sigma_{i}^{D}}{\sigma_{i}^{\mathcal{M}}} \right)\, .
\end{equation}
\vspace{-5pt}

\noindent The basic difference between the two approaches is that
while  \citet{Dwek98} and \citet{Calura08} with describes the
evolution of the abundance of the element $i$-th in the  dust,
\citet{Zhukovska08} follow the evolution of the abundance of the
$j$-th type dust in a given sample  molecules considered to
represent the situation of the  ISM. To adapt the equations to our
formulation for single elements (see Sect. \ref{Che_Evo_Mod}), we
start from:

\vspace{-7pt}
\begin{equation}
\sigma_{i}^{G}=\sigma_{i}^{\mathcal{M}}-\sigma_{i}^{D}
\end{equation}
\vspace{-7pt}

\noindent which  summing up all the types of dust  $j$-th in which a
given element is locked up the element $i$-th is locked
(independently on whether or not it is a  key element for that kind
of molecule) becomes:

\vspace{-7pt}
\begin{equation}
\sigma_{i}^{G}=\sigma_{i}^{\mathcal{M}}-\sum_{j}\nu_{i,j}^{D}\frac{A_{i,j}}{A_{j}^{D}}\sigma_{j}^{D}
\end{equation}
\vspace{-7pt} with
\vspace{-7pt}
\begin{equation}
\sigma_{i}^{D}=\sum_{j}\nu_{i,j}^{D}\frac{A_{i,j}}{A_{j}^{D}}\sigma_{j}^{D}
\end{equation}
\vspace{-7pt}

\noindent where as usual $\sigma_{i}^{D}$ is the surface mass
density of the $i$-th element in dust, while $\sigma_{j}^{D}$ is the
surface mass density of the $j$-th type dust. The above relations
says that the amount of element $i$-th contained in the dust is
given by the sum of the contributions by  all the $j$-th kinds of
dust in which $i$ is involved. For each one of the $j$-th types, we
divide $\sigma_{j}^{D}$ by the unit mass $A_{j}^{D}m_{\mathrm{H}}$
of one species $j$-th and find the number density of dust of type
$j$-th. Then, for each one of these molecules $j$-th we have
$\nu_{i,j}^{D}$ atoms of the element $i$-th. Finally, we multiply by
the atomic mass $A_{i,j}m_{\mathrm{H}}$ of the element $i$-th to get
the mass of the $i$-th element  coming from the $j$-th type dust. If
we derive respect to time we obtain

\vspace{-7pt}
\begin{equation}
\left( \frac{d \sigma_{i}^{D}}{dt}\right)_{TOT}
=\sum_{j}\nu_{i,j}^{D}\frac{A_{i,j}}{A_{j}^{D}}\left(
\frac{d\sigma_{j}^{D}}{dt}\right)_{TOT}
\end{equation}
\vspace{-7pt}

\noindent where the total evolution of dust is given by the sum of the variations
due to the star formation, the stardust injected in the ISM, the destruction
of the grains by SNa shocks, and the accretion process in the cold regions.
Taking into account only this last one:

\vspace{-7pt}
\begin{equation}
\left(\frac{d\sigma_{i}^{D}}{dt}\right)_{accr}
=\sum_{j}\nu_{i,j}^{D}\frac{A_{i,j}}{A_{j}^{D}}\left(
\frac{d\sigma_{j}^{D}}{dt}\right)_{accr}
\end{equation}
\vspace{-7pt}

\noindent The amount of element $i$-th contained in  dust is given
by the sum of the contributions by all the $j$th kinds of dust in
which $i$ is involved. For each  $j$-th type, we divide
$\sigma_{j}^{D}$ by the unit mass $A_{j}^{D}m_{\mathrm{H}}$ of one
species $j$-th of dust and find the number density of the  $j$-th
type dust. For each  $j$-th dust we have $\nu_{i,j}^{D}$ atoms of
the element $i$-th. Finally, we multiply by the atomic mass
$A_{i,j}m_{\mathrm{H}}$ of the element $i$-th to get the mass of
this coming from the $j$-th type dust. Inserting  the expression for
$\left(d\sigma_{j}^{D}/dt\right)_{accr}$ we have:

\vspace{-7pt}
\begin{flalign}
&\left( \frac{d \sigma_{i}^{D}}{dt}\right)_{accr}=
\sum_{j}\nu_{i,j}^{D}\frac{A_{i,j}}{A_{j}^{D}}\left(
\frac{d\sigma_{j}^{D}}{dt}\right)_{gg} = \nonumber \\
& = \sum_{j}\frac{\nu_{i,j}^{D}A_{i,j}}{A_{j}^{D}}\frac{\chi_{MC}}
{\left(1-\chi_{MC}\right)\tau_{MC}}\left( \overline{f}_{k,j}
\sigma_{j,max}^{D} -\sigma_{j}^{D}\right)\, .
\end{flalign}
\vspace{-7pt}

\noindent Let's now introduce $\sigma_{j,max}^{D}$, which is related
to the amount of key-element $k$-th in the dust of type  $j$-th
present in the ISM, i.e. $\sigma_{k}^{\mathcal{M}}$. We can express
$\sigma_{j}^{D}$ by means of  the key-element $k$-th of the
 $j$-th dust that is locked up into the $j$-th component itself. This
quantity is not known and not tracked by  our model. All we know is
how much of a given element is locked up into dust, but \textit{we
do not know how much of that element is locked in every kind of
dust}. For example we are able to follow how much silicon is locked
up into dust, but not how much silicon is stored  in pyroxene,
olivine and quartz. Doing the substitution we have:

\vspace{-7pt}
\begin{align}
\left( \frac{d \sigma_{i}^{D}}{dt}\right)_{accr} &=
\sum_{j}\frac{\nu_{i,j}^{D}A_{i,j}}{A_{j}^{D}\tau_{MC}}\frac{\chi_{MC}}
{\left(1-\chi_{MC}\right)} \cdot \nonumber \\
& \cdot\left[ \overline{f}_{k,j}
\frac{A_{j}^{D}}{A_{k,j}\nu_{k,j}^{D}}\sigma_{k,j}^{\mathcal{M}}-
\frac{A_{j}^{D}}{A_{k,j}\nu_{k,j}^{D}}\sigma_{k,j}^{D} \right]
\end{align}
\vspace{-7pt}

\vspace{-7pt}
\begin{align}
\left( \frac{d \sigma_{i}^{D}}{dt}\right)_{accr} &= \frac{\chi_{MC}}
{\left(1-\chi_{MC}\right)\tau_{MC}}\sum_{j}\frac{A_{i,j}\nu_{i,j}^{D}}{A_{k,j}\nu_{k,j}^{D}}\cdot
\nonumber \\ &\cdot
\left[\overline{f}_{k,j}\sigma_{k,j}^{\mathcal{M}}-\sigma_{k,j}^{D}\right]
\end{align}
\vspace{-7pt}

\noindent Posing $K=\chi_{MC}/
\left(\tau_{MC}\left(1-\chi_{MC}\right)\right)$,  we have:

\vspace{-7pt}
\begin{align} \label{FINAL_GROWTH}
\left(\frac{d \sigma_{i}^{D}}{dt}\right)_{accr} &= K
\left[\sum_{j,i=k}\left(
\overline{f}_{i,j}\sigma_{i,j}^{\mathcal{M}}
-\sigma_{i,j}^{D}\right)+ \right. \nonumber \\
&+ \left. \sum_{j,i\neq
k}\frac{A_{i,j}\nu_{i,j}^{D}}{A_{k,j}\nu_{k,j}^{D}}\left(
\overline{f}_{k,j}\sigma_{k,j}^{\mathcal{M}}
-\sigma_{k,j}^{D}\right)\right]
\end{align}
\vspace{-7pt}

\noindent where $k$ indicates the key element of the dust of type
$j$-th, because it may occur that a dust molecule contains the
element $i$-th  of which we tracking the evolution, but \textit{not}
the key element that could be of another type. For example, we could
be following the evolution of carbon and considering the silicon
carbide that contains carbon, but not as key-element. As a
consequence of this, the above has been split in two parts: in the
first one we include the part of the  $j$-th type dust grain in
which the element $i$-th is locked \textit{as key-element}; in the
second part we have the dust molecules containing the $i$-th element
but \textit{not as key element}, which instead corresponds to the
element  $k$-th. Therefore the summation goes for all the $k \ne i$.
The notation $\sigma_{k,j}^{D}$ is meant to indicate the surface
mass density of the $k$-th element locked up into the dust compounds
of type  $j$-th.

\section{Dust accretion rates for a few important elements } \label{grainevolution}

At this stage we need to specify the dust accretion rate given by
Eqn. (\ref{FINAL_GROWTH}) for some specific elements whose evolution
we intend to study. The model  follows  $16$ elements
\citep{Portinari98}: $\mathrm{H}$, $^{4}\mathrm{He}$,
$^{12}\mathrm{C}$, $^{13}\mathrm{C}$, $^{14}\mathrm{N}$,
$^{15}\mathrm{N}$, $^{16}\mathrm{O}$, $^{17}\mathrm{O}$,
$^{18}\mathrm{O}$, $^{22}\mathrm{Ne}$, $^{20}\mathrm{Ne}$,
$^{24}\mathrm{Mg}$, $^{28}\mathrm{Si}$, $^{32}\mathrm{S}$,
$^{40}\mathrm{Ca}$, $^{56}\mathrm{Fe}$.

Four  families of dust can be identified and tracked by the model:
they are the silicates, carbonaceous and iron grains, and silicon
carbide (it is worth clarifying that   $\mathrm{SiC}$ does not form
by accretion in MCs). We must distribute the 16 elements followed by
the model in the four families of dust.  First of all, $\mathrm{H}$,
$^{4}\mathrm{He}$, $^{20}\mathrm{Ne}$, $^{22}\mathrm{Ne}$ are noble
gases not involved into the dust formation process by accretion.
Second, since the hydrogen is by far the most abundant element in
the Universe, even if it may be contained in many dust molecules
(such as for instance the PAHs), its  abundance in the ISM gas will
not  be significantly affected  by dust. For what concern
$^{16}\mathrm{O}$, $^{17}\mathrm{O}$, $^{18}\mathrm{O}$,
$^{24}\mathrm{Mg}$, $^{28}\mathrm{Si}$, $^{56}\mathrm{Fe}$,
according to the simplified \citet{Zhukovska08} scheme, they are
involved in silicates (all of them) and iron grains (just some of
them). For $^{14}\mathrm{N}$, $^{15}\mathrm{N}$, $^{32}\mathrm{S}$
and $^{40}\mathrm{Ca}$ there is no specific dust family to associate
them.  However, the observations tell us that they are or could be
depleted --see for instance the case of Calcium
\citep{Whittet03,Tielens05}--, so we must set a plausible yet simple
scheme to describe them. Finally, $^{12}\mathrm{C}$ and
$^{13}\textrm{C}$ are both involved into the formation of
carbonaceous grains and silicon carbide. In the following we will
examine the case of $^{12}\mathrm{C}$, $^{13}\mathrm{C}$,
$^{16}\mathrm{O}$, $^{17}\mathrm{O}$, $^{18}\mathrm{O}$,
$^{24}\mathrm{Mg}$, $^{28}\mathrm{Si}$ and $^{56}\mathrm{Fe}$ in
some detail.

\subsection{Carbon} \label{Equations:Carbon}

Carbon is present in dust as key constituent of many molecules
containing carbon (C-molecules), otherwise named  carbonaceous
grains. It is also part of important dust molecules like silicon
carbide ($\mathrm{SiC}$) where, however, the key element ruling the
process is  silicon. $\mathrm{SiC}$ grains are mainly related to AGB
carbon stars and for them there is no process of accretion in cold
clouds of the ISM. Finally, the problem of the formation of
$\mathrm{SiC}$ in the ejecta of supernov{\ae} is still debated.

\citet{Nozawa03}, studying dust formation in population III
supernov{\ae} argues that they cannot form $\mathrm{SiC}$ grains
with radii comparable to the pre-solar $\mathrm{SiC}$ grains
identified as supernova condensates, because silicon and carbon are
firstly locked into other carbonaceous grains or silicates before
the formation of $\mathrm{SiC}$ grains can take place. However, we
observe a small but significant percentage of X-type $\mathrm{SiC}$
grains (one of the minor types A, B, X, Y and Z, that is believed to
be formed in the ejecta of type II SN{\ae} because of the isotopic
signature) with respect to the mainstream $\mathrm{SiC}$ (about 90\%
of the total and whose origin is attributed to AGB C-stars, because
of their similar $^{12}$C/$^{13}$C ratio, signatures of s-processes
and the 11.3 $\mathrm{SiC}$ $\mu$m feature in C-stars). The ratio
between X-type and mainstream type can be quantified in about 0.01
\citep{Hoppe00}. Therefore, as proposed by \citet{Deneault03},
special conditions and/or chemical pathways might have to be
considered in order to realize the formation in type II SN{\ae} of
$\mathrm{SiC}$ grains. To reproduce the observed X-type over
mainstream ratio, \citet{Zhukovska08} considers a small formation of
$\mathrm{SiC}$ in Type II SN{\ae}, however this is just an
\textit{ad-hoc} recipe based on meteorites and does not throw light
over the uncertain amount of dust, and X-type $\mathrm{SiC}$,
produced by SN{\ae}.

The equation describing the accretion of carbonaceous grains of the
ISM is:

\vspace{-7pt}
\begin{equation} \label{Carbon_evolution}
\left(\frac{d \sigma_{C}^{D}}{dt}\right)_{accr} = K
\left\{\overline{f}_{C,C} \left( 1 -
\xi_{CO} \right)\sigma_{C,C}^{\mathcal{M}} - \sigma_{C,C}^{D} \right\} \\
\end{equation}
\vspace{-5pt}

\noindent where $\sigma_{\mathrm{C,C}}^{\mathcal{M}}$ is the amount
of the available carbon,   $\xi_{\mathrm{CO}}$ a factor  accounting
for  the carbon embedded in molecules like  $\mathrm{CO}$ that do
not take part to the accretion process,  $\sigma_{\mathrm{C,C}}^{D}$
is the carbon already condensed into dust, and
$\overline{f}_{\mathrm{C,C}}$ is the average condensation factor
(see  Sect. \ref{graingrowth} model B). The accretion of C-molecules
in molecular clouds happens according to a time scale defined by
Eqn. \ref{tau_ZhukovskaB} which for the specific case of carbon is:

\vspace{-7pt}
\begin{small}
\begin{align} \label{tauC}
\tau_{C}^{gr} &=46 \frac{(A^{gw}_{C})^{\frac{1}{2}}
\nu_{C,C}^{D}}{A_{C}^{D}} \left(\frac{\rho_{C}^{D}}{3}\right) \left(
\frac{10^{3}}{n_{H}}\right) \left(\frac{3.5 \cdot
10^{-5}}{\epsilon_{C,C}}\right) Myr\, .
\end{align}
\end{small}
\vspace{-3pt}

\noindent This relation deserves some comments. First of all
$\nu_{\mathrm{C},\mathrm{C}}^{\mathrm{D}}=1$ because carbon is
present in the cold regions of molecular clouds as single atoms or
molecules with typically one atom of carbon. Second, $\rho_{C}^{D}=$
2.26 $\mathrm{g} \, \, \mathrm{cm}^{-3}$ \citep{Li01}. To determine
$A_{C}^{D}$ we should  average  the masses of the various type of
molecules present in the molecular clouds. Carbon is embedded in
$\mathrm{CO}$ (with a percentage easily  up to $20$ or $40 \%$).
There are also   lots of C-molecules like: $\mathrm{CO}_{2},
\mathrm{H}_{2}\mathrm{CO}, \mathrm{HCN}, \mathrm{CS},
\mathrm{CH}_{4}, \mathrm{C}_{2}\mathrm{H}_{2},
\mathrm{C}_{2}\mathrm{H}_{6}, \mathrm{HCOOH}, \mathrm{OCS},...$. We
consider $\mathrm{A}_{\mathrm{C}}^{\mathrm{D}} = 28$  corresponding
to $\mathrm{CO}$ as a sort of mean value fairly representing the
above group. $A_{\mathrm{C}}^{\mathrm{gw}} = 12$ assuming single
atoms as the typical accreting species. Finally, we need to specify
$\epsilon_{\mathrm{C},\mathrm{C}}$ the abundance by number with
respect to $\mathrm{H}$ of the key-species $\mathrm{C}$  driving the
process of accretion of dusty C-molecules. We have:

\vspace{-7pt}
\begin{equation}\label{EpsilonC}
\epsilon_{C,C}=\left( 1 - \xi_{CO}
\right)\frac{\sigma_{C}^{\mathcal{M}} - \sigma_{C}^{D}}{12
\sigma_{H}^{\mathcal{M}}}\, .
\end{equation}
\vspace{-7pt}

\noindent In this relation, the factor $\left( 1 - \xi_{CO} \right)$
takes into account that fact that the carbon already in
$\mathrm{CO}$ is not available to the accretion process. This
correction is not negligible. Second, we subtract from the carbon
atoms available for accretion, those ones already embedded in dust
or belonging to unreactive species like $\mathrm{SiC}$. Finally, we
obtain  for the typical accretion time:

\vspace{-7pt}
\begin{equation}
\tau_{C}^{gr} = \frac{1.80}{\left( 1 - \xi_{CO}
\right)n_{H}}\left(\frac{\sigma_{H}^{\mathcal{M}}}{\sigma_{C}^
{\mathcal{M}}-\sigma_{C}^{D}}\right)Myr\, .
\end{equation}
\vspace{-7pt}

\noindent A similar relation  is assumed  for $^{13}\mathrm{C}$, in
which the only difference is the  $2.03$ instead of $1.80$. This
assumption stands on the notion  that $^{13}\mathrm{C}$, much less
abundant than $^{12}\mathrm{C}$, undergoes  the same kind of
accretion processes and forms similar C-molecules. Clearly this is a
simplified view of the subject, however sustained by the fact that,
on average, the mean
$\mathrm{N}(^{12}\mathrm{CO})/\mathrm{N}(^{13}\mathrm{CO})$ ratio in
diffuse clouds is close or compatible with the local interstellar
isotope ratio $^{12}\mathrm{C}/^{13}\mathrm{C}$, even if the
$\mathrm{N}(^{12}\mathrm{CO})/\mathrm{N}(^{13}\mathrm{CO})$ can be
easily enhanced or reduced by a factor of two with a large spread
\citep{Liszt07,Visser09}. This allows to think that the relative
percentage of carbon isotopes in molecules and dust should not be on
average too different.

 \indent To practically use  Eqn.
(\ref{Carbon_evolution}), we need $f_{0,C,C}$ to get the average
condensation factor $\overline{f}_{C,C}$: $f_{0,C,C}$ is the initial
condensation of the key-element $\mathrm{C}$ of the carbonaceous
dust compound. If $\widetilde{\chi}_{j}^{D}$ is the mass fraction of
dust j-th contained in the part of the ISM that is not stored in
cold clouds, we have:

\vspace{-10pt}
\begin{equation} \label{f0ijChi}
f_{0,C,C}\approx\frac{\widetilde{\chi}_{C}^{D}}{\chi_{C,max}^{D}}
\approx\frac{\chi_{C}^{D}}{\chi_{C,max}^{D}}\approx\frac{\displaystyle
\sigma_{C,C}^{D}}{\displaystyle \sigma_{C,C}^{\mathcal{M}}}
\end{equation}
\vspace{-7pt}

\noindent where to derive it, we simply replaced the
mass fraction of dust in the ISM not belonging to  cold regions with
the mass fraction embedded in the cold ones:
$\widetilde{\chi}_{C}^{D}\approx \chi_{C}^{D}$. This simply means
that after every cycle of MCs there is complete restitution and
mixing with the ISM, so that on the average the ISM  dust content
always mirrors the initial mass-fraction already condensed at the
formation of the MC.

\subsection{Silicon} \label{Equations:Silicon}

\noindent Silicon is present in many types of grains generically
referred to as silicates, one of the most complicated families of
minerals. We consider only some kinds of silicates, commonly used in
models of dust formation. First of all, we list the  pyroxenes that
are inosilicates (i.e. a family containing single chain and double
chain silicates) in general indicated by
$\mathrm{XY}(\mathrm{Si},\mathrm{Al})_{2}\mathrm{O}_{6}$, where X
represents ions such as calcium, sodium, iron$^{2+}$ and magnesium,
whereas  Y represents ions of  smaller size such as chromium,
aluminum, iron$^{+3}$, magnesium. Aluminum, while commonly replacing
in other silicates, does not often do it  in pyroxenes. Typical
pyroxenes used in dust models are the end members of the
enstatite-hypersthene-ferrosilite series described by
$\mathrm{Mg}_{\mathrm{x}}\mathrm{Fe}_{1-\mathrm{x}}\mathrm{SiO}_{3}$
with $0< \mathrm{x} <1$ determining the partition between the two
possibilities. Enstatite is the magnesium end-member
$\mathrm{MgSiO}_{3}$ of the series, while ferrosilite is the iron
end-member $\mathrm{FeSiO}_{3}$: both are found in iron and stony
meteorites. Second, we have the olivines. They are a series of
minerals falling in between  two end-members, fayalite and
forsterite. They can be described by
$[\mathrm{Mg}_{\mathrm{x}}\mathrm{Fe}_{1-\mathrm{x}}]_{2}\mathrm{SiO}_{4}$
with $0 < \mathrm{x} < 1$ determining the partition. Fayalite is the
iron-rich member with $\mathrm{Fe}_{2}\mathrm{SiO}_{4}$. Forsterite
is the magnesium-rich member with $\mathrm{Mg}_{2}\mathrm{SiO}_{4}$.
The two minerals form a series where the iron and magnesium can be
interchanged  without much effect, also difficult to detect,  on the
crystal structure. Olivine can be found for example in many
meteorites like the iron-nickel ones, in the tails of comets, and in
the disks of dust around young stars. Both pyroxenes and olivines
are expected to exist in the interstellar dust because they form
from magnesium and iron that are abundant. In addition to this and
limited to the calculation of stellar dust yields, we included
quartz, the most common mineral on the surface of the Earth,
composed by $\mathrm{SiO}_{2}$ and silicon carbide, $\mathrm{SiC}$.
Silicon carbide in expected to form in the ejecta of carbon-rich
stars and type II SN{\ae}, but not by accretion in cold clouds.

Two parameters control the pyroxene-olivine mixture: the Mg fraction
$x$, assumed to be the same for pyroxenes an olivines, and the
abundance ratio between the amount of magnesium and silicon bound in
dust $A_{Mg}^{D}/A_{Si}^{D}$. This latter determines the fraction of
silicates with olivine stoichiometry
$F_{Ol}=A_{Mg}^{D}/(A_{Si}^{D}\cdot x)-1$, while for pyroxenes we
can define $F_{Pyr}=1-F_{Ol}$. The two parameters are fixed and
chosen according to present day observations of IR emission and
abundances in the diffuse ISM
\citep{Dwek97,Whittet03,Dwek05,Zhukovska08}, i.e.  $x=0.8$ and
$A_{Mg}^{D}/A_{Si}^{D}=1.06-1.07$.  Varying  these parameters does
not  significantly affect the total efficiency of dust production by
the MCs \citep{Zhukovska08}.  For the purposes of this study,  they
can be kept fixed. The equation describing the evolution of silicon
is nearly the  same no matter whether or not  silicon \textit{is the
key-element} for pyroxenes and olivines.

\textsf{The Mg/Fe case}. Let us first suppose that the key-element
is magnesium/iron and not silicon.  Using Eqn. (\ref{FINAL_GROWTH})
we get:

\vspace{-7pt}
\begin{flalign}
\left(\frac{d \sigma_{Si}^{D}}{dt}\right)_{accr} &= K
\frac{A_{Si,Pyr}\nu_{Si,Pyr}^{D}}{A_{i,Pyr}\nu_{i,Pyr}^{D}}\, \cdot
\left\{\overline{f}_{i,Pyr} \sigma_{i,Pyr}^{\mathcal{M}}- \right.\nonumber \\
&- \left.\sigma_{i,Pyr}^{D} \right\} + K
\frac{A_{Si,Ol}\nu_{Si,Ol}^{D}}{A_{i,Ol}\nu_{i,Ol}^{D}} \left\{
\overline{f}_{i,Ol}\cdot \right.\nonumber \\
& \cdot\sigma_{i,Ol}^{\mathcal{M}}- \left. \sigma_{i,Ol}^{D}
\right\}
\end{flalign}
\vspace{-7pt}

\noindent where $\overline{f}_{i,Pyr}$ and $\overline{f}_{i,Ol}$ are
the average condensation fractions of pyroxenes and olivines,
obtained according to the procedure described in Sect.
\ref{graingrowth}. Furthermore,  $\nu_{Si,Pyr}^{D}$ is the number of
silicon atoms needed for one molecule of pyroxene, i.e.
$\nu_{Si,Pyr}^{D}=1$. Since for each molecule we have just one
$\mathrm{Si}$ atom,  $\nu_{Si,Pyr}^{D} \cdot A_{Si,Pyr} = 28$. The
quantity $A_{i,Pyr} = 24$ or $56$ (magnesium or iron). We must also
define $\nu_{i,Pyr}^{D}$  (a number between $0$ and $1$). Assuming a
fixed partition $x$: $\nu_{Mg,Pyr}^{D} = 0.8$ and $\nu_{Fe,Pyr}^{D}
= 0.2$. Finally, we have the physical constants for olivines:
$\nu_{Si,Ol}^{D} = 1$, $A_{Si,Ol} = 28$, $A_{Mg,Ol} = 24$,
$A_{Fe,Ol} = 56$, $\nu_{Mg,Ol}^{D} = 1.6$,   finally
$\nu_{Fe,Ol}^{D} = 0.4$. Since $F_{Pyr}$ and $F_{Ol}$ are the
fractions of pyroxenes and olivines in which silicates are
subdivided, we have the following partition of magnesium in dust
$\sigma_{Mg}^{D}=\sigma_{Mg,Ol}^{D}+\sigma_{Mg,Pyr}^{D}=(x\cdot
F_{Pyr})/( x\cdot F_{Pyr}+2x\cdot F_{Ol})\sigma_{Mg}^{D}+(2x\cdot
F_{Ol})/( x\cdot F_{Pyr}+2x\cdot
F_{Ol})\sigma_{Mg}^{D}=Mg_{Pyr}\sigma_{Mg}^{D}+Mg_{Ol}\sigma_{Mg}^{D}
$ and of iron in silicates (once subtracted the iron embedded in
iron grains)
$\sigma_{Fe,Sil}^{D}=\sigma_{Fe}^{D}-\sigma_{Fe,Fe}^{D}=
\sigma_{Fe,Ol}^{D}+\sigma_{Fe,Pyr}^{D}=(1-x)\cdot
F_{Pyr}/((1-x)\cdot F_{Pyr}+2(1-x)\cdot
F_{Ol})\sigma_{Fe,Sil}^{D}+2(1-x)\cdot F_{Ol}/((1-x)\cdot
F_{Pyr}+2(1-x)\cdot
F_{Ol})\sigma_{Fe,Sil}^{D}=Fe_{Pyr}\sigma_{Fe,Sil}^{D}+Fe_{Ol}\sigma_{Fe,Sil}^{D}
$.

It is worth also noting that the  dust compounds of which we follow
the accretion, i.e. pyroxenes/olivines for silicates, carbonaceous
grains and iron grains (the same considered by \citet{Zhukovska08}),
are fewer in number than  the dust molecules considered by the
stellar yields of dust in the most detailed models calculated
by CNT theory  \citep[see][for more details]{Piovan11a}.
The number and variety of the dust molecules
injected into the ISM are wider than the number of accreting
species. For this reason, before splitting $\sigma_{Mg}^{D}$,
$\sigma_{Fe,Sil}^{D}$ and $\sigma_{Si}^{D}$ (see below) between
pyroxenes/olivines, we need to keep memory of the evolution of the
other dust species injected by AGB/SN{\ae} (also eventually swept up
and destroyed by shocks in the ISM), which are not included in the
accretion scheme.


\textsf{The Si case}. In alternative,  silicon can be the
key-element for olivines and pyroxenes:

\vspace{-9pt}
\begin{eqnarray}
\left(\frac{d \sigma_{Si}^{D}}{dt}\right)_{accr} &=& K
\left\{\overline{f}_{Si,Pyr} \sigma_{Si,Pyr}^{\mathcal{M}} - \sigma_{Si,Pyr}^{D} \right\} \nonumber \\
&+& K \left\{\overline{f}_{Si,Ol} \sigma_{Si,Ol}^{\mathcal{M}} -
\sigma_{Si,Ol}^{D} \right\}
\end{eqnarray}
\vspace{-9pt}

\noindent where again $\overline{f}_{Si,Pyr}$ and
$\overline{f}_{Si,Ol}$ are the average condensation fractions of
silicon in pyroxenes and olivines. The corresponding partition of
silicon in dust is
$\sigma_{Si}^{D}=\sigma_{Si,Pyr}^{D}+\sigma_{Si,Ol}^{D}=F_{Pyr}\sigma_{Si}^{D}+
F_{Ol}\sigma_{Si}^{D}$. We need to calculate the accretion time
scales of pyroxenes and olivines, to be compared with the lifetime
of molecular clouds and included into the integrations for
$\overline{f}_{Mg,Pyr}$, $\overline{f}_{Mg,Ol}$,
$\overline{f}_{Si,Pyr}$ and $\overline{f}_{Si,Ol}$. Applying Eqn.
(\ref{tau_ZhukovskaB}) we get:

\vspace{-7pt}
\begin{small}
\begin{align} \label{TauAccrPyr}
&\tau_{Pyr}^{gr} = 46 \frac{(A^{gw}_{Pyr})^{1/2}
\nu_{i,Pyr}^{D}}{A_{Pyr}^{D}} \left(\frac{\rho_{Pyr}^{D}}{3}\right)
\left( \frac{10^{3}}{n_{H}}\right) \left(\frac{3.5 \cdot
10^{-5}}{\epsilon_{i,Pyr}}\right) =\nonumber
\\ &=\frac{\tau_{0,Pyr,i}^{gr}}{\sigma_{i}^{\mathcal{M}}}\frac{\sigma_{H}^{\mathcal{M}}}{n_{H}}
\end{align}
\end{small}
\vspace{-7pt}
\begin{small}
\begin{align} \label{TauAccrOl}
&\tau_{Ol}^{gr} = 46\frac{(A_{Ol}^{gw})^{1/2} \nu_{i,d,Ol}}{A_{d,Ol}}
\left(\frac{\rho_{d,Ol}}{3}\right) \left(
\frac{10^{3}}{n_{H}}\right) \left(\frac{3.5 \cdot
10^{-5}}{\epsilon_{i,Ol}}\right) =\nonumber
\\ &=\frac{\tau_{0,Ol,i}^{gr}}{\sigma_{i}^{\mathcal{M}}}\frac{\sigma_{H}^{\mathcal{M}}}{n_{H}}
\end{align}
\end{small}
\vspace{-7pt}

\noindent where $\tau_{0,Pyr,i}^{gr}$ and $\tau_{0,Ol,i}^{gr}$ are
two constants that depend on the key-element. $\tau_{0,Pyr,i}^{gr}$
is equal to $2.05$, $1.30$ and $1.16$ for the key-elements Si, Mg
and Fe respectively, while $\tau_{0,Ol,i}^{gr}$ is equal to $2.05$,
$2.60$ and $2.32$ again for Si, Mg and Fe. $A_{Ol}^{gw}$ and
$A_{Pyr}^{gw}$ are the atomic weight of the growing species.
Finally, $A_{d,Ol} =A_{d,Pyr} = 121.41$. It is
important to note that in the  calculation of the abundances of
species available for dust accretion, we do not subtract the amount
of dust already formed $\sigma_{i}^{D}$ from the total ISM abundance
$\sigma_{i}^{\mathcal{M}}$; this is equivalent to assume that the
grains of silicates already formed act as reactive species in the
accretion process. Dropping this
assumption (and simplification), the accretion process would be too
slow. There is one exception though in the case of iron as the
key-element for the mixture of silicates mixture. In such a case  we
subtract from $\sigma_{i}^{\mathcal{M}}$ the amount of iron already
embedded in the iron dust grains (that do not react with silicates),
thus obtaining
$\sigma_{i}^{\mathcal{M}}=\sigma_{Fe}^{\mathcal{M}}-\sigma_{Fe,Fe}^{D}$.

The initial condensation fractions $f_{0,i,Pyr}$, $f_{0,i,Ol}$
needed to derive $\overline{f}_{i,Pyr}$, $\overline{f}_{i,Ol}$ can
be obtained as in Eqn. (\ref{f0ijChi}).

\subsection{Oxygen} \label{Equations:Oxygen}

Oxygen is much more abundant than the  refractory elements
$\mathrm{Si}$, $\mathrm{Fe}$, $\mathrm{Mg}$, $\mathrm{Ca}$ and
$\mathrm{S}$. Therefore,  oxygen (at least the most abundant isotope
$^{16}\mathrm{O}$)  will never become a key-element determining the
growth of the typical dust grains in the ISM. The equation
describing the evolution of the mass abundance of the
$^{16}\mathrm{O}$ is obtained from Eqn. (\ref{FINAL_GROWTH}),
dropping the terms giving the contribution of the $\mathrm{O}$ as
key-element. Therefore, considering the accretion of pyroxenes and
olivines, the temporal evolution of the   $^{16}\mathrm{O}$
abundance (we drop the mass number of the isotope) is:

\vspace{-7pt}
\begin{align}
\left(\frac{d \sigma_{O}^{D}}{dt}\right)_{accr} &= K
\frac{A_{O,Pyr}\nu_{O,Pyr}^{D}}{A_{i,Pyr}\nu_{i,Pyr}^{D}}
\nonumber\left\{\overline{f}_{i,Pyr}\sigma_{i,Pyr}^{\mathcal{M}}- \right.\nonumber \\
&- \left.\sigma_{i,Pyr}^{D} \right\}+K
\frac{A_{O,Ol}\nu_{O,Ol}^{D}}{A_{i,Ol}\nu_{i,Ol}^{D}}
\left\{\overline{f}_{i,Ol} \sigma_{i,Ol}^{\mathcal{M}} - \nonumber \right. \\
& - \left. \sigma_{i,Ol}^{D} \right\}
\end{align}
\vspace{-7pt}

\noindent where once again  the key-element could be $\mathrm{Mg}$,
$\mathrm{Si}$ or $\mathrm{Fe}$, depending on which has the lowest
abundance. The subscript $i$ indicates a generic key-element. As
usual, $\overline{f}_{i,Pyr}$ and $\overline{f}_{i,Ol}$ are the
average condensation fractions when the MC is dispersed, and
$\sigma_{i,Pyr}^{D}$ and $\sigma_{i,Ol}^{D}$ are defined as in Sect.
 \ref{Equations:Silicon}.  Introducing the Dirac delta function notation we have:
$A_{i,Pyr}=28\delta\left(A_{i}-28\right)+24\delta\left(A_{i}-24
\right)+56\delta\left(A_{i}-56\right)$ where $A_{i}$ is the mass
number of the key-element $i$. Furthermore, $A_{O,Pyr}=16$ and,
since in pyroxenes we have sulfite $\mathrm{SiO}_{3}$ with $3$
oxygen atoms,  $\nu_{O,Pyr}^{D}=3$,
$\nu_{i,Pyr}^{D}=1\delta\left(A_{i}-28\right)+0.8\delta\left(A_{i}-24
\right)+0.2\delta\left(A_{i}-56\right)$, and $A_{O,Ol}=16$.
Similarly, $\nu_{O,Ol}^{D}=4$ because in the olivines we have
sulfate $\mathrm{SiO}_{4}$,
$A_{i,Ol}=28\delta\left(A_{i}-28\right)+24\delta\left(A_{i}-24
\right)+56\delta\left(A_{i}-56\right)$, and
$\nu_{i,Ol}^{D}=1\delta\left(A_{i}-28\right)+1.6\delta\left(A_{i}-24
\right)+0.4\delta\left(A_{i}-56\right)$. The accretion time scales
$\tau_{Pyr}^{gr}$ and $\tau_{Ol}^{gr}$ that we need for
$\overline{f}_{i,Pyr}$ and $\overline{f}_{i,Ol}$, the elemental
abundances of the key-elements and the initial condensation
fractions are the same as in Sect. \ref{Equations:Silicon}.

In this context, we need  also a simple description for
$^{17}\mathrm{O}$ and $^{18}\mathrm{O}$. First of all let us examine
how they behave in the inert $\mathrm{CO}$ molecule. The way in
which the isotopes of $\mathrm{O}$ and $\mathrm{C}$ combine to form
many isotopologues other than $^{12}\mathrm{C}^{16}\mathrm{O}$ is
very complicated and thoroughly studied, because of the importance
of $\mathrm{CO}$ molecule, its easy detection and chemical stability
\citep{vanDishoeck88,Liszt07,Visser09}. The photo-dissociation of
$\mathrm{CO}$ is a line process and consequently  subject to
self-shielding that in turn depends on the column density.
Therefore, isotopologues other than $\mathrm{C}^{16}\mathrm{O}$ are
not self-shielded unless located  very deeply  into the molecular
clouds due to the  much lower abundance of $^{17}\mathrm{O}$ and
$^{18}\mathrm{O}$ \citep{Clayton03,Lee08,Visser09}. Looking at the
case of the Sun,  where the isotopic ratios  are about
$^{16}\mathrm{O}/^{18}\mathrm{O} \approx 500$ and
$^{16}\mathrm{O}/^{17}\mathrm{O} \approx 2600$
\citep{Lodders09,Asplund09}, the regions with abundances of the
isotopologues different than $C^{16}O$ significantly  reduced with
respect to other isotopologues are located very deeply in the
atmosphere. Basing on these  considerations, we set the reduction
factor of $^{17}\mathrm{O}$ and $^{18}\mathrm{O}$,
$\xi_{CO}^{\prime}$, equal to  $1/3$ of the reduction factor assumed
for $\mathrm{CO}$ in Eqn. (\ref{EpsilonC}). All this  agrees with
the observational ratios
$\mathrm{C}^{16}\mathrm{O}/\mathrm{C}^{17}\mathrm{O}$ and
$\mathrm{C}^{16}\mathrm{O}/\mathrm{C}^{18}\mathrm{O}$ in $\zeta$
Oph, $>5900$ and $\approx 1550$ respectively \citep{Savage96}).\\
\indent Another   point to note is that the ratios $^{16}\mathrm{O}$
to $^{17}\mathrm{O}$  and $^{18}\mathrm{O}$  in the interstellar
dust could be different from the same ratios in the ISM
\citep{Clayton88,Leshin97}. The subject  is a matter  of debate,
because the above ratios strongly  depend  on the site where  those
isotopes condense into  dust \citep{Meyer09}. Consequently, we
should not simply scale to $^{17}\mathrm{O}$ and $^{18}\mathrm{O}$
the results obtained for $^{16}\mathrm{O}$. However, a detailed
analysis of this issue is far beyond the purposes of this paper and
we leave it  to future improvements of  our model. For the time
being, basing on the following considerations we adopt a simple
recipe. As  in the dust grains, $^{17}\mathrm{O}$ and
$^{18}\mathrm{O}$ are much less abundant than other elements,
consider them as the key element of the dust accretion process.
Therefore, Eqn. (\ref{FINAL_GROWTH}) applied to $^{17}\mathrm{O}$
and $^{18}\mathrm{O}$ (briefly indicated as
$^{\mathrm{n}}\mathrm{O}$) becomes:

\vspace{-7pt}
\begin{flalign}
\left(\frac{d \sigma_{(^{n}O)}^{D}}{dt}\right)_{accr} &= K
\left\{\overline{f}_{(^{n}O),Pyr}\left( 1 - \xi_{CO}^{\prime}
\right) \sigma_{(^{n}O),Pyr}^{\mathcal{M}} - \right. \nonumber \\
&\left. \sigma_{(^{n}O),Pyr}^{D} \right\}+ K
\left\{\overline{f}_{(^{n}O),Ol}\left( 1 - \xi_{CO}^{\prime} \right)
\sigma_{(^{n}O),Ol}^{\mathcal{M}} \right. \nonumber \\ &\left.
-\sigma_{(^{n}O),Ol}^{D} \right\}
\end{flalign}
\vspace{-7pt}

\noindent where $\xi_{CO}^{\prime}$ takes into account  that not all
$^{17}\mathrm{O}$ and $^{18}\mathrm{O}$ in the ISM  can be used to
form new grains because part of them are locked in the
$\mathrm{CO}$.

The relative amounts of $^{17}\mathrm{O}$ and $^{18}\mathrm{O}$
available to form silicates are $\epsilon_{(^{n}O),Sil}=\left( 1 -
\xi_{CO}^{\prime}
\right)\left(\sigma_{(^{n}O)}^{\mathcal{M}}\right)/(A_{(^{n}O)}
\sigma_{H}^{\mathcal{M}})$ and the accretion timescales for
$^{17}\mathrm{O}$ and $^{18}\mathrm{O}$ are:

\vspace{-7pt}
\begin{align}
\tau_{^{17}O}^{gr} = \frac{0.97}{\left( 1 - \xi_{CO}^{\prime}
\right)n_{H}}\left(\frac{\sigma_{H}^{\mathcal{M}}}{\sigma_{^{17}O}^{\mathcal{M}}}\right)
Myr \\
\tau_{^{18}O}^{gr} = \frac{1.00}{\left( 1 - \xi_{CO}^{\prime}
\right)n_{H}}\left(\frac{\sigma_{H}^{\mathcal{M}}}{\sigma_{^{18}O}^{\mathcal{M}}}\right)
Myr
\end{align}
\vspace{-7pt}

\noindent where  we have considered a typical  silicate in which
just one atom of  $^{17}\mathrm{O}$ and/or $^{18}\mathrm{O}$ is
involved in the accretion process.

\subsection{Magnesium}  \label{Equations:Magnesium}

\noindent In our simple picture of the the grain accretion process,
magnesium intervenes in pyroxenes and olivines, often as the
key-element. Furthermore, magnesium is present also in other dust
compounds such as  MgO contained in the  SN{\ae} yields of dust. In
analogy to the case of Si, the set of equations governing the
temporal evolution of the Mg abundance is  different depending on
whether or not Mg is the key, similarly to silicon. If Mg is a  key
element:

\begin{footnotesize}
\begin{align}
\left(\frac{d \sigma_{Mg}^{D}}{dt}\right)_{accr} &= K\cdot
\left\{\overline{f}_{Mg,Pyr} \sigma_{Mg,Pyr}^{\mathcal{M}} -
\sigma_{Mg,Pyr}^{D} \right\} + \nonumber
\\ &K\cdot \left\{\overline{f}_{Mg,Ol}\sigma_{Mg,Ol}^{\mathcal{M}} -
\sigma_{Mg,Ol}^{D} \right\}
\end{align}
\end{footnotesize}

\noindent with the usual meaning of the symbols. The accretion time
scales and the initial condensation fractions of Mg can be obtained
in the same way as in Sect. \ref{Equations:Silicon}, following Eqns.
(\ref{TauAccrPyr}), (\ref{TauAccrOl}) and (\ref{f0ijChi}). If Mg is
not the key-element, but Si or Fe are playing the role, we have the
following equation for the Mg accretion:

\vspace{-7pt}
\begin{align}
\left(\frac{d \sigma_{Mg}^{D}}{dt}\right)_{accr} &= K \cdot
\frac{A_{Mg,Pyr}\nu_{Mg,Pyr}^{D}}{A_{i,Pyr}\nu_{i,Pyr}^{D}}
\left\{\overline{f}_{i,Pyr} \sigma_{i,Pyr}^{\mathcal{M}} \right.
\nonumber
\\ &\left. - \sigma_{i,Pyr}^{D} \right\} + K
\frac{A_{Mg,Ol}\nu_{Mg,Ol}^{D}}{A_{i,Ol}\nu_{i,Ol}^{D}} \cdot
\nonumber
\\ &\cdot \left\{\overline{f}_{i,Ol}\sigma_{i,Ol}^{\mathcal{M}} - \sigma_{i,Ol}^{D} \right\}
\end{align}
\vspace{-7pt}

\noindent in which $A_{Mg,Pyr}=24$, $\nu_{Mg,Pyr}^{D}=0.8$,
$A_{i,Pyr}=28$ or $56$ (silicon/iron), $\nu_{Si,Pyr}^{D}=1$,
$\nu_{Fe,Pyr}^{D}=0.2$, $A_{Mg,Ol}=48$, $\nu_{Mg,Ol}^{D}=1.6$,
$A_{Si,Ol}=28$, $\nu_{Si,Ol}^{D}=1$ and $\nu_{Fe,Ol}^{D}=0.4$. The
time scales for accretion of pyroxenes/olivines and the initial
condensation fractions are again from Eqns. (\ref{TauAccrPyr}),
(\ref{TauAccrOl}) and (\ref{f0ijChi}). For $\sigma_{i,Pyr}^{D}$ and
$\sigma_{i,Ol}^{D}$ see  Sect. \ref{Equations:Silicon}.

\subsection{Iron} \label{Equations:Iron}

Even considering the small number of accreting compounds included in
our model, iron can be locked up in grains of various type thanks to
 accretion processes in  cold regions of the ISM. First of all,
iron is locked up in  iron grains that act as the  key element.
Second, iron is also present in   pyroxene and olivine grains. In
such a case, iron may or may not be the key element.  In any case it
is removed from  the gaseous phase and stored in  grains. The
existence of two  channels for locking up iron into grains  leads to
a rather complicated equation for the evolution of the iron.  There
are the main terms to consider: in the  first and second, iron
participates  to the formation of silicates (pyroxenes and olivines)
\textit{as or not a key-element}, and in the third one  iron plays
the role of a  \textit{key-element} for the formation of iron
grains. Furthermore, the equation will be slightly different
depending on whether magnesium, silicon or iron is the  key-element
for the formation of pyroxenes/olivines. The final equation is

\vspace{-7pt}
\begin{align}
\left(\frac{d \sigma_{Fe}^{D}}{dt}\right)_{accr} &= K
\frac{A_{Fe,Pyr}\nu_{Fe,Pyr}^{D}}{A_{i,Pyr}\nu_{i,Pyr}^{D}}\left\{\overline{f}_{i,Pyr}
\sigma_{i,Pyr}^{\mathcal{M}} - \right.\nonumber
\\ &\left.-\sigma_{i,Pyr}^{D} \right\} +
K\frac{A_{Fe,Ol}\nu_{Fe,Ol}^{D}}{A_{i,Ol}\nu_{i,Ol}^{D}} \nonumber
\left\{\overline{f}_{i,Ol}\sigma_{i,Ol}^{\mathcal{M}} - \nonumber \right. \\
&- \left.\sigma_{i,Ol}^{D} \right\} \nonumber + K
\left\{\overline{f}_{Fe,Fe}\left(\sigma_{Fe}^{\mathcal{M}}-\sigma_{Fe,Sil}^{D}\right)
-\nonumber \right. \\
&\left. - \left(\sigma_{Fe}^{D}-\sigma_{Fe,Sil}^{D}\right) \right\}
\end{align}
\vspace{-7pt}

\noindent where the subscript $i$ stands for $Si$, $Mg$ or $Fe$
itself. Using the Dirac delta function notation, we have
$A_{i,Pyr}=28\cdot\delta\left(A_{i}-28\right)+24\cdot\delta\left(A_{i}-24
\right)+56\cdot\delta\left(A_{i}-56\right)$. The same expression
holds good for  $A_{i,Ol}$. We have also: $A_{Fe,Pyr}=A_{Fe,Ol}=56$,
$\nu_{Fe,Pyr}^{D}=0.2$ (according to
$\mathrm{Mg}_{\mathrm{x}}\mathrm{Fe}_{1-\mathrm{x}}\mathrm{SiO}_{3}$
with $x=0.8$), $\nu_{i,Pyr}^{D}=1\cdot
\delta\left(A_{i}-28\right)+0.8\cdot\delta\left(A_{i}-24
\right)+0.2\cdot \delta\left(A_{i}-56\right)$, $\nu_{Fe,Ol}^{D}=0.4$
(according to
$\left[\mathrm{Mg}_{\mathrm{x}}\mathrm{Fe}_{1-\mathrm{x}}\right]_{2}\mathrm{SiO}_{4}$
with $x=0.8$) and finally
$\nu_{i,Ol}^{D}=1\cdot\delta\left(A_{i,Ol}-28\right)+1.6\cdot\delta\left(A_{i,Ol}-24
\right)+0.4\cdot\delta\left(A_{i,Ol}-56\right)$. Also, with the
usual meaning, $ \overline{f}_{i,Pyr}$, $\overline{f}_{i,Ol}$ and
$\overline{f}_{Fe,Fe}$ are the average condensations before cloud
dispersion.

For internal consistency,  the system of equations must take into
account that the iron abundance used to describe the evolution of
the iron grains should be corrected for the iron already condensed
into pyroxenes and olivines, subtracting it from
$\sigma_{Fe}^{\mathcal{M}}$ and $\sigma_{Fe}^{D}$. If the
key-element is magnesium, the amount of iron $\sigma_{Fe,Sil}^{D}$
embedded into pyroxenes and olivines is given by:

\vspace{-4pt}
\begin{small}
\begin{align}
\sigma_{Fe,Sil}^{D}=\frac{\sigma^{D}_{Mg}A_{Fe}}{24}\left(
\frac{\nu_{Fe,Ol}^{D}}{\nu_{Mg,Ol}^{D}}Mg_{Ol}
+\frac{\nu_{Fe,Pyr}^{D}}{\nu_{Mg,Pyr}^{D}}Mg_{Pyr} \right)
\end{align}
\end{small}
\vspace{-4pt}

\noindent where $A_{Fe}$ is the atomic weight of the iron and
$Mg_{Ol}$ and $Mg_{Pyr}$ are defined in Sect.
\ref{Equations:Silicon}. Doing the correct substitution we get

\vspace{-7pt}
\begin{equation}
\sigma_{Fe,Sil}^{D}=\sigma_{Mg}^{D}\left(0.583\cdot
Mg_{Ol}+0.583\cdot Mg_{Pyr}\right)
\end{equation}
\vspace{-7pt}

\noindent In a similar way, if  the role of key element is played by
silicon we have:

\vspace{-7pt}
\begin{equation}
\sigma_{Fe,Sil}^{D}=\sigma_{Si}^{D}\left(0.8\cdot F_{Ol}+0.4\cdot
F_{Pyr}\right)
\end{equation}
\vspace{-7pt}

\noindent where $F_{Ol}$ and $F_{Pyr}$ have been defined in Sect.
\ref{Equations:Silicon}. If iron is the key element for silicates,
we simply have:

\vspace{-7pt}
\begin{align}
\sigma_{Fe,Sil}^{D}=\sigma_{Fe}^{D}-\sigma_{Fe,Fe}^{D}
\end{align}
\vspace{-7pt}

This clearly requires to keep track of the accretion/destruction and
injection of the iron grains by SN{\ae} and AGB stars. The accretion
time scales for pyroxenes/olivines are the same as in Eqns.
(\ref{TauAccrPyr}) and (\ref{TauAccrOl}), whereas for iron grains:

\vspace{-7pt}
\begin{align}
\tau_{Fe}^{gr} &= 46\frac{(A_{Fe}^{gw})^{1/2}
\nu_{Fe,Fe}^{D}}{A_{Fe}^{D}} \left(\frac{\rho_{D}^{Fe}}{3}\right)
\left( \frac{10^{3}}{n_{H}}\right) \cdot \nonumber
\\ & \cdot \left(\frac{3.5 \cdot
10^{-5}}{\epsilon_{Fe,Fe}}\right)=\frac{31.56}{n_{H}}\frac{\sigma_{H}}
{\left(\sigma_{Fe}^{\mathcal{M}}-\sigma_{Fe,Sil}^{D}\right)} Myr.
\end{align}
\vspace{-7pt}

\noindent where $\nu_{Fe,Fe}^{D}=1$ and $\rho_{Fe}^{D}=7.86 \,
\mathrm{g} \,\mathrm{cm}^{-\mathrm{3}}$. Concerning the accreting
elements we consider the simplest case in which they are in form of
atoms. The initial condensation fractions for magnesium/silicon as
key-elements in pyroxenes/olivines follow Eqns. (\ref{condMgPyrOli})
and (\ref{condSilPyrOli}), while for iron we have:

\vspace{-7pt}
\begin{equation}
f_{0,Fe,Fe}=\frac{\chi_{Fe}^{D}}{\chi_{Fe,max}^{D}}=
\frac{\sigma_{Fe}^{D}-\sigma_{Fe,Sil}^{D}}{\sigma_{Fe}^{\mathcal{M}}-\sigma_{Fe,Sil}^{D}}.
\end{equation}
\vspace{-7pt}

\noindent where we subtract the amount of iron already locked up in
silicates and therefore not available to for iron grains.

\subsection{Calcium}  \label{Equations:Calcium}

This element has a complicate behaviour difficult to follow. First
of all it is usually heavily depleted in the ISM
\citep{Whittet03,Tielens05}, and its  abundance in the solar system
is  quite low. The difficult arises from the total condensation
efficiency and the big fluctuations generated by the low abundance,
compared to other refractory elements. Second there is not an
average molecule that could be used to represent typical calcium
grains in a simple theoretical description. Furthermore, the
measurements of  depletion cannot be easily derived  from
observational data: in principle the ionization equilibrium equation
should be solved to derive the Ca abundance from the observations of
Ca II. Possible  estimates of Ca abundances by electron densities
and strengths of the ionizing radiation fields are not easy and
relying on ratios of ionization and recombination rates between
different elements (like CaII to those of NaI or KI for instance) is
a cumbersome affair \citep{Weingartner01c}. For all these reasons,
\citet[][and 2011, private communication]{Jenkins09} leaved Ca (and
also Na and K) aside. Trying to overcome this difficulty, we
simplify the problem as follows. Thanks to its low abundance, we
consider Ca as the key-element of the associated grains of dust; the
equation for the evolution of Ca is:

\vspace{-7pt}
\begin{equation} \label{Calcium_evolution}
\left(\frac{d \sigma_{Ca}^{D}}{dt}\right)_{accr} = K
\left\{\overline{f}_{Ca,Ca} \sigma_{Ca}^{\mathcal{M}}\left(Ca\right)
- \sigma_{Ca}^{D}\left(Ca\right) \right\} \\
\end{equation}
\vspace{-7pt}

\noindent where all the  symbols have their  usual meaning. To
derive the accretion time scale  $\tau_{Ca}^{gr}$  we do not refer
to a typical Ca dust grain  but  simply take  \textit{the shortest}
timescale among those  of  refractory elements that are most
depleted, i.e.  $\mathrm{Mg}$, $\mathrm{Si}$ and $\mathrm{Fe}$
$(\tau_{Si}^{gr},\tau_{Mg}^{gr},\tau_{Fe}^{gr})$,  and a timescale
$\tau_{Ca,Ca}^{gr}$ for the formation of dust silicates-like
molecules with one calcium atom as key atom. This last-named
timescale is multiplied for a correction factor $\mathrm{Ca}_{X}$,
eventually allowing for a fast accretion. Therefore
$\tau^{gr}_{Ca}=\min\{
\tau_{Si}^{gr},\tau_{Mg}^{gr},\tau_{Fe}^{gr},Ca_{X}\cdot
\tau_{Ca,Ca}^{gr}\}$. The initial condensation fraction at the MC
formation  is as usual
$f_{Ca,Ca}=\sigma_{Ca}^{D}/\sigma_{Ca}^{\mathcal{M}}$.

\subsection{Sulfur}  \label{Equations:Sulfur}

Sulfur is a very important element: it is often used as a reference
case of nearly zero depletion in studies of local and distant
objects \citep{Jenkins09}. However the real depletion efficiency of
this element is a matter of debate and the assumption of nearly zero
depletion cannot be  safe. \citet{Calura09} reviews data and
theoretical interpretations gathered over the past years  to
convincingly  show  that $\mathrm{S}$ can be depleted in
considerable amounts.  \citet{Jenkins09} points out  the depletion
of sulfur can be significant along some lines of sight. We take
\citet{Jenkins09}  results into account   to set upper and lower
limits to the sulfur depletion. Considering sulfur as a refractory
element, its evolution in dust  is given by

\vspace{-7pt}
\begin{equation} \label{Sulfur_evolution}
\left(\frac{d \sigma_{S}^{D}}{dt}\right)_{accr} = K
\left\{\overline{f}_{S,S} \sigma_{S}^{\mathcal{M}}\left(S\right) - \sigma_{S}^{D}\left(S\right) \right\} \\
\end{equation}
\vspace{-7pt}

\noindent where all the  symbols have their usual meaning. Current
observations do allow us to choose a grain as representative of the
accretion process.  In analogy to what made for calcium,  the
accretion time $\tau_{S,gr}$ is supposed to be  \textit{the longest}
between the timescales of the refractory elements, i.e.
$\mathrm{Mg}$, $\mathrm{Si}$ and $\mathrm{Fe}$, and a timescale
$\tau_{S,S}^{gr}$ for the formation of dust silicates-like molecules
with one sulfur atom as key atom, allowing therefore for a slow
accretion. We also introduce a multiplying scaling factor $S_{X}$ to
eventually correct this timescale. Therefore
$\tau^{gr}_{S}=S_{X}\cdot\max\{
\tau_{Si}^{gr},\tau_{Mg}^{gr},\tau_{Fe}^{gr},\tau_{S,S}^{gr}\}$. The
initial condensation fraction for $\overline{f}_{S,S}$ is
$f_{S,S}=\sigma_{S}^{D}/\sigma_{S}^{\mathcal{M}}$.

\subsection{Nitrogen}  \label{Equations:Nitrogen}

Nitrogen is known to be poorly depleted and if not depleted at all
\citep{Whittet03,Tielens05}. \citet{Jenkins09} suggests that
depletion is independent from the line of sight depletion strength
factor and that in any case depletion is very low, thus confirming
the poor ability  of nitrogen  to condense into  dust grains. As
nitrogen is included in our list of elements (both
$^{\mathrm{14}}\mathrm{N}$ and $^{\mathrm{15}}\mathrm{N}$), the
evolution of both isotopes in dust is governed by:

\begin{footnotesize}
\begin{equation} \label{Nitrogen_evolution}
\left(\frac{d \sigma_{N}^{D}}{dt}\right)_{accr} = K
\left\{\overline{f}_{N,N} \sigma_{N}^{\mathcal{M}}\left(N\right) -
\sigma_{N}^{D}\left(N\right) \right\} \\
\end{equation}
\end{footnotesize}

\noindent where for the accretion time we simply take the
\textit{longest} between $\mathrm{C}$ and $\mathrm{O}$ and the
accretion timescale with one nitrogen atom as growing species for a
nitrogen molecule. Carbon and oxygen are the nearest elements, both
do not show a strong depletion and, finally, similarly to nitrogen
both indicate (at least along some lines of sight)  low values of
depletion. Taking the longest timescale, we implicitly assume that
nitrogen very slowly accretes onto dust. As usual
$f_{N,N}=\sigma_{N}^{D}/\sigma_{N}^{\mathcal{M}}$.

\section{Supernovae: destruction time scales} \label{SNEdestruction}

Dust grains in the ISM can be destroyed by other
physical process such as the passage of shock waves by supernov{\ae}
explosions. The destruction time scale of the element $i$-h in dust
grains in the ISM because of the shocks by local SN{\ae} is defined
as the ratio between the available amount of that element locked up
into dust (the surface mass density of the element $i$-th
$\sigma_{i}^{D}\left(r,t \right)$, at a given radius and
evolutionary time) and the rate at which grains containing that
element are destroyed refueling the gaseous phase
:

\vspace{-7pt}
\begin{equation}\label{TaudestructionA}
\tau_{SNR,i}=\sigma_{i}^{D}/\left( \frac{\displaystyle
d\sigma_{i}^{D} }{\displaystyle dt}\right)_{SNR}
\end{equation}
\vspace{-7pt}

\noindent as usual, we drop the dependence from $r$ and $t$. The
destruction time scale can be expressed as  follows:

\vspace{-7pt}
\begin{equation} \label{destruction}
\left( \frac{d\sigma_{i}^{D} }{dt}\right)_{SNR} = M^{destr}_{i}\cdot
R_{SN}
\end{equation}
\vspace{-5pt}

\noindent where $M^{destr}_{i}=M^{destr}_{i} \left(r,t\right)$ is
the amount of mass destroyed by a single SNa event, while $R_{SN} =
R_{SN}\left(r,t\right)$ is the global rate of SN{\ae}, obtained by
adding together the rate of Type I and II SN{\ae}. Multiplying
numerator and denominator by $\sigma_{i}^{D}$ and combining together
Eqns. (\ref{TaudestructionA}) and (\ref{destruction}) it follows:

\vspace{-7pt}
\begin{equation}\label{TaudestructionB}
\tau_{SNR,i}=\sigma_{i}^{D}/\left(\frac{
d\sigma_{i}^{D}}{\displaystyle dt}\right)_{SNR}=
\frac{1}{R_{SN}}\frac{\sigma_{i}^{D}}{M^{destr}_{i}}\, .
\end{equation}
\vspace{-7pt}

\noindent The amount of mass destroyed by the interstellar shocks
can be defined as the amount of mass swept by the SNa shock,
multiplied by the fraction of dust mixed with the ISM (in this way
we get the total swept up mass of dust) and finally by a factor of
dust destruction $\epsilon_{i}$ that depends on the element we are
taking into account:

\vspace{-7pt}
\begin{equation}
M^{destr}_{i}=\frac{\sigma_{i}^{D}} {\sigma_{i}^{\mathcal{M}}}\cdot
M_{swept} \cdot \epsilon_{i}\, .
\end{equation}
\vspace{-7pt}

\noindent Inserting this expression into  Eqn.
(\ref{TaudestructionB}) we have:

\vspace{-7pt}
\begin{equation} \label{tauSNRfinal}
\tau_{SNR,i} = \frac{1}{R_{SN}}\frac{\sigma_{i}^{D}}{M^{destr}_{i}}=
\frac{1}{M_{swept}
\epsilon_{i}}\left(\frac{\sigma_{i}^{\mathcal{M}}} {R_{SN}}\right)\,
.
\end{equation}
\vspace{-7pt}

\noindent We need now to get an estimate of the mass swept up by a
SNa shock propagating through  the ISM. To describe the evolution of
the SNa remnant we follow \citet{Cioffi88} and \citet{Gibson94b}. In
brief, the evolution of the remnant after the SN{a} explosion is
subdivided into three main  phases: (i) free expansion that lasts
until the mass of the swept up material is comparable to the mass of
the expelled material; (ii) adiabatic expansion, or so-called
Sedov-Taylor phase, lasting until when the radiative cooling time of
the shocked gas is about equal to the expansion time of the remnant;
(iii) radiative expansion, with the formation of a cold and dense
shell behind the shock front, starting when at least some sections
of the shocked gas have radiated most of their thermal energy
\citep{Ostriker88}. Phase (i) has a simple solution that can be
obtained using the obvious relation $\left(4/3\right)\pi
R^{3}\rho_{0}=M_{SN}$ where $M_{SN}$ is the mass of the expelled
material. The duration of the phase is  $\tau = R/v \approx 200$
years (very short indeed). For the phase (ii) we have the classical
auto-similar adiabatic solution of Sedov-Taylor \citep{Ostriker88};

\vspace{-7pt}
\begin{equation} \label{SedovTaylor}
R_{s}\left(t\right)=1.15\left(\frac{E_{0}}{\rho_{g}\left(t\right)}\right)^{1/5}t^{2/5}
\end{equation}
\vspace{-7pt}

\noindent where $E_{0}$ is the energy of the blast wave in units of
$10^{50}$ erg and $\rho_{g}\left(t\right)$ is the density of the
environment. The radiative cooling leads to the formation of a thin
and dense shell at $t_{sf}$:

\vspace{-7pt}
\begin{equation} \label{tsf}
t_{sf}=3.61\cdot 10^{4}\cdot
\varepsilon_{0}^{3/14}n_{H}^{-4/7}\left(\frac{Z}{Z_{\odot}}\right)^{-5/14}
\end{equation}
\vspace{-7pt}

\noindent $n_{\mathrm{H}}$ is the number density of hydrogen atoms,
$Z$  the metallicity of the ISM and $Z_{\odot}$  the solar value.
$\varepsilon_{0}$ is the energy in units of $10^{51}$ erg. The blast
wave decelerates until when the radiative losses start  dominating.
At about $t_{PDS}\approx 0.37t_{sf}$, the  remnant enters the
so-called Pressure Driven Snowplow phase (PDS) lasting or Phase
(iii). In the early stages of the PDS phase, the radius of the
remnant changes according to the following equation (where
$t_{SN}=t-t^{\prime}$):

\vspace{-7pt}
\begin{equation}
R_{S}\left(t_{SN}\right)=R_{PDS}\left(\frac{4}{3}\frac{t_{SN}}{t_{PDS}}-\frac{1}{3}\right)^{3/10}
\end{equation}
\vspace{-7pt}

\noindent where the radii are in parsec. $R_{PDS}$ is the radius at
the start of the PDS phase:

\vspace{-7pt}
\begin{equation}
R_{PDS}=14\cdot
\varepsilon_{0}^{2/7}n_{0}^{3/7}\left(\frac{Z}{Z_{\odot}}\right)\, .
\end{equation}
\vspace{-7pt}

\noindent The inner side of the shock continuously looses energy
because of the radiative cooling and it pushes forward the shell in
the ISM. At the time $t_{merge}$, given by:

\vspace{-7pt}
\begin{equation}
t_{merge}=21.1 \cdot
t_{sf}\varepsilon_{0}^{5/49}n_{0}^{10/49}\left(\frac{Z}{Z_{\odot}}\right)^{15/49}\,
.
\end{equation}
\vspace{-7pt}

\noindent The remnant and the ISM loose their identity as single
entities and merge together. Subsequently, in the time interval  for
$t_{merge}\leq t_{SN}\leq t_{cool}$ the radius of the shell is given
by:

\vspace{-7pt}
\begin{equation}
R_{merge}=3.7
R_{PDS}\varepsilon_{0}^{3/98}n_{0}^{3/49}\left(\frac{Z}{Z_{\odot}}\right)^{9/98}
\end{equation}
\vspace{-7pt}

\noindent in parsec, where the time scale $t_{cool}$ is:

\vspace{-7pt}
\begin{equation}
t_{cool}=203t_{sf}\left(\frac{Z}{Z_{\odot}}\right)^{-9/14}
\end{equation}
\vspace{-7pt}

\noindent in years. There are four time intervals, with the
corresponding radii, to consider:

\vspace{-7pt}
\begin{small}
\begin{equation}\label{Radii}
\begin{array}{ll}
R_{S}\left(t_{SN}\right)=1.15\left(\frac{E_{0}}{\rho_{g}\left(t\right)}
\right)^{\frac{1}{5}}t_{SN}^{\frac{2}{5}}  \, \, \, \qquad \qquad 0 \leq t_{SN} < t_{PDS}; \\
R_{S}\left(t_{SN}\right)=R_{PDS}\left(\frac{4}{3}\frac{t_{SN}}{t_{PDS}}-\frac{1}{3}\right)^{\frac{3}{10}}
\,  t_{PDS} \leq t_{SN} < 1.17t_{sf}; \\
R_{S}\left(t_{SN}\right)=R_{PDS}\left(\frac{4}{3}\frac{t_{SN}}{t_{PDS}}-\frac{1}{3}\right)^{\frac{3}{10}}
\,  1.17t_{sf} \leq t_{SN} < t_{merge}; \\
R_{S}=3.7R_{PDS}\varepsilon_{0}^{\frac{3}{98}}n_{0}^{\frac{3}{49}}\left(\frac{Z}{Z_{\odot}}\right)^{\frac{9}{98}}
\qquad t_{merge} \leq t_{SN} < t_{cool}\, .
\end{array}
\end{equation}
\end{small}
\vspace{-7pt}

\noindent To obtain the shock velocity in the various time
intervals, we need the   $R_{S}$ as functions of time. Approximating
$\rho_{g}\left(t\right)\approx \overline{\rho}_{g}$, we get:

\vspace{-7pt}
\begin{small}
\begin{equation}\label{Radii}
\begin{array}{ll}
v_{S}\left(t_{SN}\right)=\frac{2}{5}\frac{R_{S}\left(t\right)}{t}  \, \, \, \qquad \qquad \qquad \qquad \qquad 0 \leq t_{SN} < t_{PDS}; \\
v_{S}\left(t_{SN}\right)=\frac{2}{5}\frac{R_{PDS}}{t_{PDS}}\left(\frac{4}{3}\frac{t}{t_{PDS}}-\frac{1}{3}\right)^{-\frac{7}{10}}
t_{PDS} \leq t_{SN} < 1.17t_{sf}; \\
v_{S}\left(t_{SN}\right)=\frac{2}{5}\frac{R_{PDS}}{t_{PDS}}\left(\frac{4}{3}\frac{t}{t_{PDS}}-\frac{1}{3}\right)^{-\frac{7}{10}}
1.17t_{sf} \leq t_{SN} < t_{merge}; \\
v_{S}\left(t_{SN}\right)=0 \qquad \qquad \qquad \qquad \qquad \qquad
t_{merge} \leq t_{SN} < t_{cool}\, .
\end{array}
\end{equation}
\end{small}
\vspace{-7pt}

\noindent After $t_{merge}$ there is no longer expansion and we
simply have $v_{S}\left(t_{SN}\right)=0$. How do we proceed in the
chemical simulation? First of all we need $n_{\mathrm{H}}$, the
number density of hydrogen atoms. This is a spatial density, in
atoms per cubic centimeter; we must derive it  from the surface mass
density of hydrogen in our flat geometry. This is done assuming as
parameters the scale height and thickness of the disk suggested by
the observational data. Once  $n_{\mathrm{H}}$ is known, we derive
the time scales $t_{sf}$, $t_{PDS}$ and $t_{merge}$,  and finally
the radii of the remnant and  velocity of the shock at the beginning
and  end of the three evolutionary phases we have described. At this
point we must fix the velocity $v_{low}$ setting the limit at which
the shock stops disrupting  the dust grains. A typical value is
about  $50$ km/s \citep[see][for a thorough discussion of this
point]{Nozawa06,Nozawa07}. Knowing the limit velocity, we can soon
determine  the time in  which the remnant expanding at the limit
velocity covers a distance equal to its own radius. The swept up
mass is:

\vspace{-7pt}
\begin{equation}\label{Msweptup}
M_{swept}=\frac{4}{3}\pi R\left(v_{low}\right)^{3}n_{H}\, .
\end{equation}
\vspace{-7pt}

The other factors of Eqn. (\ref{tauSNRfinal}) are the ratio
$\displaystyle \sigma_{i}^{\mathcal{M}}/\displaystyle R_{SN}$  and
the coefficients $\epsilon_{i}$ that describe the fraction of the
element $i$-th condensed into dust that is destroyed by the shock.
While the ratio  $\displaystyle
\sigma_{i}^{\mathcal{M}}/\displaystyle R_{SN}$ is simply provided by
the chemical model in use, to calculate $\epsilon_i$ we refer to the
studies by \citet{Nozawa06,Nozawa07}. In brief, the \citet{Nozawa06}
data are used in the first evolutionary phases of the galaxies to
derive the destruction coefficients for the ISM dust. From Table
$\left(6\right)$ of \citet{Nozawa06} we take the $a_{i}$ and $b_{i}$
to be used in:

\vspace{-7pt}
\begin{equation}
\epsilon_{j}=a_{j}E_{51}^{b_{j}}
\end{equation}
\vspace{-7pt}

\noindent where $E_{51}$ is the energy of the supernova in units of
$10^{51}$ erg. For simplicity we assume $E_{51}=1$, typical energy
of the explosion of SN{\ae} of mass not too high, that according to
the IMF are also the most numerous. In this way we get the
coefficients to be used for destruction by forward shocks in the
ISM. Furthermore, the results by \citet{Nozawa07} are used to
constrain the yields by the  SN{\ae} progenitors, before  dust is
injected into the ISM where it will interacts with the  shocks
generated by the surrounding SN{\ae}. Therefore we take into account
the amount of material injected into the ISM \citep{Nozawa03}
corrected for the effect of the internal reverse shock.

\section{Solar System Abundances} \label{AbundancesSS}

To validate the results of our model we need a reference set of
chemical abundances to compare. They are also known as the
\emph{cosmic reference abundances}.
Usually, the Solar System abundances provide the comparison set of
values, but other choices are possible
\citep{Savage96,Asplund09,Whittet10}. Since the solar system
abundances can not be determined directly from measurements of the
ISM, because of the depletion of the elements accreted into dust,
two main sources of information are usually adopted: spectroscopic
data inferred from the Sun and Meteoritic Abundances.

\indent Photospheric abundances of the Sun can not be determined
directly, but only by means of a model atmosphere of the solar
spectrum. The model must be  able to calculate lines formation
(classical 1D or hydrodynamical 3D), and takes into account radiative
transfer and non-LTE processes \citep[see][for a review about solar
abundances obtained from spectroscopy]{Basu08}. However, since there
are no lines of noble gases in the solar photosphere, coronal lines,
energetic particles or solar wind are also studied, in particular to
determine the crucial abundances of He and Ne.

 \indent Between
the tens of thousands meteorites fell on the earth, only five are C1
carbonaceous chondrites \citep{Lodders09}. They form the only group
of meteorites allowing us to determine the abundances in the
proto-solar environment that match the solar ones, and keep some
memory of the original proto-environment. However, volatile elements
and noble gases like H, He, C, N, O, and Ne, therefore
including  the most abundant elements, are heavily depleted in these
meteorites. We can not rely on mass spectrometry to safely determine
their abundances, but an additional source of information must be
adopted.

\indent It is  common practice  to combine  photospheric and
meteoritic measurements to get  the compilations of abundances
commonly used in literature and continuously updated year after year
\citep{Anders89,Grevesse93,Grevesse98,Lodders03,Asplund05a,Grevesse07,Lodders09,Asplund09}.

\renewcommand{\arraystretch}{1.3}
\setlength{\tabcolsep}{2.8pt}
\begin{table}
\scriptsize
\begin{center}
\caption[]{\footnotesize Chemical abundances for the Sun at the
present age according to the classical compilation by
\citet{Grevesse98} (GS98 - photospheric solar abundances) and the
more recent ones by \citet{Asplund09} (AGSS09 - photospheric solar
abundances) and \citet{Lodders09} (LPG09 - photospheric and
meteoritic abundances compilation). Finally, in the last column we
present the abundances we have adopted in this study. The Abundance
$\mathrm{A}(\mathrm{X})$ of the element $\mathrm{X}$ is in units of
$\log_{10}\left(\mathrm{N(X)}/\mathrm{N(H)}\right)+12$. Only the
elements included in our chemical model are reported and discussed}
\begin{tabular}{ccccc}
\hline \hline \vspace{0.1cm}
\small Element &  \small GS98   & \small AGSS09    & \small LPG09 & \small This work \\
\hline
\small H    &\small 12.00           &\small 12.00            &\small 12.00                            &\small 12.00                 \\
\small He   &\small 10.93$\pm$0.004 &\small 10.93$\pm$0.01   &\small 10.93$\pm$0.02                   &\small 10.93$\pm$0.02        \\
\small C    &\small 8.52$\pm$0.06   &\small 8.43$\pm$0.05    &\small 8.39$\pm$0.04\footnotemark[1]    &\small 8.50$\pm$0.06\footnotemark[7]                 \\
\small N    &\small 7.92$\pm$0.06   &\small 7.83$\pm$0.05    &\small 7.86$\pm$0.12\footnotemark[2]    &\small 7.83$\pm$0.05                \\
\small O    &\small 8.83$\pm$0.06   &\small 8.69$\pm$0.05    &\small 8.73$\pm$0.07\footnotemark[3]    &\small 8.73$\pm$0.06                 \\
\small Ne   &\small 8.08$\pm$0.06   &\small 7.93$\pm$0.10    &\small 8.05$\pm$0.10\footnotemark[4]    &\small 7.99$\pm$0.10                \\
\small Mg   &\small 7.58$\pm$0.05   &\small 7.60$\pm$0.04    &\small 7.54$\pm$0.06\footnotemark[5]    &\small 7.57$\pm$0.05     \\
\small Si   &\small 7.55$\pm$0.05   &\small 7.51$\pm$0.03    &\small 7.53$\pm$0.01\footnotemark[6]    &\small 7.53$\pm$0.01     \\
\small S    &\small 7.33$\pm$0.11   &\small 7.12$\pm$0.03    &\small 7.16$\pm$0.02\footnotemark[5]    &\small 7.14$\pm$0.03     \\
\small Ca   &\small 6.36$\pm$0.02   &\small 6.34$\pm$0.04    &\small 6.31$\pm$0.02\footnotemark[6]    &\small 6.34$\pm$0.04    \\
\small Fe   &\small 7.50$\pm$0.05   &\small 7.50$\pm$0.04    &\small 7.46$\pm$0.08\footnotemark[5]    &\small 7.50$\pm$0.04   \\
\hline
Z/X  &\small 0.0231$\pm$0.018  & -   & - &   -   \\
\hline \label{SSabundances}
\end{tabular}
\end{center}
\renewcommand{\arraystretch}{1}
\begin{flushleft}\footnotesize
\footnotesize$^{1}${This is the same low photosphere value as in
\citet{Lodders03}, selected from \citet{AllendePrieto02}, and
confirmed in \citet{Asplund05b} and
\citet{Scott06}.}\,\footnotesize$^{2}${N is taken from
\citet{Caffau09a}, from solar photospheric
models.}\,\footnotesize$^{3}${The O abundance is an average from
solar photospheric models by
\citet{Caffau08a,Ludwig08,Melendez08}.}\,\footnotesize$^{4}${Ne
abundance is an average from
\citet{Morel08,Landi07}.}\,\footnotesize$^{5}${Average between solar
and meteoritic values.}\,\footnotesize$^{6}${Meteoritic
value.}\,\footnotesize$^{7}${\citet{Caffau10}.}
\end{flushleft}
\end{table}
\renewcommand{\arraystretch}{1}

Two points are worth of mention. (i) According to the most recent
models, the present photospheric abundances of the Sun are lower
than the bulk abundances of the proto-Sun formed about $4.6$ Gyr ago
because of the physical processes taking place in the deep
convective region under the solar surface. The effects of thermal
diffusion, gravitational settling due to differential gravity and
radiative acceleration \citep{Turcotte02} allow  helium and heavy
elements to deposit in the interiors of the Sun.  The decay of
radioactive elements affects the isotopic compositions
\citep{Piersanti07,Lodders09}. The correction needed to obtain the
unfractioned abundances of the proto-Sun is of about $0.05$ dex for
He and $0.04$ dex for heavier elements \citep{Asplund09} or slightly
higher of $0.01$ dex for both corrections \citep{Lodders09}. With
these corrections there is agreement with the abundances observed in
nearby B stars by \citet{Przybilla08}. (ii) The He and Ne abundances
determined with coronal lines or solar wind suffer from the FIP
effect.  Therefore the most precise method to determine the He
abundance is helioseismology and the adiabatic index $\Gamma_{1}$,
whereas for Ne abundance are the photospheric ratios
$\mathrm{Ne}/\mathrm{O}$ and $\mathrm{Ne}/\mathrm{Mg}$ together with
their
systematic errors \citep{Basu08}.\\
 \indent Starting from
\citet{AllendePrieto01,AllendePrieto02}, the widely used compilation
of abundances  by \citet{Grevesse98} has been the subject of a
continuous revision of  the photospheric C, N, O abundances toward a
significantly lower value, i.e. about 0.2 dex lower than before
\citep{Asplund04,Asplund05a,Asplund05b}. These new values, derived
from 3D hydrodynamical models with updated input physics and no
macro- and micro-turbulence,  bring the Sun to better agree with the
metal content of the Galactic neighbourhood \citep{TurckChieze04}.
They  are also in good agreement with recent determinations of
B-star abundances \citep{Nieva08a,Nieva08b} and carbonaceous C1
chondrites for many elements \citep{Asplund05b,Lodders09,Asplund09}.
There is also  a better agreement between different diagnostics(OI,
$[\mathrm{\mathrm{OI}}]$ and OH vibrational and roto-vibrational
bands for Oxygen, CI, $[\mathrm{\mathrm{CI}}]$, CH, C$_{2}$ and CO
for Carbon), but a strong disagreement with the helioseismological
models \citep{Antia05,Bahcall05a,Bahcall05b,Antia06,Basu08}. In
brief, the position of the base of the convective zone, the helium
abundance, sound speed and density profiles are all affected by the
revised abundances \citep{Basu08}. Many suggestions have been
advanced to solve this discrepancy, e.g.  increased input opacities,
increased abundances of  neon or other elements, fine tuning of the
diffusion process, inclusion of other additional physical processes
etc. All this, however,  without removing the discrepancy.  The
recent revision of the abundance compilation  by \citet{Asplund09},
reversed the trend  toward higher abundances, thus in closer
agreement with helioseismology, even if the differences are still
significant. Also, a series of models calculated over  the past few
years (all based on the so-called CO$^{5}$BOLD 3D hydrodynamical
code) go toward a reconciliation with helioseismology
\citep{Caffau08a,Caffau08b,Caffau09a,Ludwig09,Maiorca09,Caffau10}.
Anyway, as pointed out in \citet{Caffau10}, results from different
3D hydrodynamical simulations may still  mutually  differ by as much
as  0.1 dex, suggesting that better validation of the hydrodynamical
models and  more and more updated solar models are needed
\citep{TurckChieze08}.

In Table \ref{SSabundances} we show the classical set of abundances
by \citet{Grevesse98} (GS98), the two most recent compilations
available in literature by \citet{Asplund09} (AGSS09) and
\citet{Lodders09} (LPG09) and, finally, in the last column, our
adopted reference set for the present-day solar abundances. The
abundances by number of the various elements are indicated by
$\mathrm{A}(\mathrm{X})$ which stands for
$\log_{10}\left(\mathrm{N(X)}/\mathrm{N(H)}\right)+12$.

 \indent Let us now examine in some detail
the  choice we made  for the abundances of the various elements
included in our model:

\textit{\textbf{Carbon}}: the carbon abundance is A(C)=8.50$\pm$0.06
according to  \citet{Caffau10}. It corresponds to a weak efficiency
of the collisions with neutral hydrogen atoms
$(\mathrm{S}_{\mathrm{H}} = 1/3)$. The uncertainty  is the sum of
statistical $(\mathrm{0.02})$ and systematic $(\mathrm{0.04})$
errors. This value is similar to the old GS98 determination and very
close to the average estimate of the carbon abundance during the
past  thirty years \citep[see Table 4 in][]{Caffau10}.

\textit{\textbf{Nitrogen}}: the abundance of Nitrogen determined by
AGSS09 does not much differ from that by LPG09, but it is
significantly lower than GS98. We adopt the  value by AGSS09
A(N)=7.83$\pm$0.05. As this is derived  from both atomic and
molecular lines,  the error is minimized.

\textit{\textbf{Oxygen}}: Oxygen, together with Neon, are the most
debated and uncertain elements \citep{Asplund09}. LPG09 adopted the
average value between the estimate from CO$^{5}$BOLD atmosphere
models by \citet{Caffau08a} (8.76$\pm$0.07), that by
\citet{Ludwig08} (8.72$\pm$0.06), and the one by \citet{Melendez08}
(8.71$\pm$0.02). AGSS09 recommends 8.69$\pm$0.05 obtained from
atmosphere models with updated physics and taking the mean value of
estimates based on  [OI], OI, OH vibration-rotation and OH pure
rotation. \citet{Whittet10} adopts 8.76$\pm$0.03 based upon
observations of nearby B-stars \citep{Przybilla08}. Averaging  all
these values, we adopt 8.73$\pm$0.06.

\textit{\textbf{Neon}}: the abundance of Neon proposed by AGSS09
(7.93$\pm$0.10) is determined from  their  O abundance and the Ne/O
ratio. LPG09 (8.05$\pm$0.10) takes the mean value  between
measurements in B-type stars and UV-flares. The proposed values are
very different. We adopt the average value of 7.99$\pm$0.10. The
corresponding  Ne/O ratio is $0.182$ in very good agreement with
\citet{Young05} and \citet{Schmelz05}.

\textit{\textbf{Magnesium}}: magnesium \textit{gf}-values for the
two ionization stages MgI and MgII suffer of well know
uncertainties, as underlined in AGSS09. For this reason we adopt the
 weighed average between the LPG09 meteoritic abundance
7.55$\pm$0.01 and the AGSS09 value, obtaining 7.55$\pm$0.05.

\textit{\textbf{Silicon}}: silicon is the element linking meteoritic
and photospheric measurements. Since  H is very depleted in C1
chondrites, meteoritic abundances are usually expressed  in the
cosmochemical scale as  number of atoms per $10^{6}$ Si atoms. If
the abundance by number of $\mathrm{Si}$ is known, we may easily obtain the
abundance by number of any element in the usual scale
$\mathrm{A(X)}=\log_{10}\left(\mathrm{N(X)}/\mathrm{N(H)}\right)+\mathrm{12}$. The
various estimates of the Si abundance listed in Table
\ref{SSabundances} agree quite well each other. Therefore we can
adopt the estimate by   LPG09  that is the average  between the
solar photospheric and meteoritic values.

\textit{\textbf{Sulfur}}: the sulfur abundance has been revised by
AGSS09, including additional lines and NLTE corrections. We take the
mean between  this value and the meteoritic abundance by LPG09 (see
their Table 4).

\textit{\textbf{Calcium}}: for the abundance of Calcium we choose
the value proposed by AGSS09, which is in good agreement with LPG09
and GS98 and is based upon recent NLTE abundance corrections by
\citet{Mashonkina07}.

\textit{\textbf{Iron}}: we adopt the value given by AGSS09, in
excellent agreement with the meteoritic measurement by LPG09  and
GS98.

\indent Once the reference abundances are assigned, we can tackle
the element depletion in the ISM, that is the under-abundance of
some elements compared to the corresponding reference value, because
a fraction of the element under consideration is  locked up in the
interstellar dust. The depletion of the element X in the ISM is
measured in the following way. Following \citet{Jenkins09} and \citet{Whittet10}
we define

\begin{equation} \label{Depletion}
\left[X_{gas}/H\right]=log\left[N\left(X\right)/N\left(H\right)\right]-log\left(X/H\right)_{ISM}
\end{equation}

\noindent where $\mathrm{N}\left(\mathrm{X}\right)$ is the column
density of the element X and
$\mathrm{N}\left(\mathrm{H}\right)=\mathrm{N}\left(\mathrm{HI}\right)+\mathrm{N}\left(\mathrm{H}_{\mathrm{2}}\right)$
the same for
 hydrogen and, finally, and $\left(\mathrm{X}/\mathrm{H}\right)_{\mathrm{ISM}}$ the assumed
reference abundance. If part of the element X is  locked up into
dust grains ($[\mathrm{X}_{\mathrm{gas}}/\mathrm{H}]\leq0$),  the
fractional abundance of the element depleted into dust is

\begin{equation} \label{DustDepletion}
\left(X_{dust}/H\right)=\left(X/H\right)_{ISM}\cdot\left(1-10^{\left[X_{gas}/H\right]}\right)\,
.
\end{equation}

Usually, establishing the  degree of depletion of an element is a
cumbersome affair, because  the choice of the reference abundance is
a difficult task  and errors and inconsistencies affect the
measurements \citep{Whittet03,Whittet10}. Recently,
\citet{Jenkins09} presented a thorough study of the depletion for
$17$ elements along $243$ lines of sight, trying to focus on the
rates of depletion and leaving aside the problem of the reference
abundance. He was able to characterize the degree of depletion
according to three parameters:  one describes the overall level of
depletion along a particular line of sight; the other two are
related to the depletion of each element along that line. Once
adopted the reference abundances proposed by \citet{Lodders03} for
the proto-sun, \citet{Jenkins09} presents the depletion parameters
for the smallest and largest  depletion efficiencies (his parameter
$F_*$), thus bracketing the region to be matched by any theoretical
model. In the context of the present study,  to correctly make use
of the \citet{Jenkins09} results, the following remark is
appropriate. \citet{Jenkins09} analysis stems from the
\citet{Lodders03} present reference abundances (see his Table 1)
which is  different from the one we have adopted (see the entries of
Table  \ref{SSabundances}. Therefore, a suitable shift has to be
applied.  Furthermore, the \citet{Lodders03} reference abundances
for  the proto-solar environment have been  obtained from the
present-day solar+meteoritic ones by adding $\sim 0.07$, which is
likely  too a correction \citep{Jenkins09,Lodders09,Przybilla08}.
Therefore, we derive our present-day reference abundances applying
the more realistic correction suggested by \citep{Asplund09}.

\renewcommand{\arraystretch}{1.4}
\setlength{\tabcolsep}{2.8pt}
\begin{table*}
\scriptsize
\begin{center}
\caption[]{\footnotesize  In  column (1) we list the elements we
have considered. In columns (2) and (3) we show the chemical
abundances for the proto-solar environment, according to the
\citet{Lodders03} compilation adopted by \citet{Jenkins09} and the
values we have adopted. In columns (4) and (5)  we give
$\left[\mathrm{X}_{\mathrm{gas}}/\mathrm{H}\right]_{0}$ and
$\left[\mathrm{X}_{\mathrm{gas}}/\mathrm{H}\right]_{1}$, the
smallest and largest  depletion values for the various elements are
obtained from \citet{Jenkins09}, with a correction to account for
the difference between the reference set of elemental abundances. In
column (6) we list the abundances in the ISM and in columns (7) and
(8)   we show the abundances of elements in dust in units of
$10^{6}\left(\mathrm{X}_{\mathrm{dust}}/\mathrm{H}\right)$, for the
smallest and largest depletions, respectively. Finally columns (9)
and (10) show the abundances for the Warm Disk and the Cool Disk,
according to the fit made by  \citet{Jenkins09} of the
\citet{Savage96} data}
\begin{tabular}{cccccccccc}
\hline \vspace{0.1cm} \small Element &\small
$A\left(X\right)$\footnotemark[1]   &\small
$A\left(X\right)$\footnotemark[2]  & \small $\left[{X_{gas}\over
H}\right]_{0}$ & \small $\left[{X_{gas}\over H}\right]_{1}$
& \small $\left({X_{ISM} \over H} \right)_{\odot}$
& \small $\left({X_{dust}\over H} \right)_{0}$  & \small $\left({X_{dust}\over H} \right)_{1}$
& \small $\left[{X_{gas}\over H} \right]_{WD}$ & \small $\left[{X_{gas}\over H} \right]_{CD}$\\
\small  &  &   &  &  &\small ($\cdot10^{6}$)   & \small ($\cdot10^{6}$)   & \small ($\cdot10^{6}$) & & \\
\hline
(1)  &  (2) & (3)& (4) & (5) & (6) & (7) & (8) & (9) & (10) \\
\hline
\small H    &\small 12.00           &\small 12.00            &\small  -                        &\small -                     &\small 10$^{6}$    &\small -                        &\small -                  &\small -     &\small -     \\
\small He   &\small 10.984$\pm$0.02 &\small 10.98$\pm$0.02   &\small  -                        &\small -                     &\small 9.5$\cdot10^{4}$ &\small -                        &\small -                  &\small -     &\small -    \\
\small C    &\small 8.46$\pm$0.04   &\small 8.54$\pm$0.06    &\small  -0.192$\pm$0.194         &\small -0.293$\pm$0.075      &\small 347   &\small 124$^{+80}_{-124}$       &\small 170$^{+28}_{-33}$  &\small -0.204   &\small -0.283  \\
\small N    &\small 7.90$\pm$0.11   &\small 7.87$\pm$0.05    &\small  -0.079$\pm$0.119         &\small -0.079$\pm$0.119      &\small 74    &\small 12$^{+15}_{-19}$         &\small 12$^{+15}_{-19}$   &\small -0.079   &\small -0.079   \\
\small O    &\small 8.76$\pm$0.05   &\small 8.77$\pm$0.05    &\small  -0.020$\pm$0.060         &\small -0.246$\pm$0.055      &\small 588   &\small 27$^{+73}_{-82}$         &\small 255$^{+40}_{-45}$  &\small -0.047   &\small -0.222  \\
\small Ne   &\small 8.08$\pm$0.06   &\small 7.97$\pm$0.10    &\small  -                        &\small -                     &\small 93     &\small -                        &\small -                  &\small -     &\small -    \\
\small Mg   &\small 7.62$\pm$0.02   &\small 7.61$\pm$0.05    &\small  -0.260$\pm$0.030         &\small -1.257$\pm$0.029      &\small 41    &\small 18$^{+2}_{-2}$           &\small 38$^{+1}_{-1}$     &\small -0.380    &\small -1.157   \\
\small Si   &\small 7.61$\pm$0.02   &\small 7.57$\pm$0.01    &\small  -0.180$\pm$0.035         &\small -1.319$\pm$0.052      &\small 37    &\small 13$^{+2}_{-2}$           &\small 35$^{+1}_{-1}$   &\small -0.320   &\small -1.205   \\
\small S    &\small 7.26$\pm$0.04   &\small 7.18$\pm$0.03    &\small   0.243$\pm$0.092         &\small -0.635$\pm$0.206      &\small 15    &\small -11$^{+6}_{-6}$\footnotemark[3]          &\small 12$^{+2}_{-2}$   &\small 0.137    &\small -0.548   \\
\small Ca   &\small 6.41$\pm$0.03   &\small 6.38$\pm$0.04    &\small  -2.090$\pm$0.200         &\small -3.930$\pm$0.200      &\small 2.4   &\small 2.38                     &\small 2.398&\small -2.311 &\small -3.746  \\
\small Fe   &\small 7.54$\pm$0.03   &\small 7.54$\pm$0.04    &\small
-0.951$\pm$0.038         &\small -2.236$\pm$0.041      &\small 35
&\small 31$^{+1}_{-1}$           &\small 34$^{+0}_{-0}$   &\small
-1.105 &\small -2.107
\\
\hline
 \label{Jenkins}
\end{tabular}
\end{center}
\renewcommand{\arraystretch}{1}
\begin{flushleft}\footnotesize Abundances $A(X)$ of the element $X$
are in units of
$\log_{10}\left(\mathrm{N(X)}/\mathrm{N(H)}\right)+12$.
\footnotesize$^{1}${Proto-solar abundances adopted in
\citep{Jenkins09}.}\,\footnotesize$^{2}${Proto-solar abundances
adopted in this work.}\,\footnotesize$^{3}${The positive depletion
of sulfur means that instead of observing a depletion of the element
the gas seems to be enriched of sulfur atoms respect to the
reference set.}\,
\end{flushleft}
\end{table*}
\renewcommand{\arraystretch}{1}

Table \ref{Jenkins}  shows the reference abundances for the SoNe
used by \citet{Jenkins09} and those adopted in our set for the
proto-solar environment, together with the  upper and lower limits
for the smallest $\left(F_{*}=0\right)$ and heaviest
$\left(F_{*}=1\right)$ depletion efficiencies  adapted from
\citet{Jenkins09} to our case. Calcium deserves some remarks. It is
not included in the \citet{Jenkins09} list, whereas we take it into
account. Furthermore, as this element is  heavily depleted in the
ISM \citep{Whittet03,Tielens05}, we consider the values proposed by
\citet{Crinklaw94} for the smallest and heaviest depletion. The
value proposed by \citet{Whittet03} for the diffuse clouds falls in
the middle. \\
\indent In Fig. \ref{Depletion} we finally show the range of
depletions adopted in our work to test the simulations versus the
SoNe. Since the depletion is line-of-sight dependent, we show for
each element the range allowed taking into account the smallest and
largest depletion with relative error bars. The values of depletion
for cool and warm disk are also shown. For the sake of clarity, we
expanded the depletion range for C, N, O in the lower left corner of
the figure.

\begin{figure}
\includegraphics[height=7.5cm,width=7.8truecm]{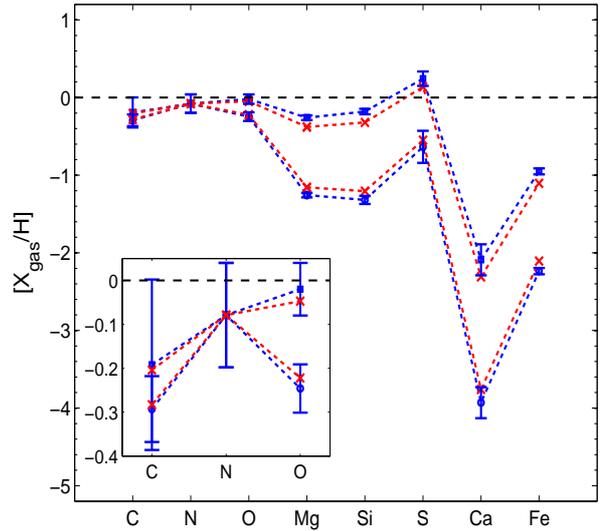}
\caption{The range of depletions is shown for those elements in our
list that  intervene in the dust formation or are present in the
ejecta by AGB stars and SN{\ae} (C, N, O, Ms, Si, S, Ca and Fe). The
insert  shows in detail the depletions for C, N and O. Four values
are plotted for each element. Filled squares and circles represent
the smallest and largest depletion, respectively. The crosses
determine the range for the depletion in the Warm and Cool
components of the Galactic Disk, according to \citet{Jenkins09} and
\citet{Savage96}. See Table \ref{Jenkins} for more details.}
\label{Depletion}
\end{figure}

\section{Models of the ISM with dust: results} \label{Dustymodels}

The chemical model for disk galaxies we are proposing  extends the
the original dust-free, multi-zone model with radial flows developed
by the Padova group over the years
\citep{Chiosi80,Chiosi86,Portinari98,Portinari99,Portinari00,Portinari04a,Portinari04b}
to which the reader should refer for all the details not mentioned
here. The model is quite complicate and obviously contains many
parameters. Therefore it would be wise to suitably select the
parameters to vary guided by some general considerations to be kept
in mind: (i) This study is  mainly devoted to highlight the role of
dust in chemical models rather than perfectly reproducing the
properties of gas and stars of the MW in the local pool. Again dust
properties will be the target of a forthcoming paper, just as
function of the galacto-centric distance along the MW Disk
\citep{Piovan11c}.  For this reason various recipes for dust
formation and evolution in the ISM  must be tested. (ii) A
reasonable agreement between model results and observational data
for gas, dust and stars (both in the solar vicinity and across the
galactic disk) must be reached for the sake of physical consistency
of the whole model. (iii) Finally, we will take advantage of the
results already obtained by
\citet{Chiosi80,Chiosi86,Portinari98,Portinari99,Portinari00,Portinari04a,Portinari04b}
as far as the fine tuning of several important parameters is
concerned.

\begin{figure*}
\centerline{\hspace{-0.9cm}
\includegraphics[height=7.8cm,width=18.2cm]{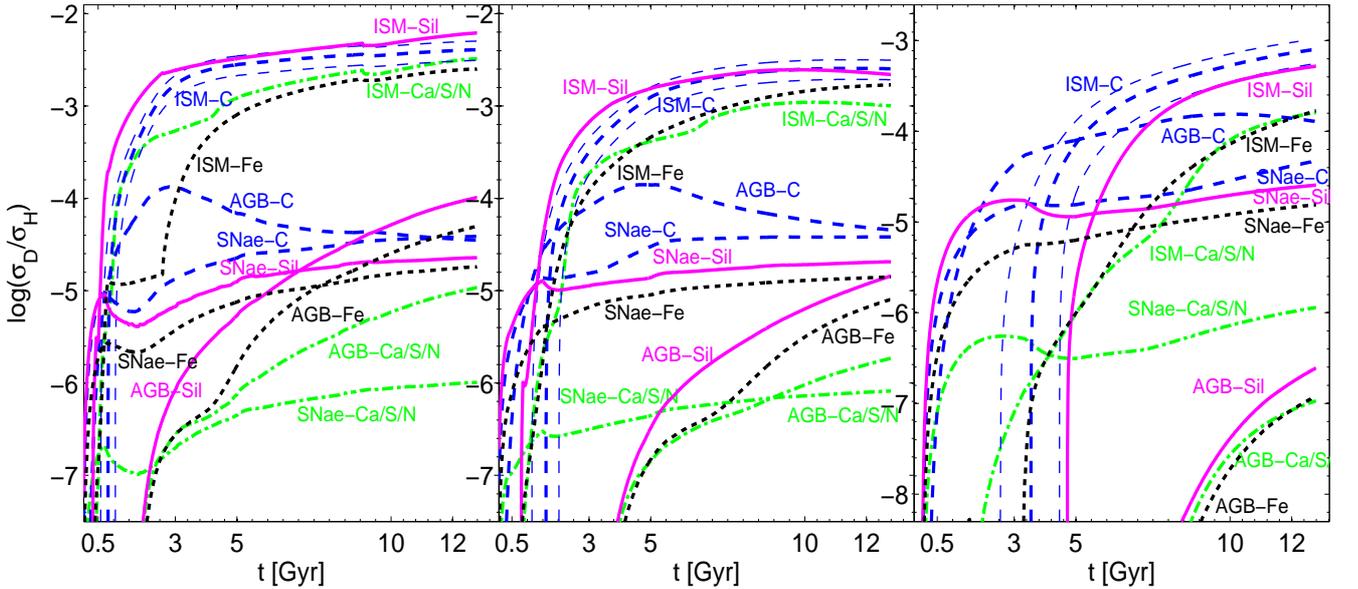}}
\caption{Temporal evolution of the contribution to the abundance of
dust by the four types of grain (on which we distributed the single
elements) and the three sources. All the contributions have been
properly corrected for the destruction of dust. \textbf{Left panel}:
results for an inner ring of the MW centered at 2.3 kpc. We show:
silicates (continuous lines), carbonaceous grains (dashed lines),
iron dust (dotted lines) and, finally, other grains bearing
$\mathrm{S}$, $\mathrm{Ca}$ and $\mathrm{N}$ (dot-dashed lines). For
each group we distinguish the net contributions from the ISM
accretion, AGB and SN{\ae}, that is: ISM-C, AGB-C and SN{\ae}-C for
carbon grains, ISM-Sil, AGB-Sil and SN{\ae}-Sil for silicates,
ISM-Fe, AGB-Fe and SN{\ae}-Fe for the iron dust and finally,
ISM-Ca/S/N, AGB-Ca/S/N and SN{\ae}-Ca/S/N for the other grains. The
two thin dashed lines represent the same $\mathcal{GDABBCBBB}$ model
but with 15\% (upper dashed line) and 45\% (lower dashed line) of
$\mathrm{CO}$. In all cases  $\xi_{\mathrm{CO}}=0.30$ as default
value. \textbf{Central panel}: the same as in the left panel but for
the SoNe at 8.5 kpc. \textbf{Right panel}: the same as in the left
panel but for an outer ring at 15.1 kpc. Since the ratio
$\sigma_{D}/\sigma_{H}$ is lower, the scale of the y-axis is shifted
respect to the scales for the inner region and the SoNe in the left
and central panels.} \label{EvolCOThree}
\end{figure*}

\indent  In the following each model is identified by a string of
nine letters (the number of parameters) in italic face whose
position in the string and  the alphabet corresponds to a particular
parameter and choice for it. The position in the string is the same
as in the list below. Let's now shortly comment on the parameters we
have considered and the choices we have made for each of them
together with the identification code.

\begin{description}
\item[(1)] \textsf{The IMF with its lower and upper limits and the fraction
$\zeta$  of stars with mass $M > 1M_{\odot}$}. Eight IMF are
considered as described in Sect. \ref{SFRandIMF}: Salpeter
($\mathcal{A}$), Larson ($\mathcal{B}$), Kennicutt ($\mathcal{C}$),
Kroupa original ($\mathcal{D}$), Chabrier ($\mathcal{E}$), Arimoto
($\mathcal{F}$), Kroupa 2002-2007 ($\mathcal{G}$), Scalo
($\mathcal{H}$) and, finally, Larson adapted to the SoNe
($\mathcal{I}$). Since some of them are similar, we will examine in
particular only the results obtained for some interesting cases
useful to understand the influence of the IMF. The Upper and lower
mass limits (and $\zeta$) are selected according to the default
values already given  in Sect. \ref{SFRandIMF}.

  \item[(2)] \textsf{The Star Formation law}. Five SFR are  considered (Sect.
\ref{SFRandIMF}: constant SFR ($\mathcal{A}$), Schmidt law
($\mathcal{B}$), Talbot \& Arnett (T\&A - $\mathcal{C}$), Dopita \&
Ryder (D\&R - $\mathcal{D}$) and, finally, Wyse \& Silk (W\&S -
$\mathcal{E}$).

 \item[(3)] \textsf{The fraction of MCs in which dust accretion takes place}.
This quantity must be specified because the chemical model in use
does not contain a real multi-phase description of the ISM. Two
cases are included:  a constant fraction ($\mathcal{A}$) based on
the SoNe data $(\chi_{MC}=0.2)$ and a varying $\chi_{MC}$
($\mathcal{B}$) related to the local SFR and total gas density (see
Sect. \ref{modelB}).

  \item[(4)] \textsf{The model for the accretion of grains in cold molecular
regions}. Two choices are available: the simple model by
\citet{Dwek98}  based on typical accretion timescales for dust
grains as modified by \citet{Calura08} (see Sect. \ref{modelA})
($\mathcal{A}$) and the recent and more refined model by
\citet{Zhukovska08} (see Sect. \ref{modelB}) ($\mathcal{B}$).

  \item[(5)] \textsf{The condensation efficiencies for dust in Type Ia SN{\ae}}.
Two cases are possible: \citet{Dwek98} ($\mathcal{A}$) and
\citet{Zhukovska08} ($\mathcal{B}$). In the former they contribute
to the dust budget in a away comparable to that of Type II SN{\ae},
whereas in the latter their role is negligible except for a small
amount of iron in agreement with the observations. The different choices for this parameter
have been discussed in \citet{Piovan11a}.

  \item[(6)] \textsf{The condensation efficiencies for dust in Type II
SN{\ae}}. Three choices are possible. The first one by
\citet{Dwek98} who suggests a high condensation efficiency
($\mathcal{A}$), the efficiencies
by \citet{Nozawa03,Nozawa06,Nozawa07} based on dust nucleation models
and taking into account the effects of the reverse and forward
shocks ($\mathcal{B}$), and finally those by \citet{Zhukovska08} who
favor a low condensation efficiency in SN{\ae} based upon pre-solar
grain observations. The impact of these different choices on the results
has been discussed in \citet{Piovan11a}.

  \item[(7)] \textsf{The condensation efficiencies for dust from  AGB stars}. Two choices
are available: the simple recipes by \citet{Dwek98} ($\mathcal{A}$ )
and the condensation efficiencies obtained by full calculations of
dust formation in synthetic AGB models by \citet{Ferrarotti06}
($\mathcal{B}$). Again, for more details on these different possibilities
for AGB stardust production, see \citet{Piovan11a}.

  \item[(8)] \textsf{The age at which a bar is introduced} to reproduce the
radial distribution of the gas in the Galactic disk, in particular
in the region of the molecular ring around 4 kpc. For the purposes
of this work we do not play with the bar effect: the pattern of
velocity to simulate the bar effect in the radial flows mechanism is
simply taken from \citet{Portinari00},(to which the reader should
refer for all the details) and is suitably chosen for every SFR law.
There is no bar-effect on the SoNe, for this reason between the
three cases included (no bar effect ($\mathcal{A}$), an onset of the
bar 4 Gyr ago ($\mathcal{B}$) and 1 Gyr ago ($\mathcal{C}$), we will
simply just fix the bar effect at the case ($\mathcal{B}$). For more
details and discussion on this point, see \citet{Piovan11c}, where
the dust formation and evolution on the whole disk is examined.

  \item[(9)] \textsf{The parameters of the SFR laws}: these are chosen according to
the discussion by \citep{Portinari99}. While the exponents $k$ or
$m$ and  $n$ are fixed (see Sect. \ref{SFRandIMF}) we let the
efficiency of the star formation $\nu$ vary and assume three
values more or less in the ranges suggested by \citep{Portinari99}
for every SF law. The minimum value for $\nu_{min}$ is case
$\mathcal{A}$ and the maximum value for $\nu_{max}$ is case
$\mathcal{C}$). The average case is $\mathcal{B}$.
\end{description}

\noindent Table \ref{Parameters} summarizes  the parameters we have
just described  together with their associated identification code:
the sequence must  be read from top to bottom. For example, the
string $\mathcal{DBAABABAB}$ corresponds to Kroupa 1998 IMF, Schmidt
SFR, ANN model for $\chi_{MC}$, \citet{Dwek98} accretion model,
\citet{Zhukovska08} type Ia SN{\ae} recipe for dusty yields,
\citet{Dwek98} condensation efficiencies for type II SN{\ae},
\citet{Ferrarotti06} yields for AGB stars, no bar and high
efficiency $\nu$ of the SFR. If not otherwise specified radial flows
and bar effect will always be included by default.

\renewcommand{\arraystretch}{1.3}
\setlength{\tabcolsep}{2.8pt}
\begin{table*}
\scriptsize
\begin{center}
\caption[]{\footnotesize Parameters of the models. Column (1) is the
parameter number, column (2) the associated physical quantity, and
column (3)  the source and the italic symbols are the identification
code we have adopted. See the text for some more details and
\citet{Piovan11b} for a detailed description.}
\begin{tabular}{ccl}
\hline \noalign{\smallskip}
n$^{o}$         & \small Parameter     &\small Source and identification label  \\
\hline \noalign{\smallskip} \small 1 & \small IMF           &\small Salpeter\footnotesize$^{1}$
$(\mathcal{A})$,  Larson\footnotesize$^{2}$ $(\mathcal{B}$), Kennicutt\footnotesize$^{3}$ $(\mathcal{C})$
Kroupa orig.\footnotesize$^{4}$ $(\mathcal{D})$, \\
&&\small  Chabrier\footnotesize$^{5}$ $(\mathcal{E})$, Arimoto\footnotesize$^{6}$ $(\mathcal{F})$, Kroupa
2007\footnotesize$^{7}$ $(\mathcal{G})$,
Scalo\footnotesize$^{8}$ $(\mathcal{H})$, Larson SN\footnotesize$^{9}$ $(\mathcal{I})$ \\

\small 2 &\small SFR law              &\small Constant SFR
$(\mathcal{A})$, Schmidt\footnotesize$^{10}$ $(\mathcal{B})$,
Talbot \& Arnett\footnotesize$^{11}$ $(\mathcal{C})$, Dopita \& Ryder\footnotesize$^{12}$ $(\mathcal{D})$, Wyse \& Silk\footnotesize$^{13}$ $(\mathcal{E})$ \\
\small 3 & \small $\chi_{MC}$ model    &\small Artificial Neural
Networks model\footnotesize$^{14}$ $(\mathcal{A})$, Constant $\chi_{MC}$
as in the Solar Neigh.\footnotesize$^{15}$ $(\mathcal{B})$               \\

\small 4 &\small Accr. model          &\small Modified
\citet{Dwek98} and \citet{Calura08} $(\mathcal{A})$;
adapted \citet{Zhukovska08} model $(\mathcal{B})$         \\

\small 5 &\small SN{\ae} Ia model &\small Dust injection adapted
from: \citet{Dwek98}, \citet{Calura08} $(\mathcal{A})$,
\citet{Zhukovska08}  $(\mathcal{B})$                   \\

\small 6 &\small SN{\ae} II model     &\small Dust injection adapted
from: \citet{Dwek98} $(\mathcal{A})$, \citet{Zhukovska08}
$(\mathcal{B})$, \\
&& \small \citet{Nozawa03,Nozawa06,Nozawa07}
 $(\mathcal{C})$\\

\small 7 &\small AGB model            &\small Dust injection adapted
from: \citet{Dwek98}
$(\mathcal{A})$, \citet{Ferrarotti06}  $(\mathcal{B})$  \\

\small 8 &\small Galactic Bar\footnotesize$^{16}$   &\small No onset
$(\mathcal{A})$, onset at $t_{G}-4$ Gyr
$(\mathcal{B})$, onset at $t_{G}-1$ Gyr $(\mathcal{C})$ \\

\small 9 &\small Efficiency SFR\footnotesize$^{17}$       &\small Low efficiency
$(\mathcal{A})$, medium efficiency $(\mathcal{B})$
, high efficiency $(\mathcal{C})$  \\
\hline \label{Parameters}
\end{tabular}
\end{center}
\renewcommand{\arraystretch}{1}
\footnotesize$^{1}${\citet{Salpeter55}}.\, \footnotesize$^{2}${\citet{Larson86,Larson98}}.\,
\footnotesize$^{3}${\citet{Kennicutt83,Kennicutt94}}.\,\footnotesize$^{4}${\citet{Kroupa98}}.\,
\footnotesize$^{5}${\citet{Chabrier01}}.\,\footnotesize$^{6}${\citet{Arimoto87}}.\,
\footnotesize$^{7}${\citet{Kroupa02a,Kroupa07}}.\,\footnotesize$^{8}${\citet{Scalo86}}.\,
\footnotesize$^{9}${\citet{Larson86,Scalo86,Portinari04a}}.\,\footnotesize$^{10}${\citet{Schmidt59}}.\,
\footnotesize$^{10}${\citet{Talbot75}}.\,\footnotesize$^{11}${\citet{Talbot75}}.\,
\footnotesize$^{12}${\citet{Dopita94}}.\,\footnotesize$^{13}${\citet{Wyse89}}.\,
\footnotesize$^{14}${\citet{Piovan11c}}.\,\footnotesize$^{15}${\citet{Zhukovska08}}.\,
\footnotesize$^{16}${\citet{Portinari00}}.\,\footnotesize$^{16}${\citet{Piovan11b}}.\,
\end{table*}
\renewcommand{\arraystretch}{1}

\begin{figure}
\centerline{\hspace{-5mm}
\includegraphics[height=7.5cm,width=8.5cm]{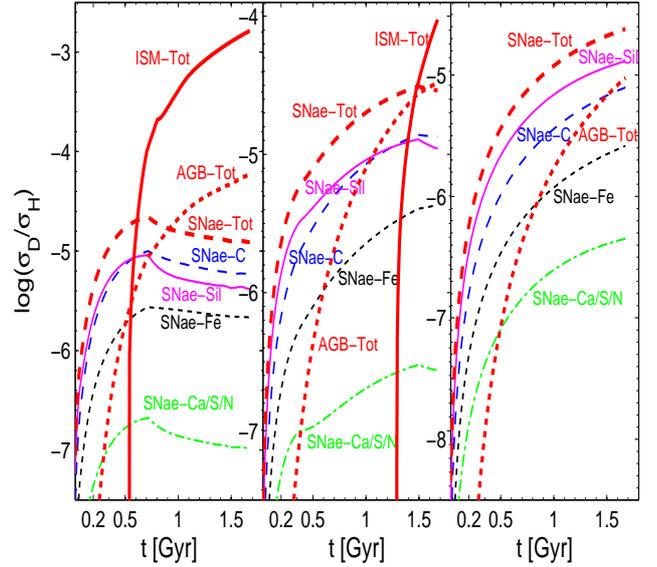}}
\caption{Temporal  evolution of the contribution to the abundance of
dust during the \textit{first 1.5 Gyr-2 Gyr}. All the contributions
have been properly corrected for the destruction of dust.
\textbf{Left panel}: results for the inner ring of the MW at 2.3
kpc. We show for the SN{\ae} injection: silicates (thin continuous
line SN{\ae}-Sil), carbonaceous grains (thin dashed line,
SN{\ae}-C), iron dust (thin dotted line, SN{\ae}-Fe) and, finally,
other grains bearing $\mathrm{S}$, $\mathrm{Ca}$ and $\mathrm{N}$
(thin dot-dashed line, SN{\ae}-Ca/S/N). The \textit{thick} lines
represent the total contribution from one source to the dust budget
sub-divided in: SN{\ae} (thick dashed line, SN{\ae}-tot), ISM (thick
continuous line) and AGB (thick dotted line). \textbf{Central
panel}: the same as in the left panel but for the SoNe at 8.5 kpc.
\textbf{Right panel}: the same as in the left panel but for the
outer ring at 15.1 kpc.} \label{EvolCOThree1Gyr}
\end{figure}

Obviously, these are not the only parameters of chemical models:
even the classical ones are themselves very rich of parameters and
when including also radial flows, bars and dust formation/evolution
the parameter space acquires many more dimensions so that a full
exploration of it is a cumbersome affair. However, it may  happen
that: (1) some variables play a secondary role, thus not influencing
that much the results (in particular we focus on the effects related
to dust, the target of the work) and just adding second order
corrections; (2) some variables, even if playing an important  role
and influencing significantly the results, have a well defined
and/or restricted sphere of influence therefore are not  of much
interest here because their effect is clear. The parameters
belonging to these two classes are not varied in the models, but
kept fixed to a suitable value. In the following we examine the
effect of some primary and secondary parameters and the general
behaviour of the dust model. To this purpose we adopt the
$\mathcal{GDABBCBBB}$ model as the `default' one, with radial flows
and bar included \citep{Portinari00} as
the reference case. This model uses the most detailed theoretical recipes for
the yields of dust and the amounts of MCs, while the SFR and the IMF
are simply selected between the available ones.

\subsection{Fraction of $\mathrm{CO}$ in the ISM}\label{COFraction}

As already discussed in Sects. \ref{Equations:Carbon} and
\ref{Equations:Oxygen}, a percentage from  20\% to 40\% of the
Carbon and up to 20\% of the Oxygen is locked in the CO molecules
\citep[see ][]{vanDishoeck93,vanDishoeck98}. This affects the amount
of Carbon available as key-element for the growth process owing to
the lowered abundance of this element in the gaseous phase.
Furthermore, in the dense and cold regions of the ISM, CO tends to
condense onto dust grains \citep{Goldsmith01,Bacmann02,Whittet10},
leaving the gas phase and sticking efficiently to dust
\citep{Whittet10}. The depletion of the CO can reach in pre-stellar
cores even a factor of 10 \citep{Tafalla02,Walmsley04}. All this is
parameterized by the fraction $\zeta_{\mathrm{CO}}$ of
$\mathrm{CO}$. The effect of $\zeta_{\mathrm{CO}}$ on the
key-elements $^{12}$C and $^{13}$C (and of
$\zeta_{\mathrm{CO}}^{\prime}$ on $^{17}$O and $^{18}$O) is
straightforward: the higher is the amount  in CO, the lower is the
amount of free atoms in the gas-phase available for the growth, thus
implying longer timescales and slower dust formation. Therefore, we
expect a lower contribution  by the accretion to the dust budget of
the ISM when high values of $\zeta_{\mathrm{CO}}$ are adopted.\\
\indent In Fig. \ref{EvolCOThree}, we present the time evolution of
the contributions to the dust budget for three regions of the MW: a
central region $(r_{k}=2.3$kpc), the SoNe $(r_{k}=8.5$kpc) and an
outer region $(r_{k}=15.1$kpc). First of all, we do not show the
temporal  evolution of the single elements, but of  important groups
of elements  representative of the main typical ISM dust types.
These are the silicates (olivines+pyroxenes+quartz+silicon in SiC),
carbonaceous grains (carbon grains+carbon in SiC), iron grains and,
finally, other grains containing S, N and Ca. All the dust
abundances have been normalized to the local hydrogen density and
corrected for  dust destruction so as to represent the effective net
contribution to the dust budget. The contributions are split   in
three main sources: AGBs, SN{\ae} and ISM accretion. In order to
explore the effect of $\mathrm{CO}$,   in the $\mathcal{GDABBCBBB}$
model we vary the $\mathrm{CO}$ abundance from 15\% up to 45\%: the
two thin dashed lines in Fig. \ref{EvolCOThree} bracket the region
of variation of the ISM contribution. Carbon accretion  in the ISM
is the only process varying with  the $\mathrm{CO}$ abundance:  the
lower $\xi_{\mathrm{CO}}$  the higher is the ISM contribution. This
effect can be significant and we cannot easily get rid of it when
evaluating carbon depletion: however, it is clear and
straightforward and limited to the only carbon. Even if
$^{16}\mathrm{O}$ is a component  of the  $\mathrm{CO}$ molecule,
due to its high abundance and  never being a key-element, it is
scarcely affected by  variations of  $\xi_{\mathrm{CO}}$. The same
holds true for $^{17}\mathrm{O}$ and $^{18}\mathrm{O}$ but in this
case due to their low abundance: their budget depends on
$\xi_{\mathrm{CO}}$, but their contribution to the global budget is
negligible. Therefore, the variations of $\xi_{\mathrm{CO}}$  do not
affect the silicates budget.\\
\indent From the analysis of Fig. \ref{EvolCOThree} we can notice
several general features of the dust evolution common to all the
models that are  worth to be underlined: (i) the main contribution
to the dust enrichment during most of the  Galaxy lifetime is due to
the accretion process in the ISM. Furthermore, dust production is
much higher in the inner regions of the Galactic Disk compared to
the outer ones where the weight of the ISM  gets smaller so that
during many Gyr the stardust injected from AGB stars and SN{\ae}
drives the total yields. This is ultimately due to the low number
densities of metals that do not favour the accretion process; (ii)
the inner regions reach higher metallicities than the outer ones:
therefore, all physical processes depending on the metallicities are
much enhanced in the central regions of the Galaxy, e. g. the yields
of dust from oxygen rich M-stars; (iii) even if some sources (like
AGB-C stars in the inner regions) may vary their contributions with
time, in general the \textit{total} amount of dust keeps growing
monotonically. In some way, this mirrors  the metallicity enrichment
of the ISM and the fact that dust formation is very sensitive to metals.\\
\indent In Fig. \ref{EvolCOThree1Gyr} we show in detail the
evolution of the contributions of AGB stars, SN{\ae} and ISM to the
total budget of dust during the first 1.5-2 Gyr for the three
selected regions of the Galactic disk. For AGB, SN{\ae} and ISM we
show the total budget. However, limited to SN{\ae},  we also
distinguish the various types of grains. During the early stages,
stardust dominates the scene: going from the innermost to the
outermost regions it takes more and more time for the ISM accretion
to overcome the stellar contribution. This is ultimately due to the
higher densities and metal content of the inner regions that favors
the onset of accretion. For the outer regions, with low density and
low SFR, the ISM starts to be important only at t$>$4 Gyr (see Fig.
\ref{EvolCOThree}). It is interesting to note how the dust-to-gas
ratio for net yield by SN{\ae} is nearly constant going from the
innermost regions (left panels of Fig. \ref{EvolCOThree1Gyr} and
\ref{EvolCOThree}) to the external ones (right panels of the same
figures), even if the associated timescales  are much different.
Toward the center of the Galaxy, we have more hydrogen, higher SFR,
higher SN{\ae} rate and higher  yields of dust. Both gas (via
infall) and SN{\ae} dust (via infall, SFR and SN{\ae} explosions)
grow in absolute value moving from outside to inside, whereas their
ratio does not change in the same proportions (considered that in
any case the yields of dust by SN{\ae}  do not depend on the
metallicity, at least according the the kind of theoretical results
to our disposal).

\subsection{The effect of the IMF}\label{IMFeffect}

Different IMFs influence in a crucial way the injection of dust into
the ISM. For every generation of stars, the
relative amount of newly born massive objects able to produce dust
via the SNa channel and  the amount of low and intermediate mass
stars refueling the ISM with the dust produced through the TP-AGB
phase, both depend on the IMF. This last one also determines the timescales
of stardust injection and the relative contribution by SN{\ae} and
AGB stars.

\begin{figure*}
\centerline{\hspace{-2cm}
\includegraphics[height=6.5cm,width=18.0cm]{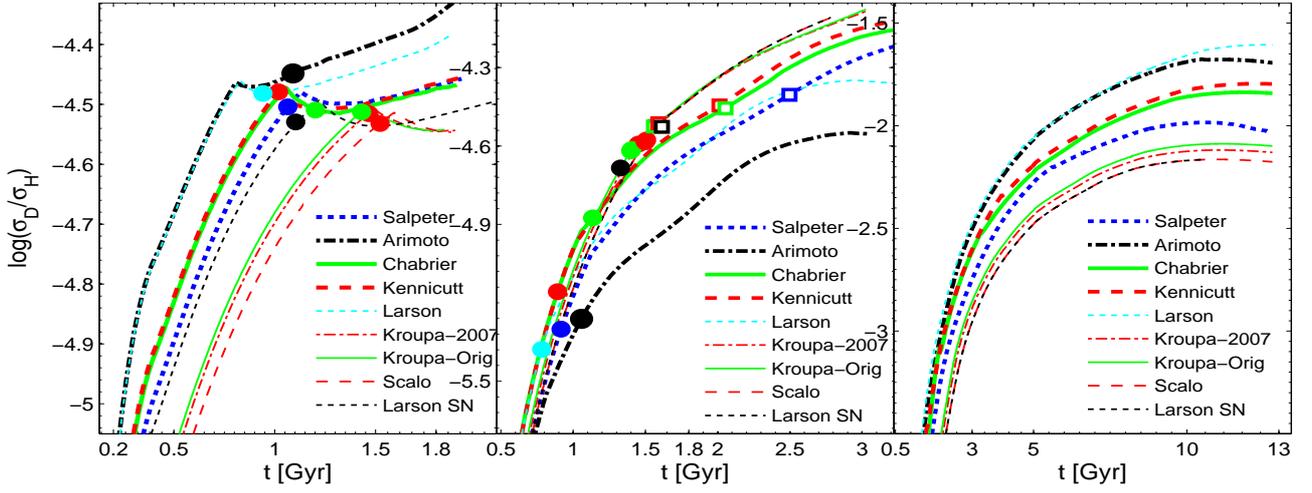}}
\caption{Temporal evolution of the contribution to the abundance of
dust during  the \textit{first 2 Gyr-3 Gyr} in the SoNe at 8.5 kpc
and for \textit{different IMFs}. All the contributions have been
corrected for the destruction of dust. Nine IMFs have been
considered: Salpeter (thick dotted line), Arimoto (thick dot dashed
line), Chabrier (thick continuous line), Kennicutt (thick dashed
line), Larson (thin dotted line), Kroupa-2007 (thin dot-dashed
line), Kroupa original (thin continuous line), the Scalo IMF (thin
dashed line) and, finally, the Larson IMF adapted to the SoNe (thin
dashed line). See Sect. \ref{SFRandIMF} for more details.
\textbf{Left panel}: temporal evolution of the total contribution to
the dust budget by SN{\ae} in the MW SoNe. The filled symbols
represent the instant when the contribution by ISM accreted dust
\textit{equalizes} the SN{\ae} injected dust amount. \textbf{Middle
panel}: time evolution of the total contribution to the dust budget
by AGB stars for the MW SoNe. The filled symbols represent the
instant when the contribution by AGB injected dust
\textit{equalizes} the ISM dust production by accretion, while the
empty symbols represent the instant (if eventually it happens) when
AGB dust equalizes the SN{\ae} injected dust. \textbf{Right panel}:
time evolution of the total dust budget for different IMFs.}
\label{DifferentIMF_SN}
\end{figure*}

\begin{figure*}
\centerline{\hspace{-2cm}
\includegraphics[height=6.5cm,width=18.0cm]{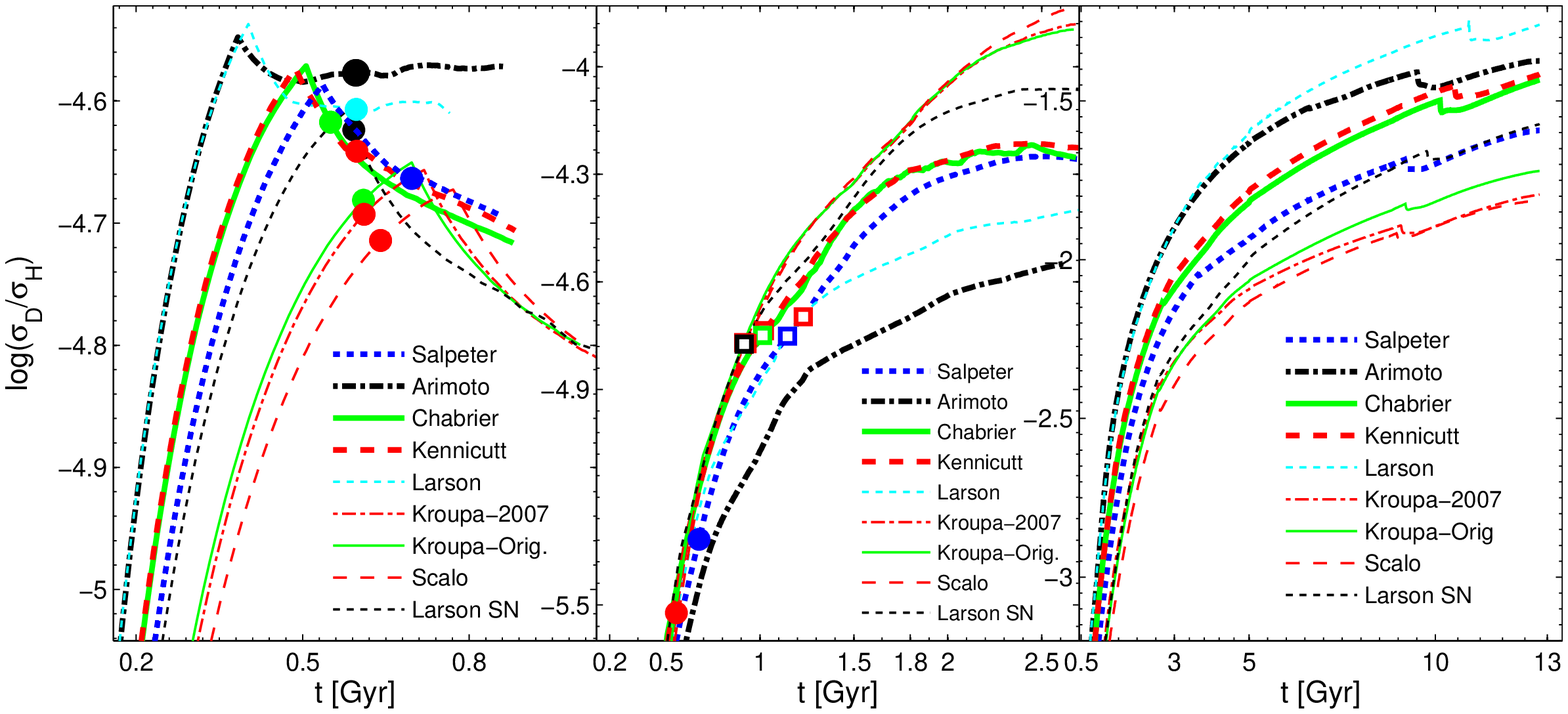}}
\caption{Temporal evolution of the contribution to the abundance of
dust during the \textit{first 1-2 Gyr} in the \textit{inner part} of the MW
disk at 2.3 kpc from the centre and for \textit{different IMFs}. All
the contributions have been corrected for the destruction of dust.
Nine IMFs have been considered as in Fig. \ref{DifferentIMF_SN}.
See Sect. \ref{SFRandIMF} for more details. The meaning of all the
symbols is the same as in Fig. \ref{DifferentIMF_SN}.}
\label{DifferentIMF_SN_INNER}
\end{figure*}

In Fig. \ref{DifferentIMF_SN} we compare the degree of dust
enrichment obtained using nine different IMFs chosen among those
widely used in literature. The reference case is always given by the
model $\mathcal{GDABBCBBB}$. All the others are obtained from this
by varying the  first parameter of the list from case $\mathcal{A}$
to case $\mathcal{I}$. First of all, let us examine the effects of
the IMF during the first evolutionary stages, when the dust
enrichment is mainly due to SN{\ae}, which also supply the seeds and
metals for the accretion process in the ISM. We begin with the the
solar vicinity displayed in the left panel of Fig.
\ref{DifferentIMF_SN}. We note that, at varying the IMF and keeping
fixed all the other parameters of the model, the age at which the
dust enrichment by the accretion in the ISM becomes comparable to
that by SN{\ae} can vary by about $\sim 0.5$ Gyr. The time
difference can be easily explained as due to the different
percentage of massive stars exploding as SN{\ae} (and thus refueling
the ISM) with the different IMF (see the entries of Table
\ref{IMFfractions}).

\renewcommand{\arraystretch}{1.3}
\setlength{\tabcolsep}{2.8pt}
\begin{table}
\scriptsize
\begin{center}
\caption[]{\footnotesize Mass fractions in different mass intervals
predicted by different IMFs. All IMFs are normalized to unity over
the mass range of validity. The following nine IMFs are considered:
Salpeter, Larson, Kennicutt, original/old Kroupa, Chabrier, Arimoto,
new Kroupa multi-slope power law 2002-2007, Scalo and, finally,
Larson adapted to the Solar Neighbourhood  (See Sect.
\ref{SFRandIMF} for more details). The upper mass limit is always
100\, M$_{\odot}$, whereas the lower limit is chosen according to
the discussion made in Sect. \ref{SFRandIMF}. Three mass intervals
are  considered according to the different stardust factories. All
masses are in solar units.}
\begin{tabular}{cccc}
\hline \vspace{0.1cm}
\small IMF &  \small M$<$1 \footnotemark[1]   &
\small 1 $\leq$ M $<$ 6 \footnotemark[2]    & \small M $\geq$ 6 \footnotemark[3]  \\
\hline
\small Salpeter         &\small 0.6075           &\small 0.2285            &\small 0.1640       \\
\small Larson           &\small 0.3470           &\small 0.3568            &\small 0.2962       \\
\small Kennicutt        &\small 0.4094           &\small 0.3883            &\small 0.2023       \\
\small Kroupa (old)     &\small 0.5948           &\small 0.3016            &\small 0.1036       \\
\small Chabrier         &\small 0.4550           &\small 0.3517            &\small 0.1933       \\
\small Arimoto          &\small 0.5000           &\small 0.1945            &\small 0.3055       \\
\small Kroupa 2002-2007 &\small 0.6198           &\small 0.2830            &\small 0.0972       \\
\small Scalo            &\small 0.6802           &\small 0.2339            &\small 0.0859       \\
\small Larson SoNe      &\small 0.5614           &\small 0.3130            &\small 0.1256       \\
\hline \label{IMFfractions}
\end{tabular}
\end{center}
\renewcommand{\arraystretch}{1}
\begin{flushleft}
\,\footnotesize$^{1}${Fractional mass of stars that do not
contribute to the dust budget of stellar origin.}\,\footnotesize$^{2}${Fractional
mass of stars that contribute to the stardust budget via the AGB
channel.}\,\footnotesize$^{3}${Fractional mass of stars that
contribute to the stardust budget via the type II SN{\ae} channel.}
\end{flushleft}
\end{table}
\renewcommand{\arraystretch}{1}

\noindent Some IMFs (like those by Kroupa, Larson SoNe and Scalo)
predict a small number of SN{\ae} compared to others (like those by
Larson, Kennicutt or Chabrier) that are more generous in the number
of massive stars and hence Type II SN{\ae}. Therefore, in the former
case a small injection of dust by SNa explosions and a slow
accretion (fewer seeds to disposal) are expected. The opposite holds
true  with the latter case case  favouring the formation of massive
stars. Furthermore, with the former case the time at which the ISM
gets dust-rich by accretion is delayed with respect to the other
case. This is also shown in the panels of Fig. \ref{DifferentIMF_SN}
where the time when the ISM accretion equalizes the dust enrichment
by SN{\ae} at decreasing the relative percentage of massive stars in
the IMF is marked.  It is also evident that IMFs skewed  toward
massive stars produce much more dust of stellar origin.
Consequently,  before  accretion in the ISM starts driving the
evolution of the dust, large  differences brought
by the IMF are possible.\\
\indent So far we have examined the solar vicinity with a relatively
mild star formation efficiency. What about the innermost region of
the MW characterized by a much higher SFR? The situation is shown in
the left panel of  Fig. \ref{DifferentIMF_SN_INNER} which displays
the dust  enrichment due to SN{\ae} as in Fig.
\ref{DifferentIMF_SN}. Compared to the solar vicinity, we note that
SN{\ae} produce many more seeds and metals, accretion in the ISM
develops faster and becomes important very earlier on. However, in
the ISM the dust production  by accretion becomes more or less
comparable to that by  SN{\ae} at the same time
\textit{independently} of the IMF (see the large dots marked in the
left panel of Fig. \ref{DifferentIMF_SN_INNER}). In the case of the
solar vicinity, the cross-over stage hardly occurred below 1 Gyr
extending up to 1.5 Gyr, whereas now  they all fall in the age range
0.6-0.7 Gyr. In a medium rich of seeds, dust accretion grows faster
and the effect of the IMF somehow loses importance. \\ The
differences both in the amounts of dust of stellar origin injected
and the timescale of earlier enrichment in dust by SN{\ae} are very
large. In general, the effects induced by variations in the IMF  can
be very large during the earliest stages of evolution. Along this
line of thought, we can expect that in high-redshift obscured
galaxies with high SFR (easily even higher than the early SFR of the
inner regions of the MW) some IMFs may  not be able  to produce the
amounts of observed dust of stellar origin before the dust accretion
process has become significant. The immediate implication of this
for primeval galaxies can be easily foreseen.  According
\citet{Draine09},  some accretion in the ISM to explain the amount
of dust in observed primeval galaxies is required. The picture
should be as follows: (i) before dust by accretion in the ISM and
dust injected by stars become comparable, the effect of the IMF
prevails and determines the amounts of dust present in the Galaxy;
(ii) if the SFR is high, the IMF does not play an important role in
determining the onset of the dust accretion process in the ISM,
whereas if (iii) the SFR is low, different IMFs cause an important
spread in the ages at which dust by accretion becomes important.\\
\indent We pass now to examine the role played by AGB stars. In the
middle panels of Figs. \ref{DifferentIMF_SN} and
\ref{DifferentIMF_SN_INNER} we see for the same regions of the left
panels the contribution of the AGB stars  to the total dust budget.
The AGB stars contribute significantly over  a longer time-scale:
the filled symbols show the age at which dust produced by AGB stars
and dust accreted in the ISM become comparable, whereas the empty
ones show the same but for AGB stars versus SN{\ae}. In both cases
the age is marked when and if the equality among the three
contributions can be established. There are indeed some extreme
IMFs, where massive stars are favored with respect to the
intermediate mass ones progenitors of AGB stars (like the Arimoto
IMF), in which during  the first Gyrs  AGB stars never reach SN{\ae}
in the inner MW regions (Fig. \ref{DifferentIMF_SN_INNER} - middle
panel). Furthermore,  IMFs richer in  intermediate mass stars
produce  bigger amounts of dust by  AGB stars. This may somehow be
correlated with the delayed appearance in a galaxy spectral energy
distribution of the  PAHs features: the delay may depend on
efficiency of dust injection by AGB stars.\\
\indent Finally, in the right panels of Figs.
(\ref{DifferentIMF_SN}) and (\ref{DifferentIMF_SN_INNER}) we show
the evolution of the total dust content in the ISM: it is
interesting to note how the differences between the various IMFs in
the early stages are the same as those expected (observed)  at the
present age, keeping constant all other parameters. The IMFs
producing more dust by SN{\ae} (and of course seeds for accretion)
in the early stages are the same for which we get higher amounts of
dust at the current age. The differences among the various IMFs can
be significant. Indeed, a fast enrichment of dust during the early
stages caused by  SN{\ae}, (keeping fixed all other parameters, the
condensation coefficients in particular) goes together with a strong
enrichment in metals. More metals means more atoms available for
dust to grow in the ISM. For this reason IMFs skewed toward  massive
stars favor the accretion of dust  in the ISM and lead to higher
final contents of dust.

\subsection{The effect of the SF law}\label{SFeffect}

\begin{figure}
\centerline{\hspace{-5mm}
\includegraphics[height=7.5cm,width=8.5cm]{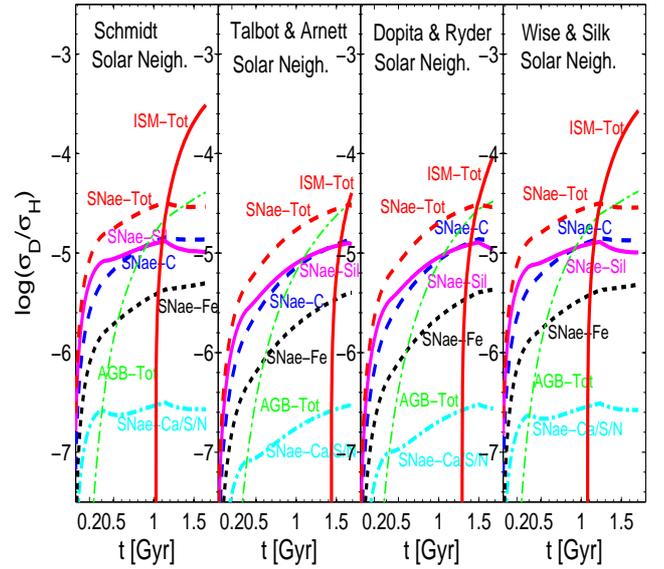}}
\caption{Temporal evolution of the dust budget in the SoNe during
the early stages  until 1.5-2.5 Gyr. Four SF laws are shown in the
four panels: Schmidt, Talbot \& Arnett, Dopita \& Ryder and Wyse \&
Silk. We display: the amount of dust grains accreted in the ISM
(continuous lines), the amount of dust injected by AGB stars (thin
dot-dashed line), the total amount of SN{\ae} stardust (dashed line)
also subdivided into the various grain families, respectively, i.e.
dotted line (iron-dust), silicates (continuous line), carbonaceous
grains (dashed line), and S/Ca/N based grains (dot-dashed line).
\textbf{Left panel}: Temporal evolution of the dust budget injected
in the SoNe by the various sources and  the Schmidt law.
\textbf{Central-left panel}: the same as in the left panel but for
the Talbot \& Arnett law. \textbf{Central-Right panel}: the same as
in the left panel but for the Dopita \& Ryder SF law. \textbf{Right
panel}: the same as in the left panel but for the Wyse \& Silk SF
law.} \label{FourSFRlaw1Gyr}
\end{figure}

As presented in Sect. \ref{SFRandIMF}, four SF laws have been
considered. The results are shown in the four panels of Fig.
\ref{FourSFRlaw1Gyr} limited to the early stages of the evolution.
The  reference model is $\mathcal{GDABBCBBB}$, in which the SFR is
changed as indicated.

The parameters  $k$ and $\nu$ of the SFR are chosen in agreement
with the analysis made by \citet{Portinari99}, in such a way that
the sole effect of the SFR law  is isolated. All other parameters of
the model are kept fixed. At given $k$ and $\nu$, the different SF
laws widely adopted for the MW disk produce a similar dust budget.
Some differences can be noted (i) in the amount of dust before the
onset of the ISM accretion, with the Schmidt and/or Wyse \& Silk
laws favouring a higher amount of dust by SN{\ae} and (ii) in the
age at which the production of dust by accretion in the ISM becomes
more important than that by SN{\ae}. Depending on the SF law, the
time interval in which SN{\ae} dominate the total dust budget gets
short or long.

\begin{figure}
\centerline{\hspace{-5mm}
\includegraphics[height=7.5cm,width=8.5cm]{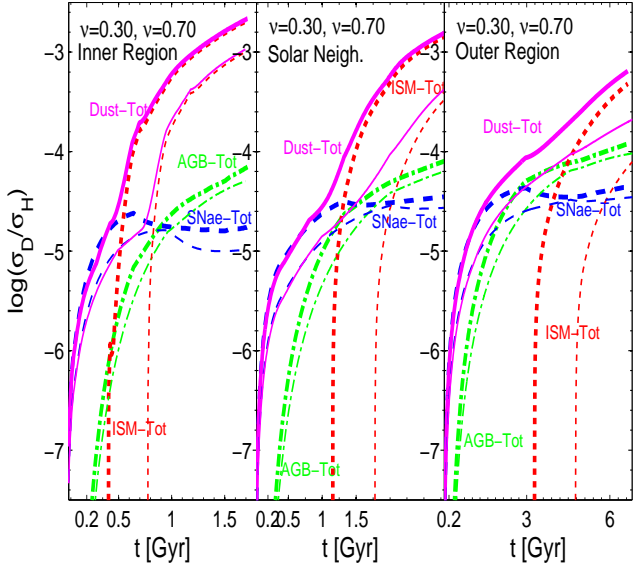}}
\caption{Temporal evolution of the contribution to the dust budget
during  the first Gyrs at varying the coefficient $\nu$ of the star
formation law, namely the Dopita \& Ryder SF law, for \textit{three
regions} of the MW. The thin line is  $\nu = 0.30$ (low efficiency),
while the thick line is for $\nu = 0.70$ (high efficiency).  We
show: the total amount of dust (from accretion in the ISM plus dust
ejected by SN{\ae} and AGB stars - continuous line); the
contribution by accretion of dust grain in the ISM (dotted line);
the total contribution by SN{\ae} (dashed line); the contribution by
AGB stars (dot-dashed line). \textbf{Left panel}: the results for an
inner ring  of the MW. \textbf{Central panel}: the same as in the
left panel but for the SoNe. \textbf{Right panel}: the same as in
the left panel but for an outer ring of the MW.} \label{SFRtwoNU}
\end{figure}

Given that the specific expression for the SF law is not of primary
importance here (at least choosing among the  ones we included in
this study), we turn the attention to the efficiency of SF
represented by the parameter $\nu$.  In Fig. \ref{SFRtwoNU} we show
the contribution to the dust budget in the usual three significant
rings of the MW at varying $\nu$ from $\nu = 0.30$ to $\nu = 0.70$.
As expected lower values of $\nu$ imply a smaller rate of SN{\ae}
and number of AGB stars, therefore  a delay in  the ISM dust
accretion process  because there are  less seeds/metals injected by
stars into the ISM from the stars, and finally a lower total
production of dust. As expected, varying the SF parameters affects
the system in the early phases of the evolution when the SF is
strong. This can be noticed once comparing thin and thick lines in
Fig. \ref{SFRtwoNU}.

\begin{figure}
\centerline{\hspace{-3mm}
\includegraphics[height=7.5cm,width=8.5cm]{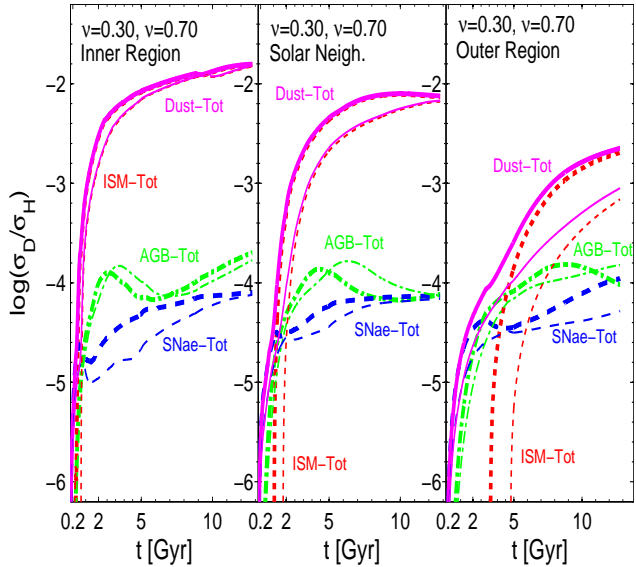}}
\caption{Temporal evolution of the contribution to the dust budget
up to the present age at varying the coefficient $\nu$ of the star
formation law from 0.30 to 0.70. The meaning of the symbols is the
same as in Fig. \ref{SFRtwoNU}. \textbf{Left panel}: the results for
an inner ring of the MW. \textbf{Central panel}: the same as in the
left panel but for the SoNe. \textbf{Right panel}: the same as in
the left panel but for an outer MW ring.} \label{SFRtwoNU13Gyr}
\end{figure}

Does the efficiency $\nu$ affect also the dust budget at the present
epoch? In Fig. \ref{SFRtwoNU13Gyr} we show the evolution of the same
three regions of Fig. \ref{SFRtwoNU} up to the present age of the
MW. As we can see, in the inner regions and the solar vicinity, even
if there is a significant difference in the past, at the present
time  the difference gets  negligible, whereas  in the outer regions
the difference in the  dust budgets (both  total and partial ones)
remains remarkable. This is an effect of the adopted SF laws that
are all scaled to the current star formation at the solar
neighbourhood and for this reason tend to ultimately produce the
same result as the evolution proceeds. However, we expect that if
the SF law is not tied up to  normalization or scaling factor, the
adoption of different SF laws would have a strong impact on the
whole evolution. The difference in the outer regions is explained by
the long delay in onset of the ISM dust accretion process in the
case of low efficiencies $\nu$  (very low star forming environment):
when eventually  the ISM accretion process becomes important, there
is not enough time  to reach the dust budget produced in the high
$\nu$ case. This behaviour is also strengthened by the inward radial
flows that remove gas  from the  outermost regions.

\subsection{The effects of different models for dust accretion  in the ISM}
\label{Accreffect}

\noindent In this study we consider two models of dust accretion in
the ISM (see  Sects. \ref{modelA}, \ref{modelB} and
\ref{grainevolution} for details): model $\mathcal{A}$ based on
\citet{Dwek98} where accretion is simply included with a general
timescale depending on the destruction timescale. Indeed, the
accretion timescale is half of the destruction timescale, and the
same value is used for all the elements \citep{Dwek98,Calura08}.
Model $\mathcal{B}$ based on \citet{Zhukovska08} in which a
different timescale for each  element is adopted and the evolution
of the dust abundances for a number of elements, supposedly
contained in a several species of dust grains, is followed.
Furthermore, there is no \textit{a priori} connection between the
destruction and accretion timescales.

\begin{figure}
\centerline{\hspace{-3mm}
\includegraphics[height=7.5cm,width=8.5cm]{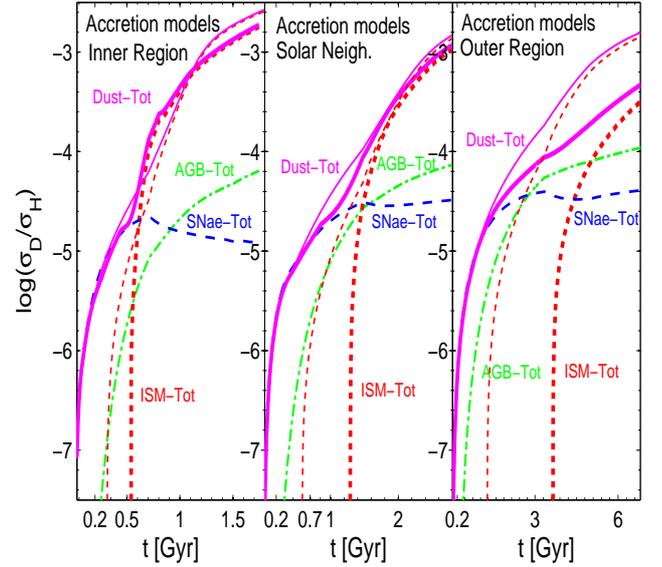}}
\caption{Temporal evolution of the dust production during the early
phases of the MW evolution at varying the accretion model used to
describe dust formation in the ISM. Three regions have been
considered as usual: an inner one (left panel), the SoNe (middle
panel), and an outer region (right panel). Thick lines represent
model $\mathcal{A}$ based upon \citet{Dwek98} and \citet{Calura08},
whereas thin lines represent model $\mathcal{B}$ based upon
\citet{Zhukovska08}. For AGB (dot-dashed line) and SN{\ae} (dashed
line) we only show one line as the contribution is fixed. We show:
the total amount of dust grains in the ISM (continuous lines) and
the total amount of accreted dust in the ISM (dotted lines).
\textbf{Left panel}: The results for an inner region.
\textbf{Central panel}: The results for the SoNe. \textbf{Right
panel}: The results for an outer region.} \label{Accretion1Gyr}
\end{figure}

In Fig. \ref{Accretion1Gyr}, we show the evolution of the dust
budget for the two accretion models. The amounts of dust produced by
SN{\ae} and AGB stars and all the other parameters are kept fixed.
In particular, the description of the dust destruction process is
the same in both cases: therefore we can examine the sole effect of
accretion. In the very early stages there is no difference: the
budget is dominated by SN{\ae}. The accretion process however starts
to be significant very fast in model $\mathcal{A}$, because it
simply depends on the adopted timescale, while in model
$\mathcal{B}$ it is more sensitive to the physical conditions of the
environment:  in general it tends  to slow down at decreasing SFR
and densities (i.e. passing from the innermost to the outermost
regions).  Even if there is some difference between  model
$\mathcal{A}$ and  $\mathcal{B}$,  the behaviour of the
\textit{total} dust budget is  similar. This finding means that if
we are interested in   the total amount of dust produced, we can do
it simply choosing a suitable  timescale for dust formation. This
holds everywhere but the  outer regions where for model
$\mathcal{A}$ we probably need  a longer timescale of dust formation
to closely agree with model $\mathcal{B}$. This sounds reasonable
because it is likely that the accretion timescale varies with the
environment. In any case even the simple model of dust accretion
with constant timescale is fully adequate to follow the evolution of
the \textit{total} dust budget in  the early stages of  evolution.

\begin{figure}
\centerline{\hspace{-3mm}
\includegraphics[height=8.0cm,width=8.5cm]{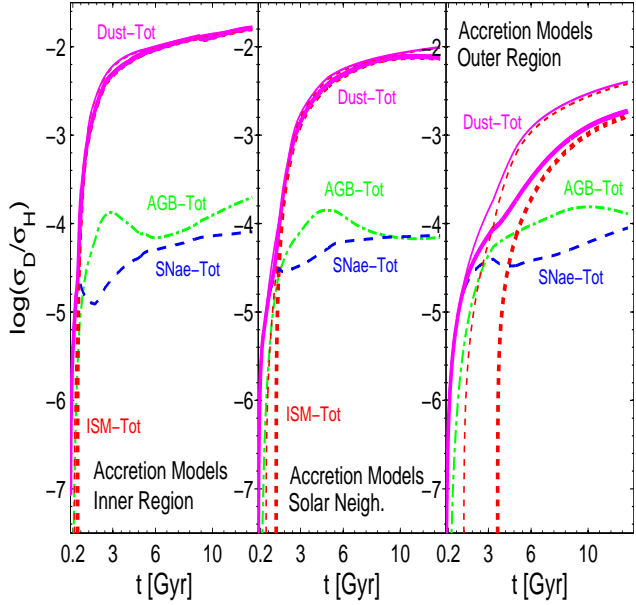}}
\caption{Temporal evolution of the dust production in the MW up to
the present age at varying the accretion model used to describe dust
formation in the ISM. \textbf{Left panel}: The results for the inner
ring.  \textbf{Central panel}: The same but for the Solar
Neighborhood. \textbf{Right panel}: The same but for the outer ring.
} \label{Accretion13Gyr}
\end{figure}

In Fig. \ref{Accretion13Gyr} we extend  the evolution up to the
present  time: in the inner regions and solar vicinity of the MW the
differences  between the two models are quite small, whereas as
expected they are large in the external regions of the disk. In
brief,  in model $\mathcal{A}$ the accretion in the ISM starts very
early independently from the environment and  consequently  it gives
rise to a higher dust content  compared to model $\mathcal{B}$.

\begin{figure}
\centerline{\hspace{-3mm}
\includegraphics[height=7.6cm,width=8.5cm]{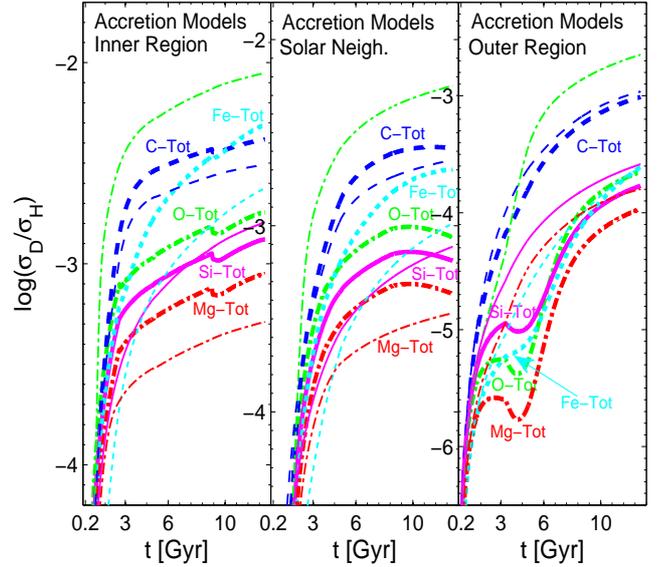}}
\caption{Temporal evolution of the dust production in the MW up to
the present age at varying the accretion model used to describe dust
formation in the ISM. In this plot we put into elements the
contribution of individual element to the total dust budget.
\textbf{Left panel}: The results for an inner ring of the MW.
\textbf{Central panel}: The same but for the  SoNe. \textbf{Right
panel}: The same but for an outer ring of the MW. }
\label{Accretion13GyrELEMENTS}
\end{figure}

A deeper insight of the differences brought by the models of dust
accretion is possible looking at the evolution of the single
elements composing the dust. In Fig. \ref{Accretion13GyrELEMENTS} we
show the evolution of some depleted elements up to the present time,
namely C, O, Mg, Si and Fe. As expected there is a strong
disagreement between the two accretion models in all the regions. In
the case of oxygen the difference is striking: model $\mathcal{A}$
with a fixed timescale produces a lot of oxygen in dust. Since there
is no description of how the various elements enter the different
dust grains, with a fixed timescale the most abundant element is
also the most abundant in dust. With a simple model we can not
follow, for instance, the different ways in which oxygen is bound in
silicates, or iron is bound both in iron-dust and silicates. Of
course, only the comparison with the observational depletion factors
can highlight the issue.  Most likely,  model $\mathcal{B}$ taking
into account the physical conditions of the medium and following in
detail the evolution of typical dust grains should better reproduce
the observational depletion factors.

\subsection{A final note on the model  parameters}\label{Finalnote}

We have just discussed the main parameters entering the problem of
the calculation of the dust enrichment of the ISM. However we must
underline the following point: classical chemical models are widely used in literature to follow
the  metal enrichment  in  galaxies of different morphology. They
are already complicated because of the many physical ingredients
entering the problem, like the law of SF, the IMF, the stellar
yields and the geometrical description of the galaxy. When the dust
is added to the  problem, the parameter space  literally blows up,
because in addition to the classical parameters we have to consider also
those governing the dust content.  The amount of dust of stellar
origin injected in the ISM, the way in which the dust content
increases/decreases by accretion/destruction and other physical
processes, all these concur to extend the list of parameters.
In addition to those we have already discussed, a couple of them
deserves some discussion.

\begin{description}
\item[(1)] The chemical model we are working with has a single-phase
description of the gas content and in particular of the MC component
in which the dust accretion takes place. For this reason we have
included two possibilities for the MC fraction $\chi_{MC}$. Either
$\chi_{MC}$ is assumed to be constant and equal to the present day
value in the solar vicinity, or it stands on the data for the MW and
it correlates the fraction of MCs to the SFR and the gas density
through ANNs. In this latter option, that is described in detail in
\citet{Piovan11c}, we are therefore assuming a variable amount of
MCs. Both the recipes tend to the same value for the current time
and produce similar results for the SoNe, that is for our target,
since the differences in $\chi_{MC}$, most striking in the early
stages of the evolution, are not enough to produce significant
differences. The reason for this can be traced in the SFRs and
densities of the SoNe that are never characterized by extreme
values. Instead, quite different results between the two recipes are
expected for the outer or inner regions of the MW, because of the
more significant excursion of the physical variables
\citep{Piovan11c}.

\item[(2)] All our models includes radial flows of matter and the effects
of a galactic bar. All the details and typical parameters are taken
from \citet{Portinari99} and \citet{Portinari00}. Radial flows and
bar have been taken into account to better reproduce a wide number
of properties of the MW disk, in particular to simultaneously
reproduce the radial gradients and the peak of gas observed around 4
kpc. The main motivation for including the radial flows and the bar
is therefore the consistency between theory and observation as far
as the depletion factors, radial/local abundances and gas masses are
concerned. However, the bar influences the innermost properties of
the MW and it is of no interest in this work. Again, the influence
of the radial flows of matter mainly applies to the radial
properties of the disk \citep{Piovan11c}. If limited to the SoNe,
their effect could be mimicked with a slightly different choice of
the parameters. Concerning the radial flows, it must be finally
observed that in principle the pattern of velocities of gas and dust
could be different. For the sake of simplicity, we assume here that
gas and dust are moving with the same velocities. Thanks to this gas
and dust should have similar radial behaviour \citep[see][for all
details]{Piovan11c}.
\end{description}

\begin{figure*}
\centerline{\hspace{-2cm}
\includegraphics[height=16cm,width=18cm]{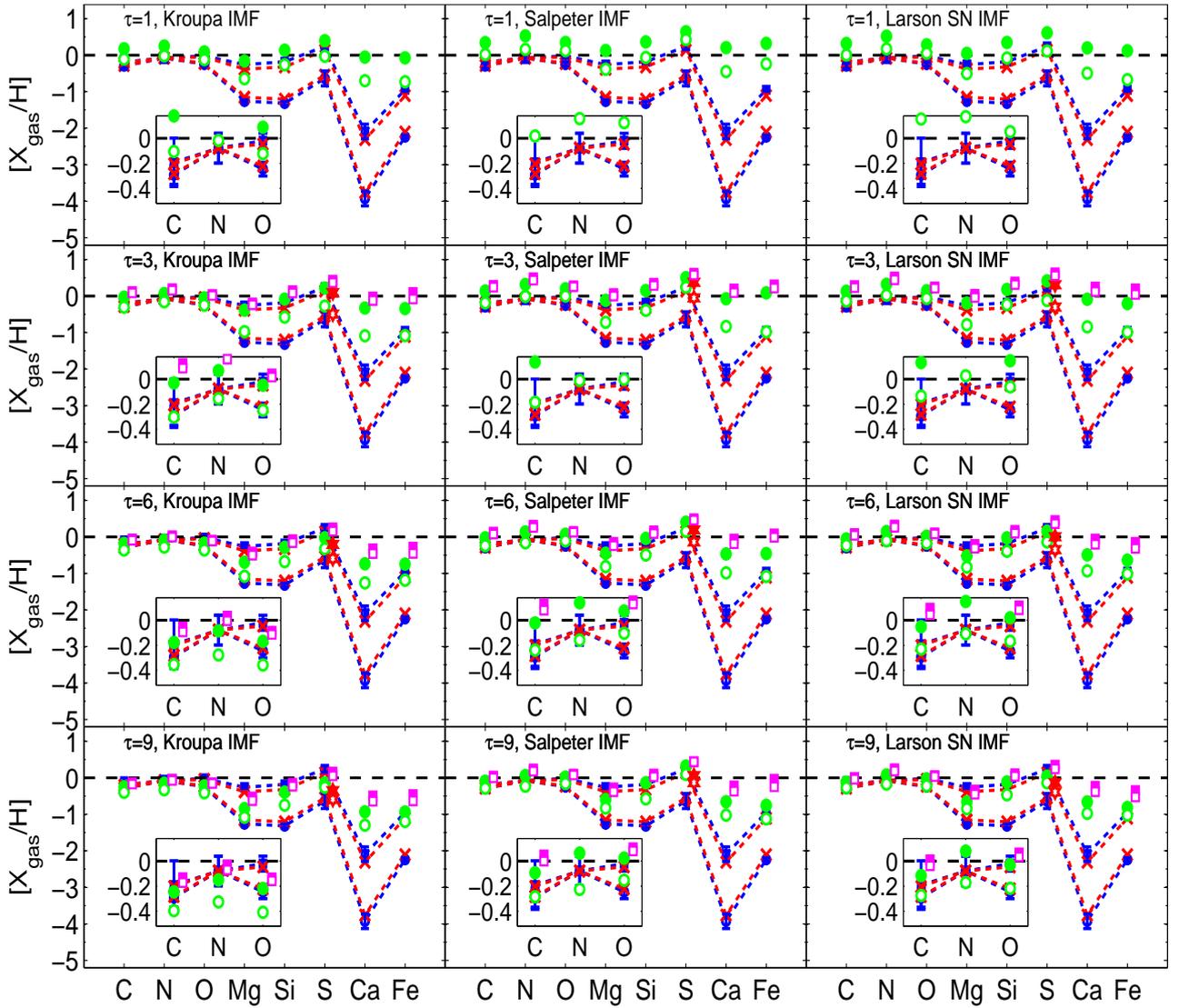}}
\vspace{-1.2cm} \caption{Depletion of C, N, O, Mg, Si, S, Ca and Fe
in the ISM as observed in the SoNe. The observations are compared
with the models at varying three important parameters, namely the
IMF, the SF efficiency $\nu$, and the mass accretion time scale
$\tau$. For the IMF we consider : the recent Kroupa IMF (left
 panels), the classical Salpeter one (central panels) and the
Larson IMF adapted to the SoNe (right panels). These IMFs are all
described in Sect. \ref{SFRandIMF}). Four cases are considered for
$\nu$, that is $\nu = 0.30$ (empty circles), $\nu = 0.70$ (filled
circles), $\nu = 1.10$ (empty squares) and $\nu = 1.50$ (filled
squares). Finally, four values are used for the accretion time
scale, namely  $\tau = 1$, $\tau = 3$, $\tau = 6$ and $\tau = 9$ Gyr
, from the top line to the bottom line of each panel.}
\label{DepletionFIT}
\end{figure*}

\begin{figure*}
\centerline{\hspace{-2cm} \vspace{-1.2cm}
\includegraphics[height=13.5cm,width=18cm]{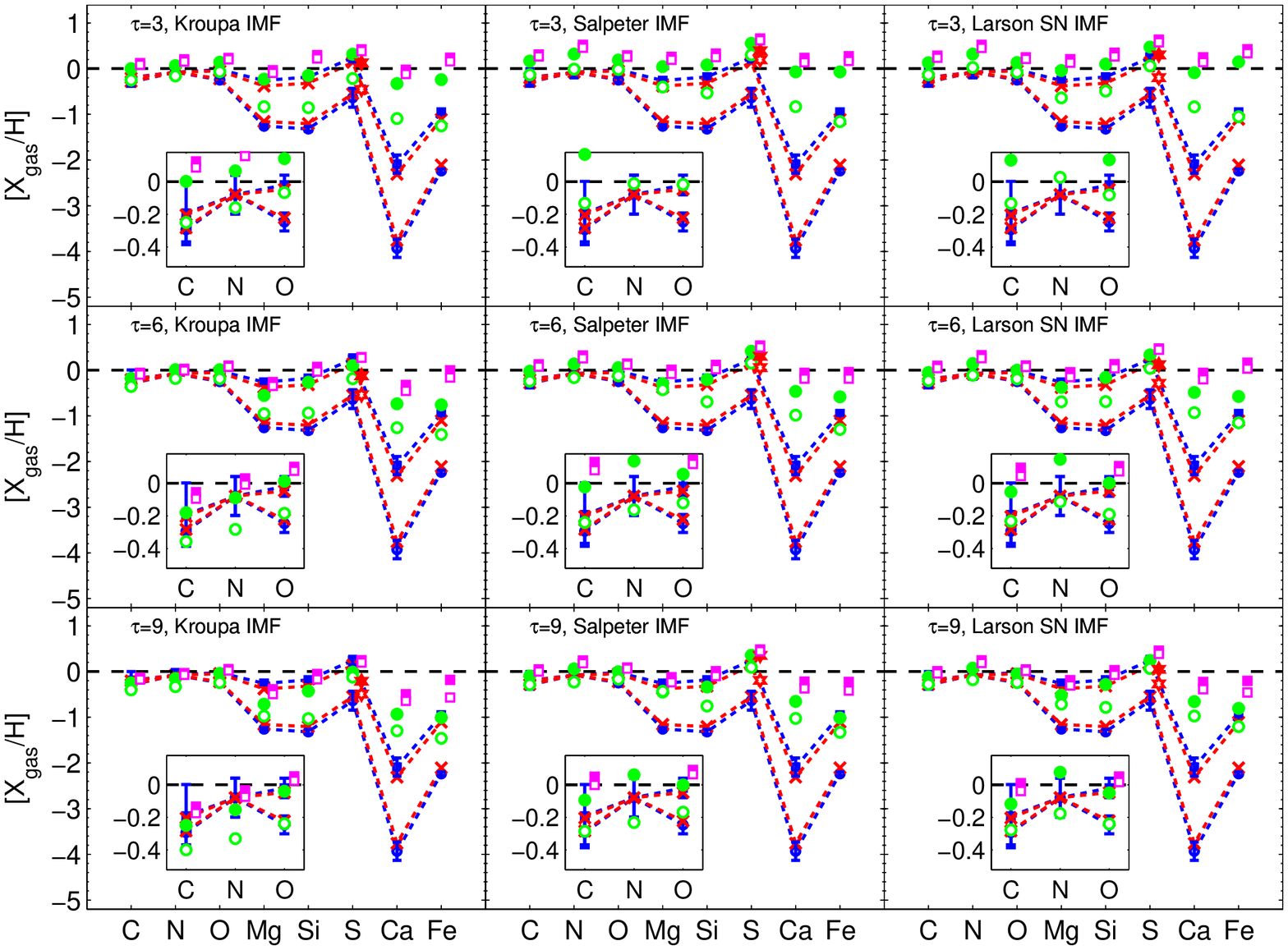}}
\caption{Depletion of C, N, O, Mg, Si, S, Ca and Fe in the ISM as
observed in the Solar Neighbourhood compared with models calculated
 with enhanced yields of  Mg. The observational data  is compared with the
models at varying of the IMF (three choices), the efficiency of the
star formation $\nu$ (four choices) and the infall timescale $\tau$
(three choices). The meaning of the symbols is the same as in Fig.
\ref{Depletion}.}\label{DepletionFITB}
\end{figure*}

\section{Models and observations}\label{Models_Observ}

The final step of our study is  to compare the theoretical results
with the data in the Solar Vicinity,
thus ultimately validating the model we have built up to
describe the dust enrichment  of the ISM. As we already discussed,
the model has many parameters, which together with the many
observational data to match  would make the search of best-fit
solution a huge task to be accomplished. However, pursuing this
strategy would not lead us to get a deeper physical insight
of the dust formation in the MW Disk and Solar Vicinity in particular.
Therefore, instead of looking for the absolute best-fit model, we
are more interested in the model response to thoughtful choices of
the parameters and in the comparison of model results with data on
element depletion, local abundances. The following observational data for the MW are
taken in consideration:

\begin{description}
\item[(1)] The depletion of the elements in the local ISM (see
Sect. \ref{AbundancesSS}) is the main check for a dust accretion
model. This one to be physically consistent must reproduce the
observational depletion of many elements.  Obviously the depletion
is line-of-sight dependent and we can only estimate the range of
plausible values (See Fig. \ref{Depletion}).
\item[(2)] The local evolution of the elemental abundances
in the Solar Vicinity. This is observationally indicated by many
diagnostic planes such as  $[\mathrm{El}/\mathrm{H}]$ vs.
$[\mathrm{Fe}/\mathrm{H}]$
 and $[\mathrm{El}/\mathrm{Fe}]$ vs.
$[\mathrm{Fe}/\mathrm{H}]$ for  some elements that are also involved
in the dust formation process, derived from large samples of F and G
stars. The models need to match those diagnostic planes.
 This allows us to check
that not only the process of dust formation/injection is properly
simulated but also that the total enrichment process, as observed in
the different generations of stars, is realistically reproduced.
\item[(3)] The large scale
properties of the MW in the Solar Vicinity  like the surface
densities of stars and gas, rate of SN{\ae},  dust-to-gas ratio, and
present day  SFR must be reproduced.
\item[(4)] The age-metallicity relation in the $[\mathrm{Fe}/\mathrm{H}]$
vs. Age plane as observed in stars of the Solar Vicinity.
\item[(5)] The metallicity of the Sun at the current age and of the
proto-Sun about $4.56$ Gyr ago.
\end{description}
\vspace{-2mm}

\begin{figure*}
\centerline{
\includegraphics[height=6.5cm,width=8.0cm]{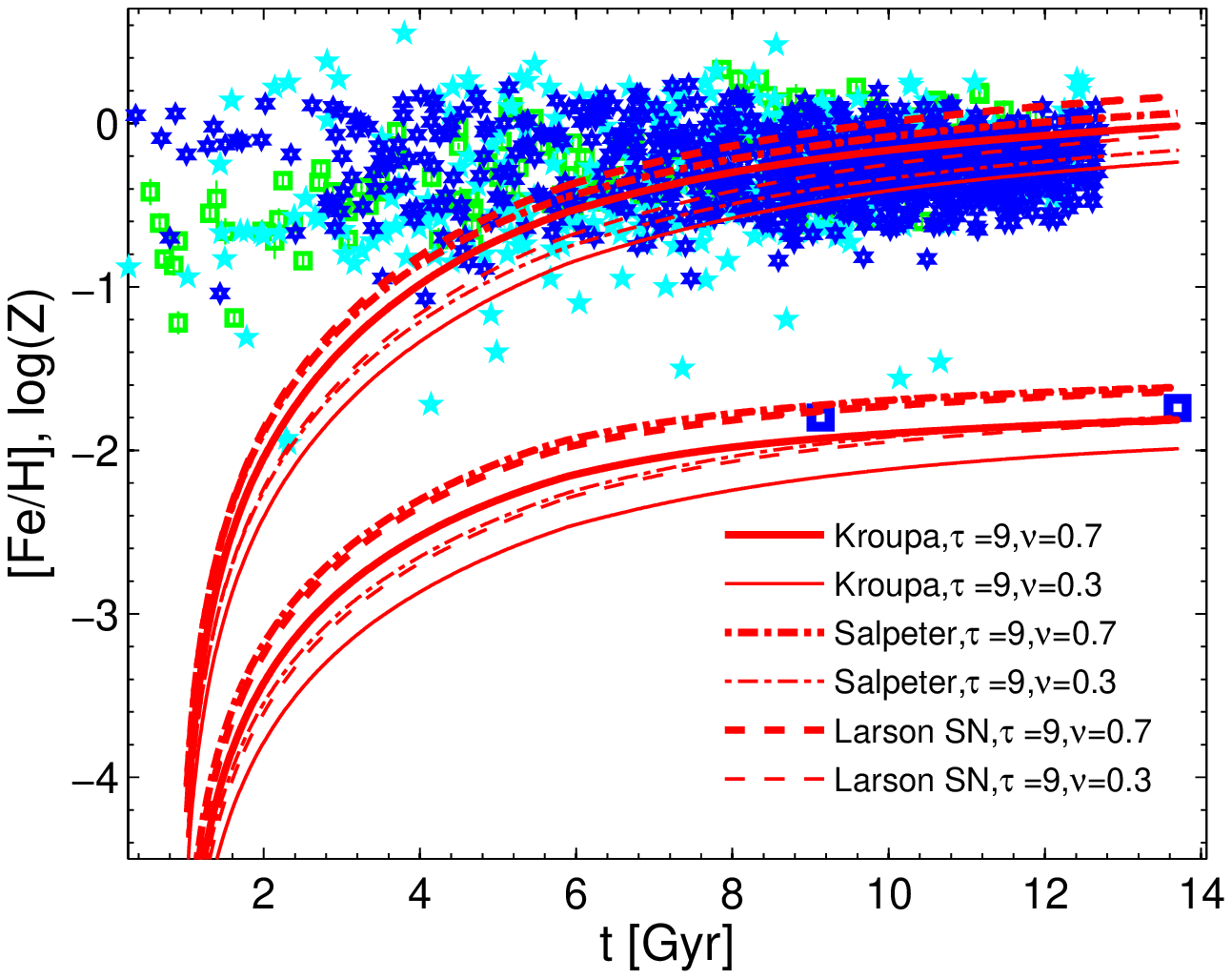}
\hspace{-20pt}
\includegraphics[height=6.5cm,width=8.0cm]{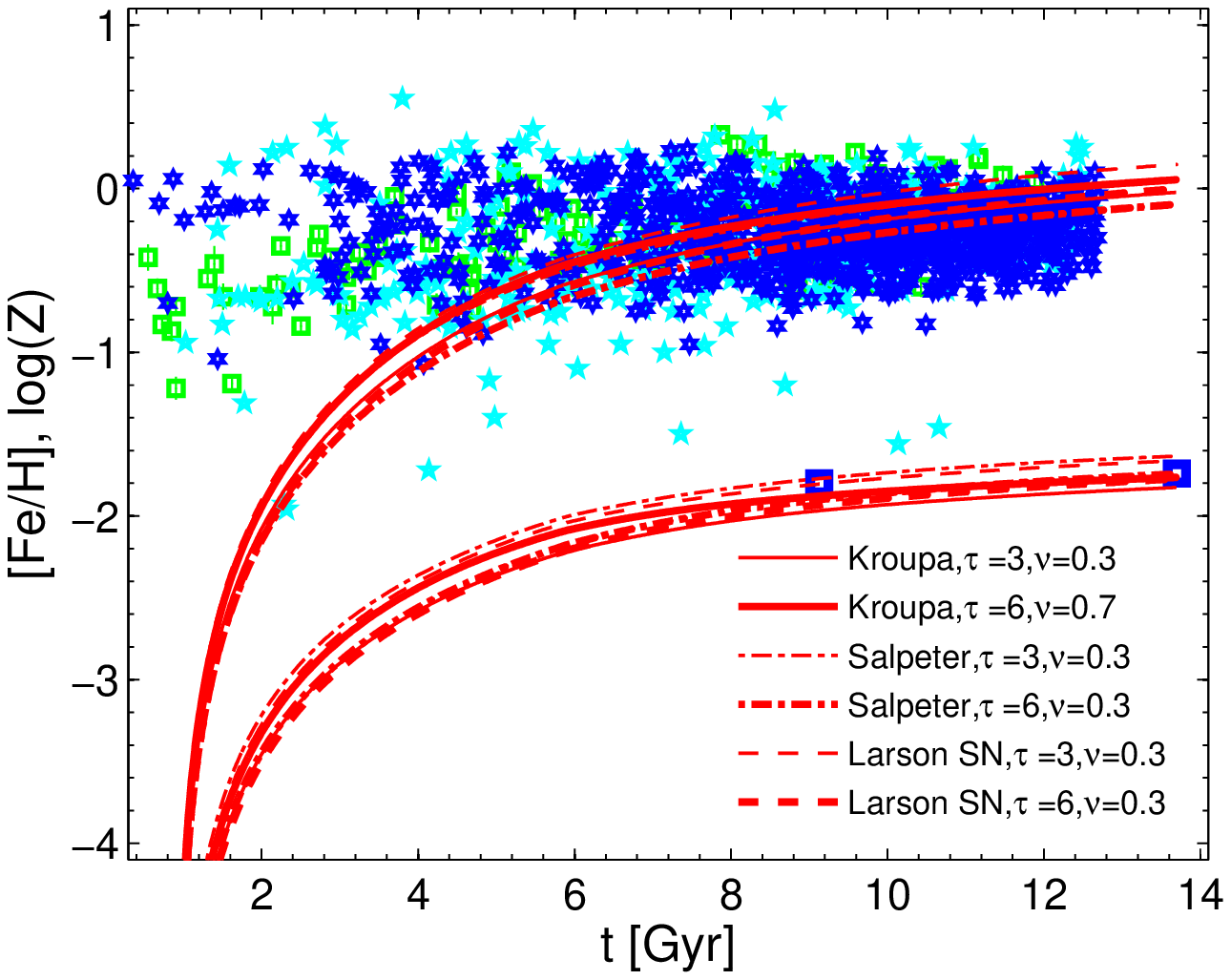}}
\caption{Temporal evolution of the metallicity $Z$ and the iron
abundance $[Fe/H]$ in the Solar Neighborhood  for a wide sample of
models. Data are taken from \citet{Ibukiyama02} (dark stars:
photometric $[Fe/H]$;  light stars: spectroscopic $[Fe/H]$),
\citet{Ramirez07} (squares: sample of thin and thick disk stars).
Open squares represent the metallicity of the solar system and of
present-day ISM \citep{Gail09}. \textbf{Left panel}: six models are
represented, at varying the IMF, the efficiency of the star
formation $\nu$ and the infall timescale between the values $\tau =
3$ and $\tau = 6$ Gyr. \textbf{Right panel}: six models are shown at
varying the IMF and the efficiency $\nu$ between $0.3$ and $0.7$
with a fixed infall timescale $\tau = 9$ Gyr.}
\label{FeHandZradiale}
\end{figure*}

\begin{figure*}
\centerline{\hspace{-30pt}
\includegraphics[height=6cm,width=6.5truecm]{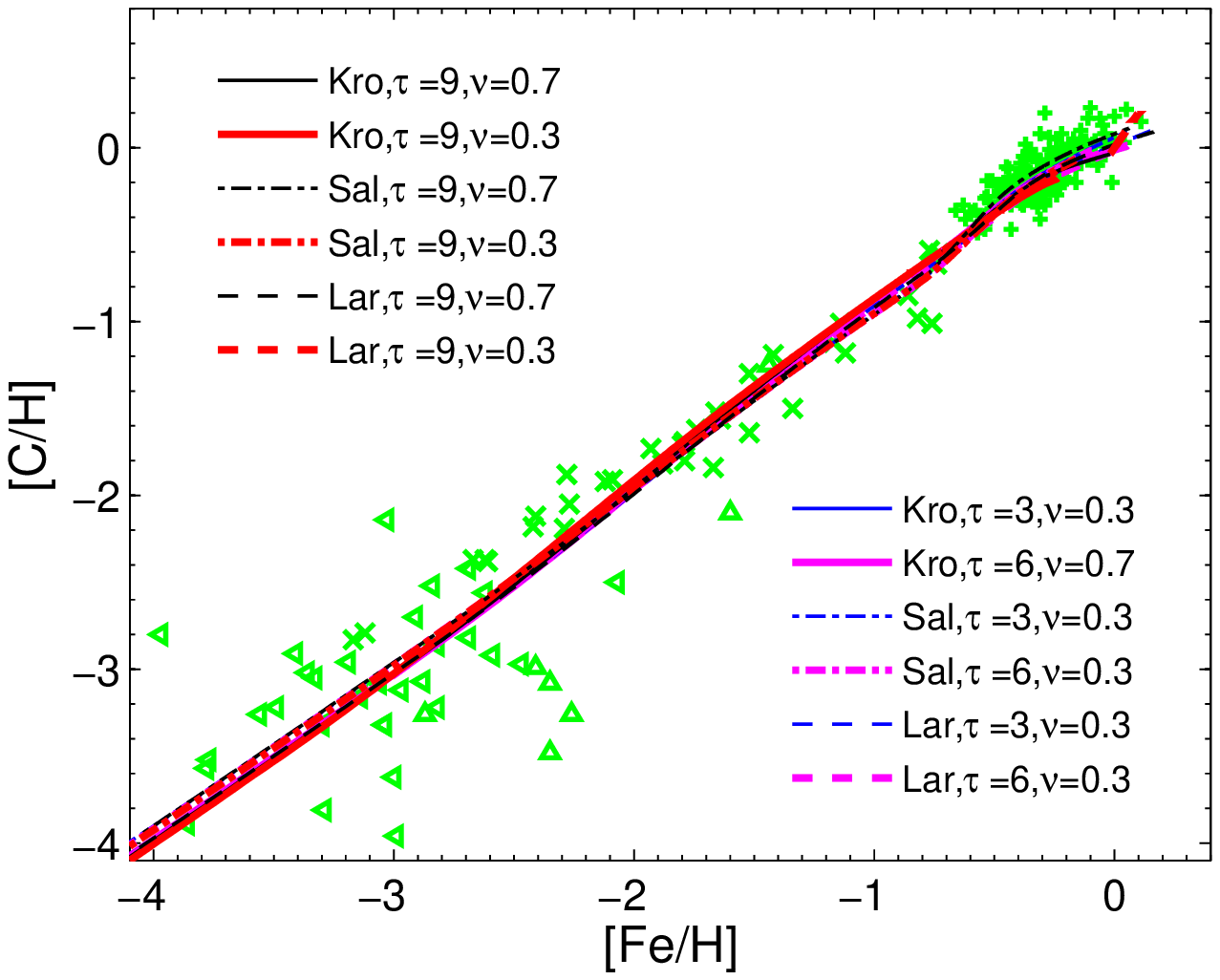}
\hspace{-20pt}
\includegraphics[height=6cm,width=6.5truecm]{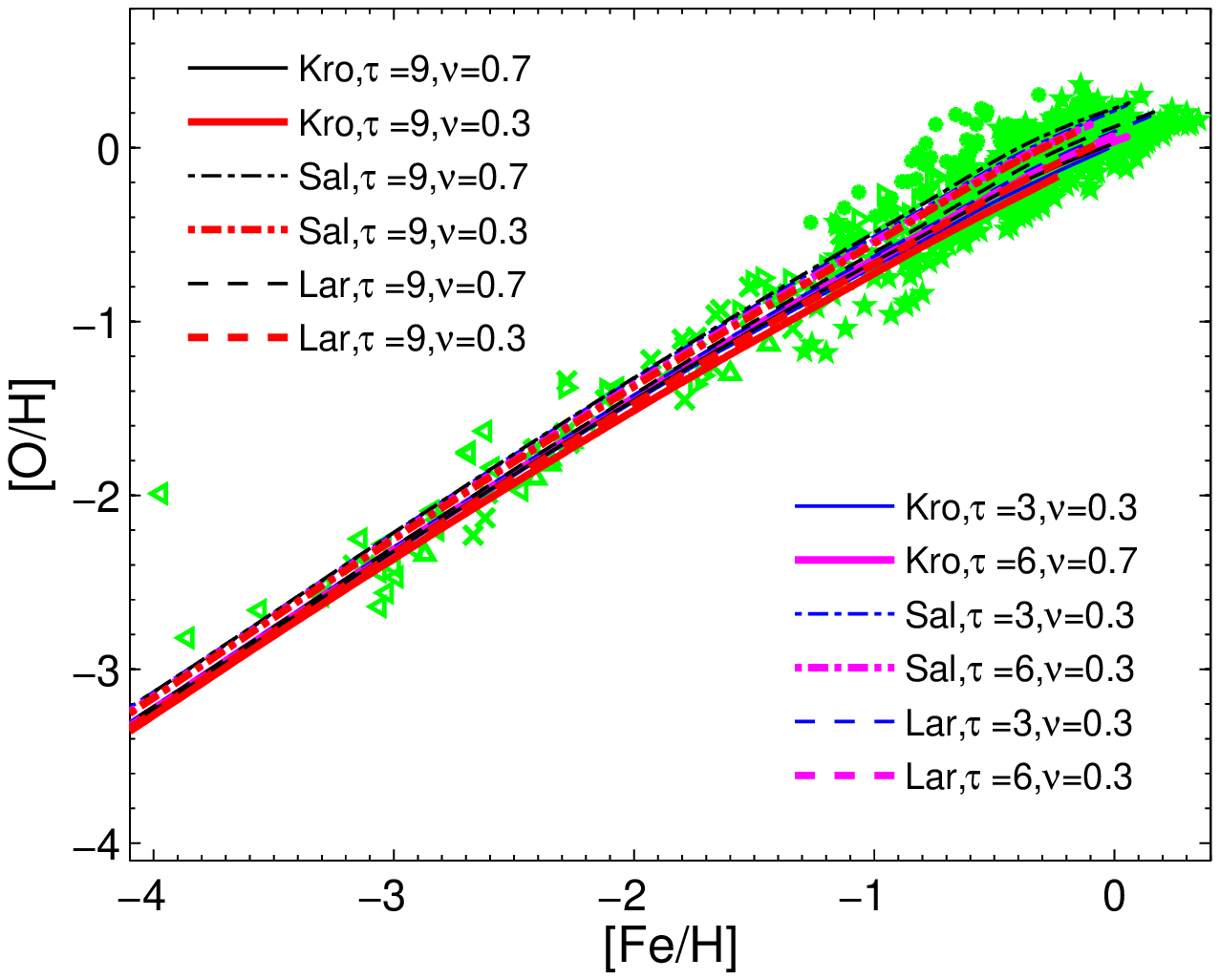}
\hspace{-20pt}
\includegraphics[height=6cm,width=6.5truecm]{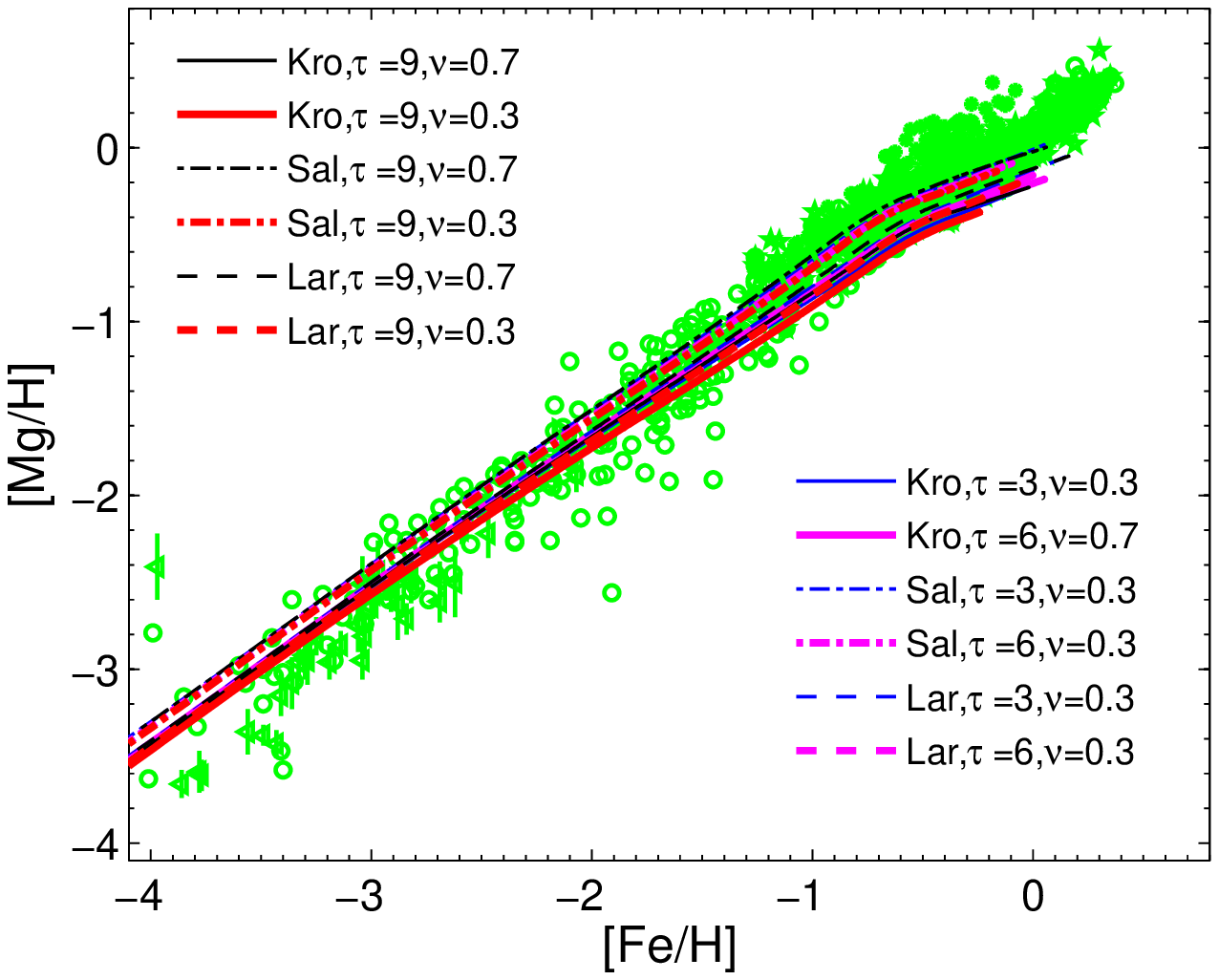}
\hspace{-30pt}} \vspace{-2pt} \centerline{\hspace{-30pt}
\includegraphics[height=6cm,width=6.5truecm]{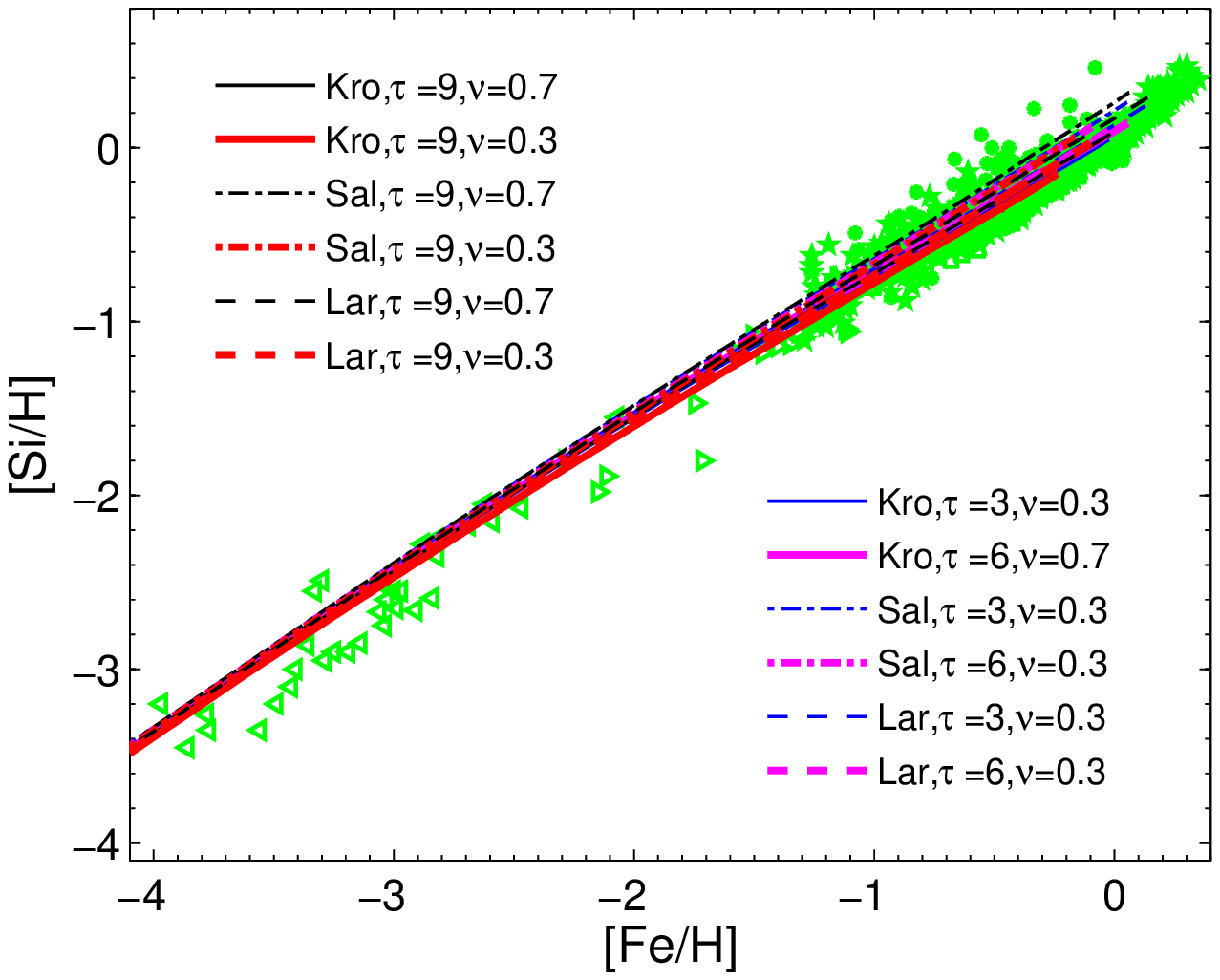}
\hspace{-20pt}
\includegraphics[height=6cm,width=6.5truecm]{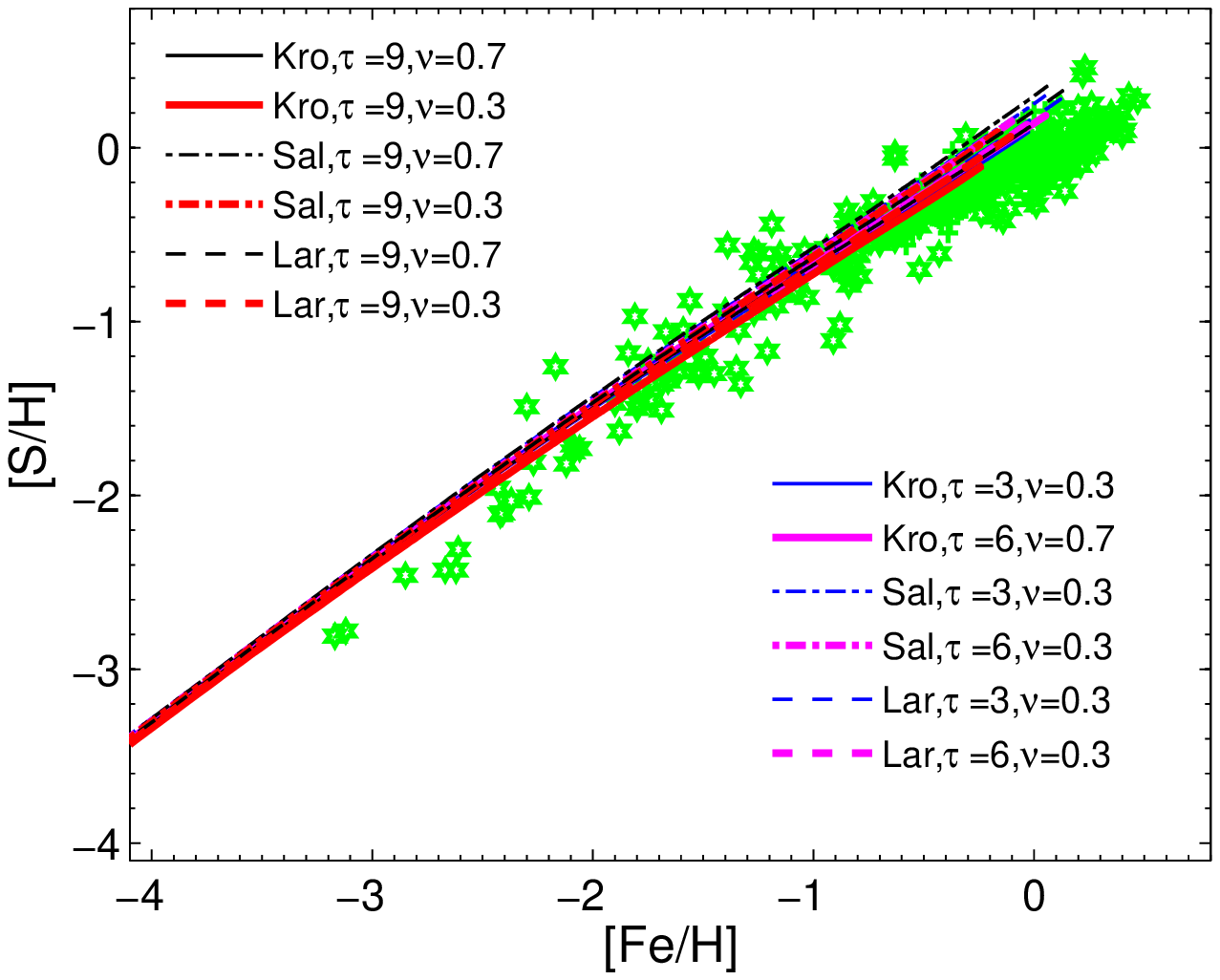}
\hspace{-20pt}
\includegraphics[height=6cm,width=6.5truecm]{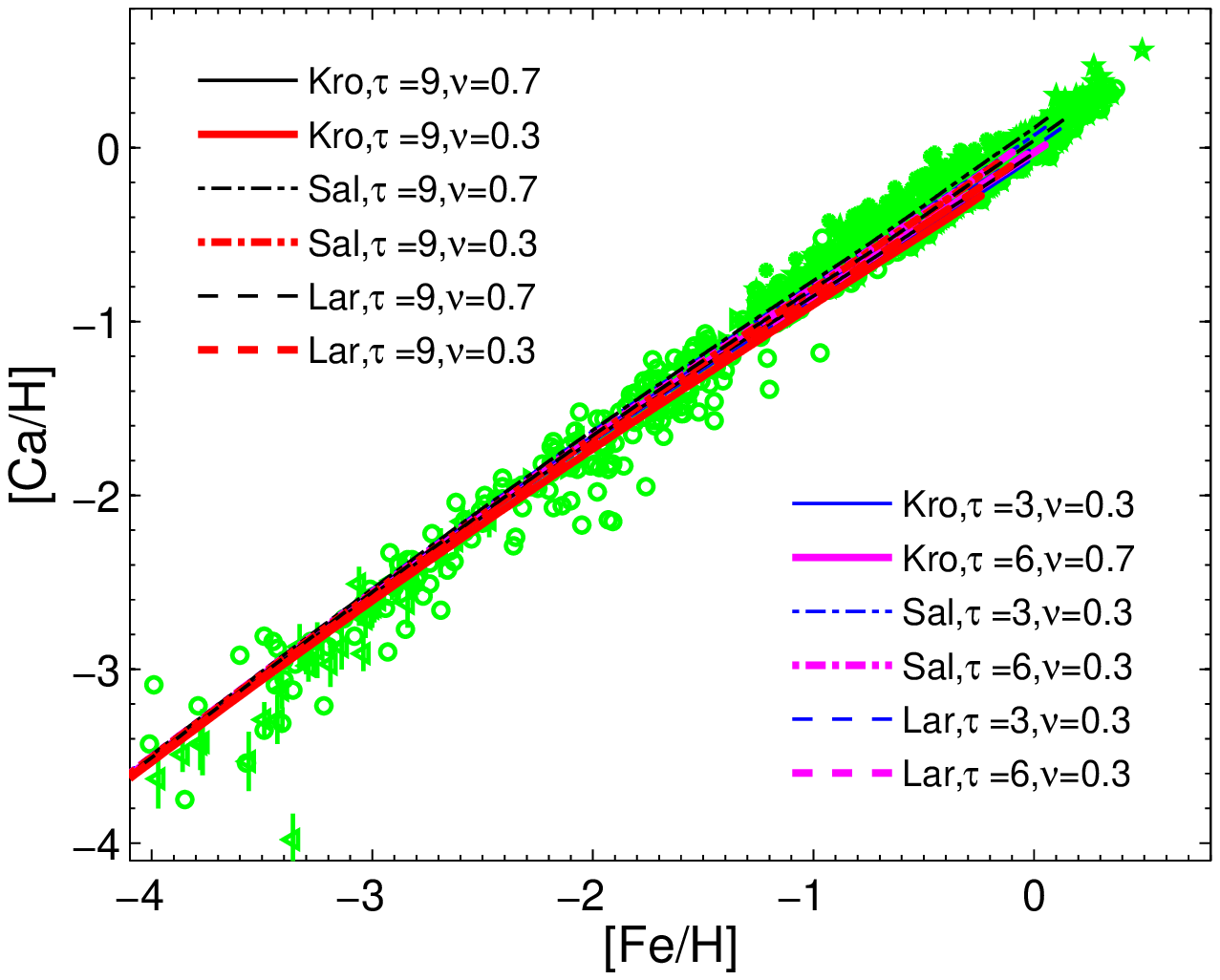}
\hspace{-30pt}}
\caption{Evolution of the elemental abundances in
the Solar Neighborhood as measured by a sample of F and G stars and
compared with the simulations. The evolution of
$[\mathrm{A}/\mathrm{H}]$ vs. $[\mathrm{Fe}/\mathrm{H}]$ is
represented for six elements of interest, namely C, O, Mg, Si, S and
Ca. The data are taken from the following databases: Oxygen from
\citet{Chen00,Melendez02,Reddy03,Gratton03,Akerman04,Cayrel04,Jonsell05,Soubiran05}
; Carbon from \citet{Melendez02,Reddy03,Akerman04,Cayrel04};
Magnesium from \citet{Chen00,Gratton03,Reddy03,Soubiran05} and
\citet{Venn04} (circle); Silicon from
\citet{Chen00,Gratton03,Reddy03,Cayrel04,Jonsell05,Soubiran05};
Sulphur from \citet{Reddy03} and \citet{Caffau05} (six-point star)
and, finally, Calcium from
\citet{Chen00,Gratton03,Reddy03,Venn04,Cayrel04,Soubiran05,Jonsell05}.
Twelve models are superposed to the data, at varying the IMF (three
cases are considered:  Kroupa, Salpeter and Larson adapted to the
Solar Neighborhood), the efficiency of the star formation $\nu$ (two
values: $\nu = 0.3$ and $\nu = 0.7$) and the infall timescale $\tau$
(three values: $\tau=3$, $\tau =6$ and $\tau =9$ Gyr). Not all the
combinations are shown but only the most interesting ones according
to Figs. \ref{DepletionFIT} and \ref{DepletionFITB}.}
\label{EvolutionAbundancesAH}
\end{figure*}

The reference model is always $\mathcal{GDABBCBBB}$ according to
the parameter list contained in Table \ref{Parameters} and it includes
the radial flows and the bar.  The main parameters we are going to
play with  are the star formation efficiency $\nu$, the infall
timescale $\tau$ and the IMF.  For this latter we consider those by
 Kroupa, Larson (however adapted to the Solar Vicinity in the high
 mass tail as discussed in  Sect. \ref{SFRandIMF}) and by  Salpeter (usually taken as
the reference case for comparison). As already listed in Table
\ref{IMFfractions}, we limit ourselves to the IMFs that do not
predict a high number of SN{\ae}.  IMFs of this type  are perhaps
more suited  to other dust-rich environments, like starburst
galaxies or ellipticals \citep{Valiante09,Gall11a,Gall11b,Pipino11}.  The  MW
disk and the Solar Vicinity in particular  seem to require IMFs somewhat poor in
massive stars \citep{Dwek98,Calura08,Zhukovska08}. Four infall
timescales are considered, that is $\tau=1$, $\tau=3$, $\tau=6$ and
$\tau=9$, from the shortest to the longest, and four values of the star formation
efficiency, $\nu =0.3$, $\nu=0.7$, $\nu =1.1$ and $\nu=1.5$, from
the lowest to the highest efficiencies.\\

\renewcommand{\arraystretch}{1.3}
\setlength{\tabcolsep}{2.8pt}
\begin{table*}
\scriptsize
\begin{center}
\caption[]{\footnotesize Comparison of the properties of the MW in
the SoNe with model results\footnotesize$^{1}$ and with the
\citet{Zhukovska08} model from which the  observational data  is
taken:}
\begin{tabular}{ccccc}
\hline \hline \vspace{0.1cm}
\small Observable & \small Observed & \small ZGT08\footnotesize$^{2}$ & \small This work & \small Reference \\
\hline
\small Total surface density $\sigma\left(r_{\odot},t_{G}\right)$ $\left[\textrm{M}_{\odot}\textrm{pc}^{-2}\right]$ & \small 50 - 62 & \small 56  & \small 52 & \small \citet{Holmberg04}  \\
\small ISM surface density $\sigma_{\mathcal{M}}\left(r_{\odot},t_{G}\right)$ $\left[\textrm{M}_{\odot}\textrm{pc}^{-2}\right]$&\small 7 - 13 &\small 9.7 &\small 10 - 19.5 &\small \citet{Dickey93}  \\
    & \small $\sim$ 8   & \small 9.7  &\small 10 - 19.5  &  \small \citet{Dame93} \\
    & \small 13 - 14   & \small 9.7  &\small 10 - 19.5  &  \small \citet{Olling01} \\
\small Gas fraction $\sigma_{\mathcal{M}}\left(r_{\odot},t_{G}\right)/\sigma\left(r_{\odot},t_{G}\right)$  & \small 0.05 - 0.2   & \small 0.17  & \small 0.2 - 0.39  &  \\
\small Surface density of visible stars $\sigma_{*}\left(r_{\odot},t_{G}\right)$ $\left[\textrm{M}_{\odot}\textrm{pc}^{-2}\right]$ & \small 30 - 40 & \small 38.6 &\small 23 - 35 & \small \citet{Gilmore89}\\
\small Surface density of stellar remnants $\left[\textrm{M}_{\odot}\textrm{pc}^{-2}\right]$ & \small 2 - 4 &
\small 7.7 & \small 6 - 10 & \small \citet{Mera98} \\
\small Star Formation Rate $\left[\textrm{M}_{\odot}\textrm{pc}^{-2}\textrm{Gyr}^{-1}\right]$ & \small 3.5 - 5 &
\small 3.1 & \small 1.6 - 4 & \small \citet{Rana91} \\
\small SN{\ae} II rate $\left[\textrm{pc}^{-2}\textrm{Gyr}^{-1}\right]$ & \small 0.009-0.0326 & \small 0.016 & \small 0.01 - 0.04 & \small \citet{Tammann94}\\
\small SN{\ae} Ia rate $\left[\textrm{pc}^{-2}\textrm{Gyr}^{-1}\right]$ & \small 0.0015-0.0109 & \small 0.0024 & \small 0.0035 - 0.0078 & \small \citet{Tammann94}\\
\small Infall Rate $\left[\textrm{M}_{\odot}\textrm{pc}^{-2}\textrm{Gyr}^{-1}\right]$ & \small 0.5 - 5 &\small 1.45 &\small 0.25 - 1.80 & \citet{Braun04}\\
\noalign{\smallskip} \hline \label{Observables}
\end{tabular}
\end{center}
\renewcommand{\arraystretch}{1}
\begin{flushleft}\footnotesize$^{1}${The range of values
for the entries of column (4) refers to the models discussed in this
section, for the case with no correction to Mg abundance. Of all the
models shown  in Fig. \ref{DepletionFIT}, we select only the twelve
 models presented in Fig. \ref{FeHandZradiale}.} \,
\footnotesize$^{2}${\citet{Zhukovska08}}.\,
\end{flushleft}
\end{table*}
\renewcommand{\arraystretch}{1}

\indent The results of these models  are presented in the various panels of
Fig. \ref{DepletionFIT}. In general, moving from the upper right to
the bottom left corner of the figure, the depletion of the elements
gets easier to obtain.  We find that: (i) in the case with the Kroupa
IMF  the observational range of abundances is easily reproduced over
ample ranges for the remaining  two parameters ($\nu$ and $\tau$);
(ii) models with the other two IMFs  lead  good results only if the
infall timescale $\tau$ and the efficiency $\nu$ are properly
chosen. About the depletion of the individual elements we  note what
follows:

\textit{\textbf{Carbon}}: carbon depletion is simulated assuming
that the fraction of carbon hidden in the CO molecules amounts to
$0.3$. This parameter plays a key role as shown in Fig.
\ref{EvolCOThree}. At varying $\xi_{CO}$ it is possible to allow for
less or more dust embedded in C grains, whose accretion depends on
the C atoms free to accrete. Carbon depletion is, in any case, well
reproduced for the most common sets of the parameters.

\textit{\textbf{Nitrogen}}: to reproduce the small depletion of
nitrogen we find that the simple choice of the longest timescale
between the oxygen and carbon, and the use of a mean nitrogen dust
grain  were not enough to keep low the amount of nitrogen condensed
into dust. To this aim, we introduce a multiplicative factor $N_{X}$
for the accretion time scale $\tau_{N,N}^{gr}$ that enters Eqn.
(\ref{Nitrogen_evolution}) and tune it so that the  low
observational depletion is reproduced. The results shown in Figs.
\ref{DepletionFIT} and \ref{DepletionFITB} include the scaling
factor $N_{X}$. We get that $N_{X} \gtrsim 10$ is required to fetch
the small accretion of  nitrogen.

\textit{\textbf{Magnesium and Silicon}}: magnesium is embedded into
dust thank to the presence of  olivines/pyroxenes and in our best
simulations it is found to be easily very depleted. The preference
goes toward the   maximum allowed depletion rather than the minimum
one. The opposite happens for silicon which is more often found
close to the minimum depletion limit (see Fig. \ref{DepletionFIT}).
This common behavior for  Mg and Si  can be explained with the
under-production of Mg  characterizing  the SN{\ae} yields by
\citet{Portinari98} that are ultimately based on \citet{Woosley95}:
This is a long  known problem \citep{Timmes95,Francois04} that
becomes crucial in our case, because Mg is one of the possible key
elements driving the accretion process of the silicates. If the
amount of Mg atoms available to form dust is under-abundant, this
will set up an upper limit to the amount of olivines/pyroxenes that
can be formed. In this study, first  we calculate models with  the
original  yields by \citet{Portinari98}\footnote{The  latest version
of the yields is used in which   the under-abundance of Mg is
partially corrected (Portinari, 2011 private communication).}
(filled and empty circles and squares in Fig. \ref{DepletionFIT}).
This choice leads to  a dust mixture where Mg (already
under-abundant)  is usually more depleted than Si. The case is well
illustrated  in Fig. \ref{DepletionFIT}. According to
\citet{Zhukovska08}, the yields of Mg  based on \citet{Woosley95}
could be  suited to reproduce the Milky Way once scaled  by a factor
of 2-3 in order to obtain Mg/Si/Fe ratios in agreement with the
observational ones for the MW, thus leading to a dust mixture closer
to reality. Keeping in mind this suggestion, we  calculate the same
models of Fig. \ref{DepletionFIT} slightly modifying the yields of
Mg for Type II SN{\ae}, in practice changing the tabulations of
\citet{Portinari98} according to the suggestion by
\citet{Zhukovska08}. These new calculations  allow us  to test
the model response to variations of  the Mg yields.  The results are
shown in Fig. \ref{DepletionFITB} where  we note that the agreement
with the observational data  is better for all the IMFs, in
particular with the Kroupa IMF the results are very good.
Furthermore,  the whole range of  observational  depletions is now
better covered by varying the star formation efficiency than with
the original yields of Mg: changing  $\nu$ from 0.3 to 0.7 all the
range of observational values is obtained. However, as in the
previous case of  Fig. \ref{DepletionFIT},  the highest star
formation rates enrich too much the ISM in metals and the process of
dust formation is hardly able to deplete the ISM. The models clarify
that  the process of dust formation \textit{crucially} depends on
the the amounts of metals injected by the stars via mass loss and/or
SN{\ae}. They also make clear that the set of abundances at the base
of the theoretical models for metal enrichment and dust production
of the ISM must be strictly identical, otherwise there would be no
consistency between the two descriptions (consistency is of course
always secured in reality).

\textit{\textbf{Iron}}: in normal circumstances iron is highly
depleted. In most of the models in Fig. \ref{DepletionFIT} the best
we can obtain at varying the parameters is to reach the upper limit
of the observational depletion range, i.e. $-2 \leq
[\mathrm{Fe}_{\mathrm{gas}}/H] \leq -1$. Even if two possible
processes for the formation of dust (iron dust and silicates) are
included, the mechanism of iron formation in cold regions of our
models is not able to reach such lower  values. Probably a more
complex model,  for instance with a spectrum of MCs with different
lifetimes and/or some processes of accretion in other parts of the
ISM different from MCs,  is needed  to reach the severe $-2$
depletion limit. With the modified Mg abundances, the results
improve a bit due to the higher amount of silicates and hidden iron,
but still we are not able to fully span the interval from $-1$ to
$-2$.

\textit{\textbf{Calcium}}: some of the  considerations we made for
the iron apply to  calcium, for which the observed depletion is even
more severe than that of  iron. For calcium, there is no overlap at
all with the observational interval. As described in Sect.
\ref{Equations:Calcium}, we  introduce a multiplicative  factor
$Ca_{X}$ in the accretion time scale $\tau_{Ca,Ca}^{gr}$  and try to
calibrate it against the observational values. Even using $Ca_{X}
\lesssim 0.05$ it is not possible to describe the strong  depletion
of Ca. The results presented in Figs. \ref{DepletionFIT} and
\ref{DepletionFITB} are for a small scaling factor $\sim 0.1$.

 \textit{\textbf{Sulfur}}: sulfur can be very depleted along some
lines of sight, much less along other. Since a typical sulfur dust
grain is not available to simulate the accretion process, we used a
simple prescription for the average accretion  with sulfur atoms
accreting on themselves (see Sect. \ref{Equations:Sulfur} for more
details). A  multiplicative  factor $S_{X}$ has been introduced to
calibrate the timescale $\tau_{S}^{gr}$, the longest between the timescales
of the refractory elements and the accretion of sulfur atoms on themselves.
The results displayed in
Figs. \ref{DepletionFIT} and \ref{DepletionFITB} show the sulfur
depletion with (filled and empty hexagons) and without (filled and
empty circles) the effect of $S_{X}$ on the accretion time scale
$\tau_{S}^{gr}$. With $S_{X}$ going from 1 (no correction - circles)
to 0.6-0.4 (small correction - hexagons) the theory fairly agrees
with observational data.

 \textit{\textbf{Oxygen}}: this element does not participate as a
key-element in any accretion process of dust formation.  However it
takes part to the formation of the silicates and it is an ingredient
of the dust yields from stars of different masses. Because of its
high abundance, the depletion is small, even if lots of oxygen atoms
are contained in olivines/pyroxenes. In Figs. \ref{DepletionFIT} and
\ref{DepletionFITB} we can see that the theoretical predictions well
reproduce the observational data, in particular  when Mg is
corrected for under-abundance in the yields. The agreement  simply
follows from using the correct Mg/Si ratio.

\textsf{Abundances and Depletions in the Solar Vicinity.} To check
the internal consistency of the depletion models, we must secure
that the theoretical evolution of the abundances of the various
elements in the Solar Neighborhood is able to reproduce the
observational data. It would be a point of strong contradiction if
the models can reproduce the elemental abundances  in the dust, but
fail to reproduce the pattern of abundances in the gas and stars. In
Fig. \ref{FeHandZradiale} we show the evolution of the metallicity Z
and  iron abundance $[\mathrm{Fe}/\mathrm{H}]$ for a wide selection
of models, taking into account more or less the combinations of IMF,
$\nu$ and $\tau$ that best reproduce the depletion measured in the
SoNe. In such a case, one should  discard the models with $\tau = 1$
that hardly fit the data for depletion in the SoNe. Most of the
remaining models fairly agree with the observational data with the
exception of the case with the Kroupa IMF, slow infall ($\tau = 9$),
and low SFR ($\nu = 0.3$) that is not able to grow fast enough in
metals. In Table \ref{Observables} we present the check of the
minimum/maximum values obtained in our models vs. the observations
in the SoNe of various physical quantities of interest. The same
simulations presented in Fig. \ref{FeHandZradiale} are compared with
both the results obtained by \citet{Zhukovska08} and the
observational data. The general agreement between our results and
the observations is good and it allows to conclude that we are
employing a correct modelling of the SoNe. Once that the more
general quantities, like the mass of stars, gas, the SNa rates and
the total surface mass are satisfactorily reproduced we can proceed
to examine the evolution of the abundances of the single elements in
the SoNe and how they match up with the local data. \\

In the panels of Fig. \ref{EvolutionAbundancesAH} we show the time
evolution in the Solar Neighborhood of the abundances of six
elements heavily involved in  dust formation, namely C, O, Mg, Si, S
and Ca,  in the diagnostic planes   $[\mathrm{A}/\mathrm{Fe}]$ vs.
$[\mathrm{Fe}/\mathrm{H}]$. The observational data refers  to  F and
G stars (see Fig. \ref{EvolutionAbundancesAH} for more details on
the data and associated legend). Twelve models are displayed at
varying the IMF (Larson SoNe, Salpeter and Kroupa), efficiency of
star formation ($\nu=0.3$ and 0.7), and the infall timescale $\tau$
(3, 6 and 9 Gyr). Not all the cases are shown for $\tau = 3$ and
$\tau = 6$ but only the most interesting ones as far as the degree
of depletion is concerned (see  Figs. \ref{DepletionFIT} and
\ref{DepletionFITB}). In general, the agreement is good for  $\tau =
3$ and $\tau = 6$, while for $\tau = 9$, during the first 2 Gyr (in
practice until $[\mathrm{Fe}/\mathrm{H}] \sim -1.5$) the enrichment
in metals is too slow with respect to the observations. The only
exception is carbon  which always agrees with the observations. It
must be pointed out that we are using infall models with a single
timescale. This simplified picture leads to models that in the
$[\mathrm{A}/\mathrm{Fe}]$ vs. $[\mathrm{Fe}/\mathrm{H}]$ planes
cannot reach regions of very low or very high
$[\mathrm{Fe}/\mathrm{H}]$. In reality, during the lifetime of the
MW different time scales could be involved in the evolution of the
describe the Solar Neighborhood, e.g.  a fast early enrichment
followed by a much slower one. Double infall models
\citep{Chiosi80,Chiappini97a} or more complicated scenarios  are
however beyond the aims of this study. Finally, it is worth
noticing: (i) the under-abundance of Mg in the top right panel; (ii)
the marginal agreement of the models with sulfur evolution.

\begin{figure}
\centerline{
\includegraphics[height=6.5cm,width=8.0cm]{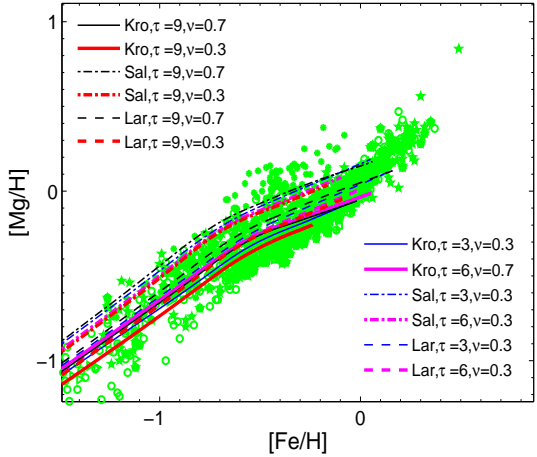}}
\caption{Evolution of the Mg abundance   in the Solar Neighborhood
as measured in a sample of F and G stars. The yields have been
slightly corrected for the Mg  under-abundance. This corrected
abundance is the same as in Fig. \ref{DepletionFITB}.}
\label{MgCORR}
\end{figure}

\begin{figure*}
\centerline{\hspace{-30pt}
\includegraphics[height=6cm,width=6.5truecm]{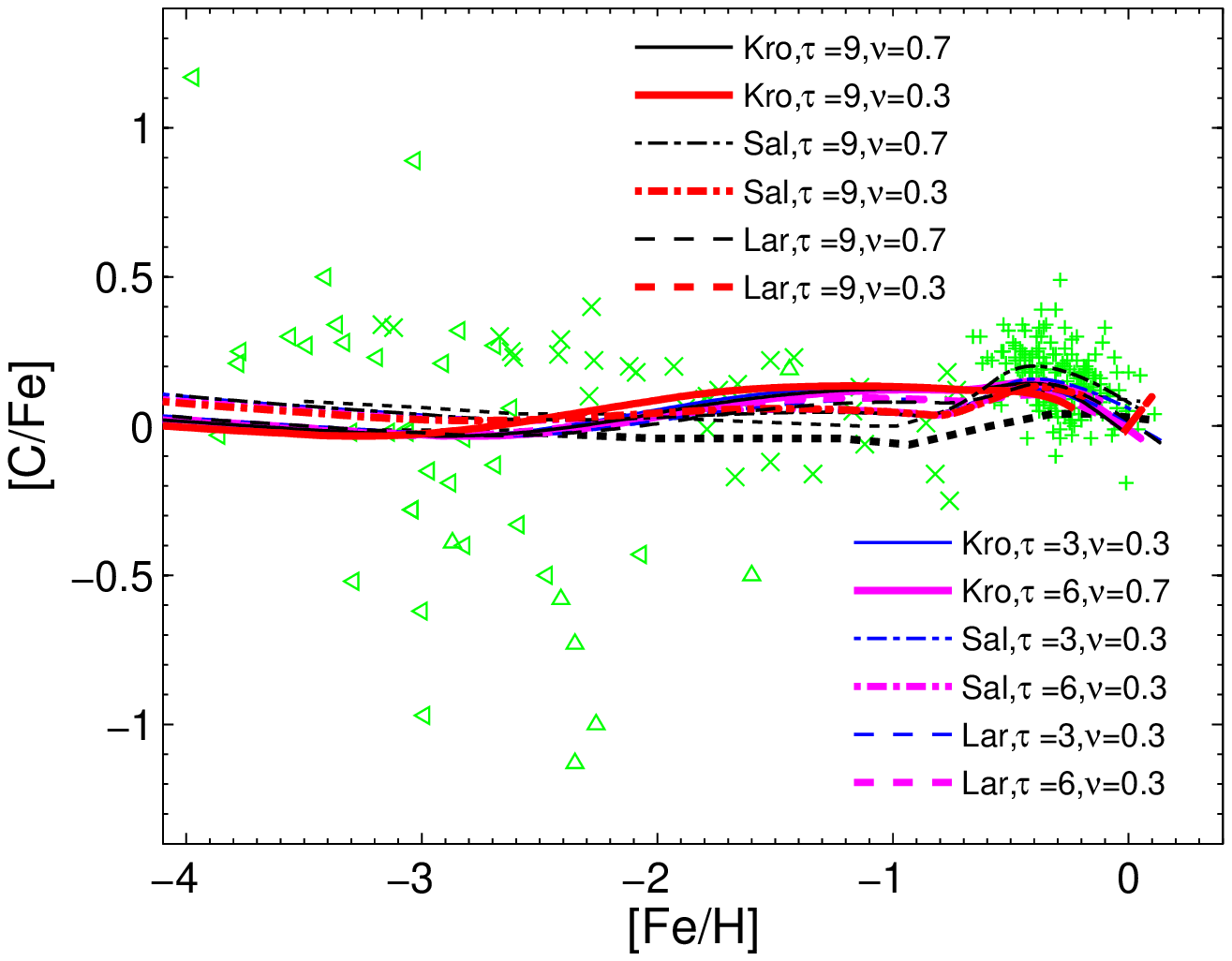}
\hspace{-20pt}
\includegraphics[height=6cm,width=6.5truecm]{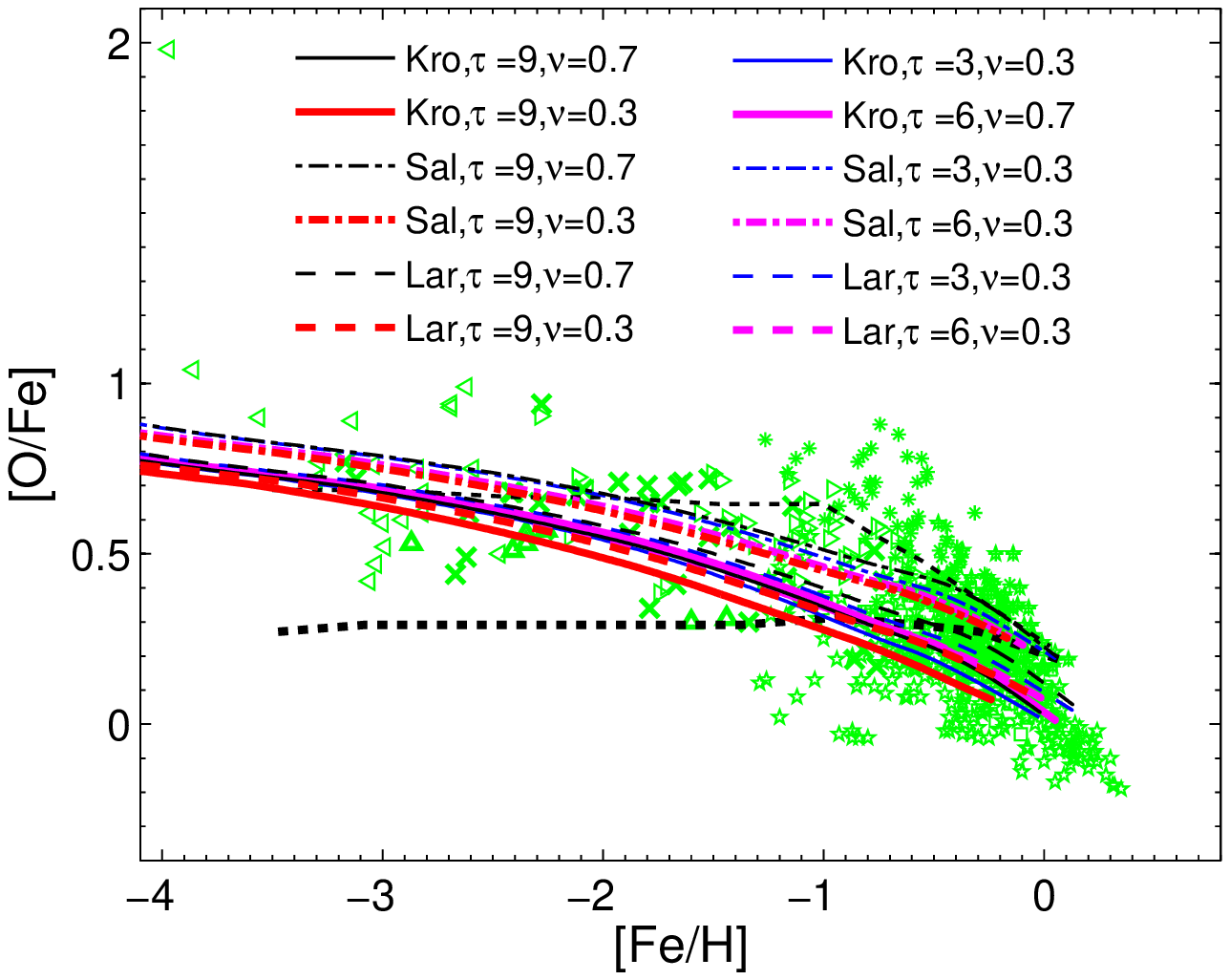}
\hspace{-20pt}
\includegraphics[height=6cm,width=6.5truecm]{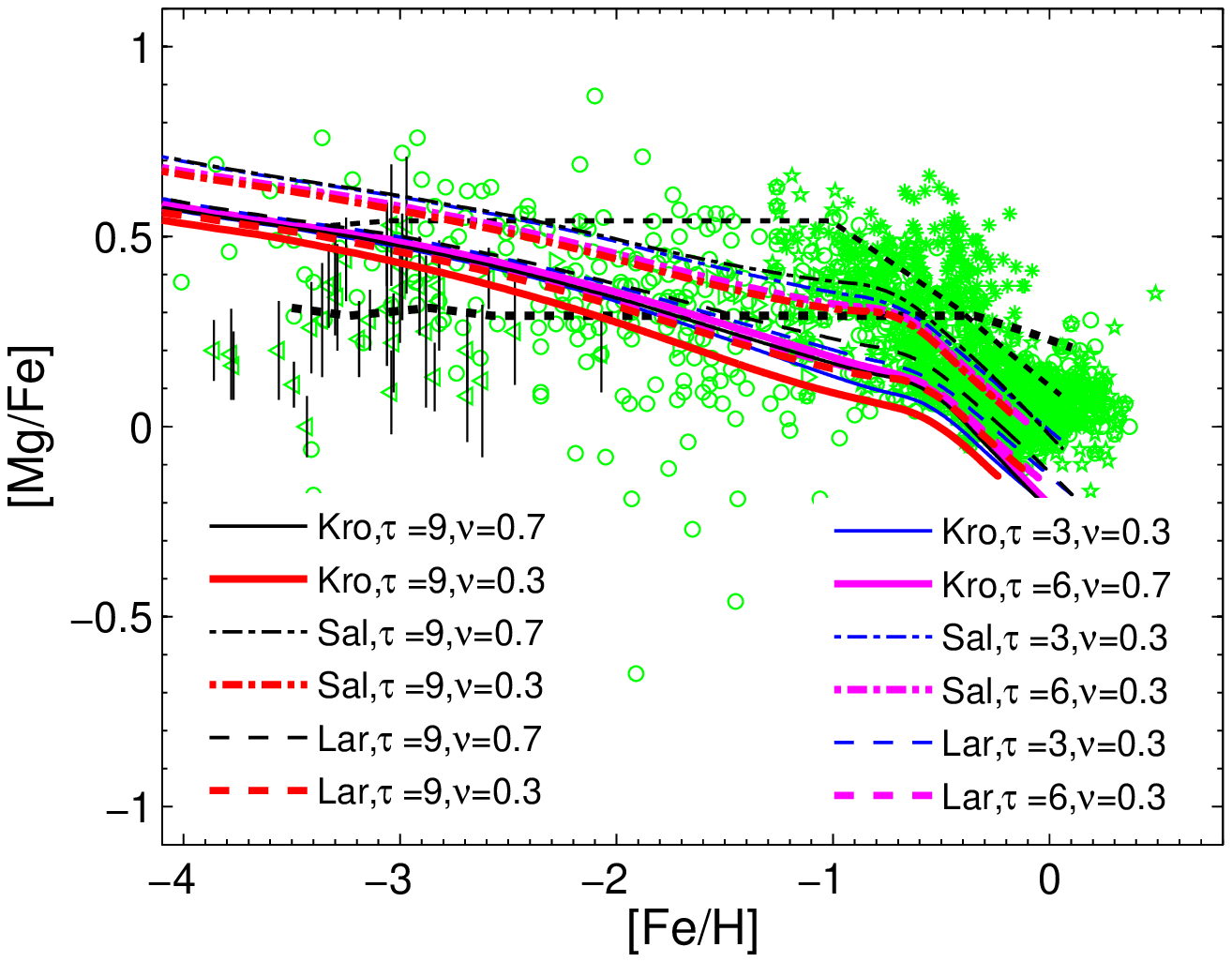}
\hspace{-30pt}} \centerline{\hspace{-30pt}
\includegraphics[height=6cm,width=6.5truecm]{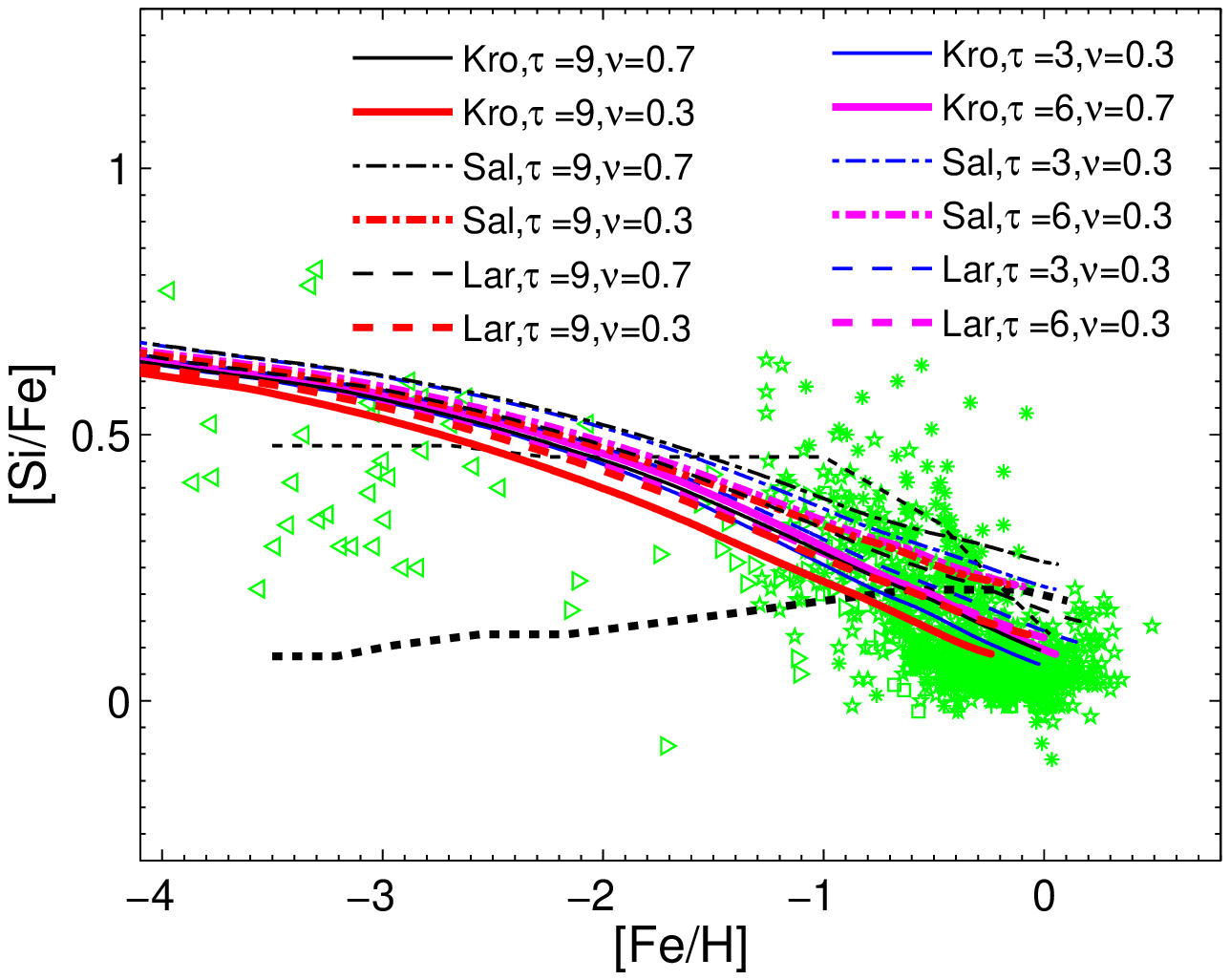}
\hspace{-20pt}
\includegraphics[height=6cm,width=6.5truecm]{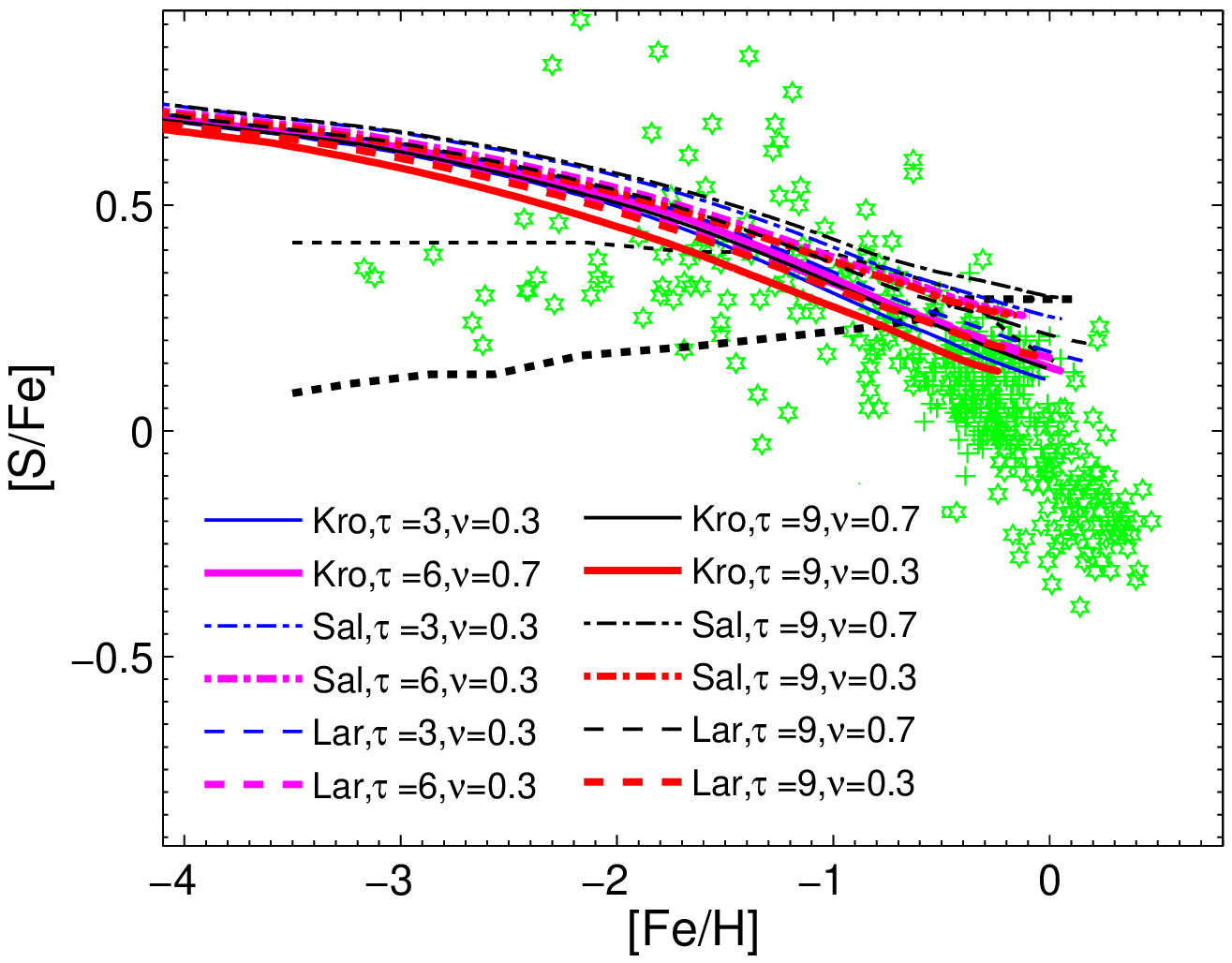}
\hspace{-20pt}
\includegraphics[height=6cm,width=6.5truecm]{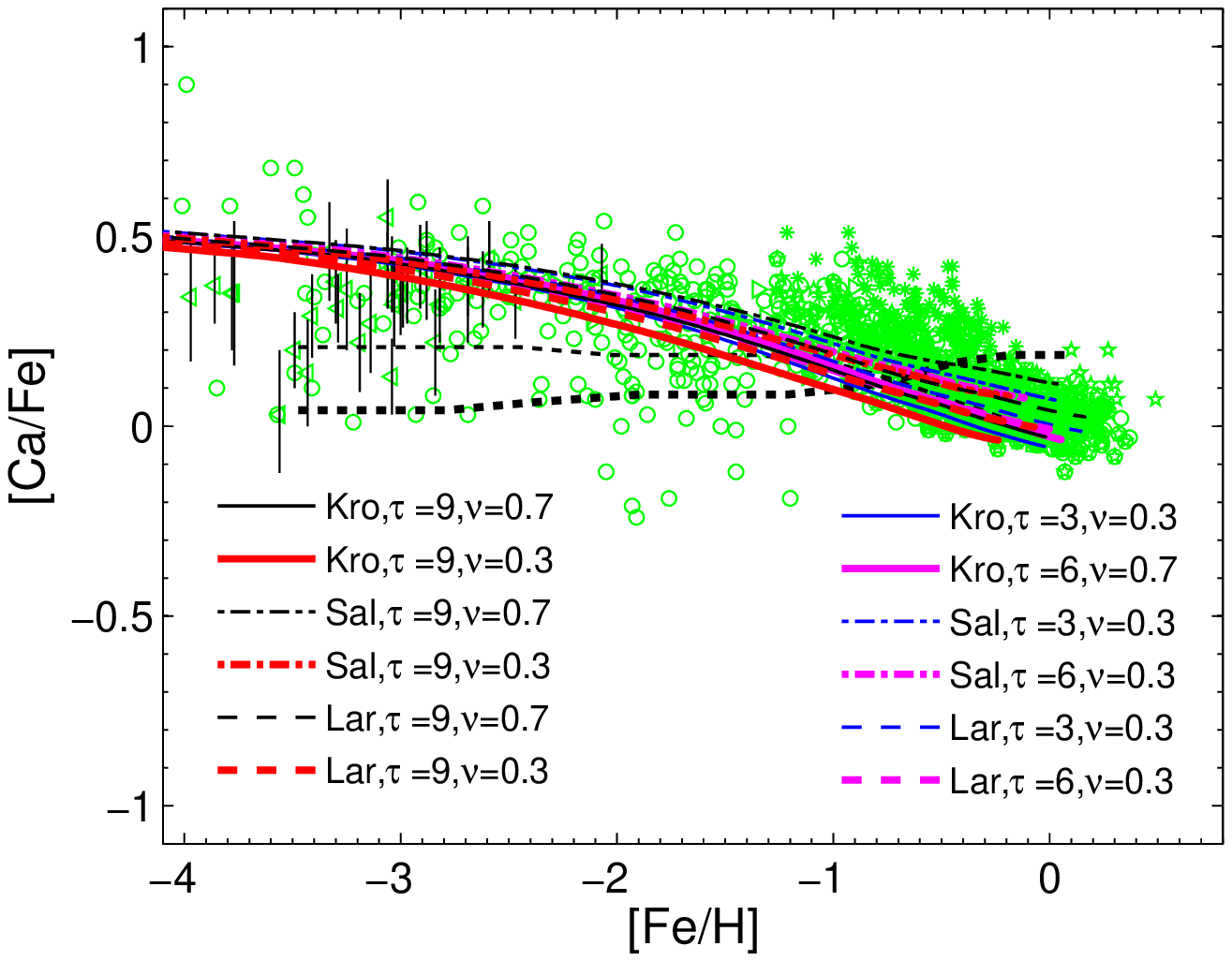}
\hspace{-30pt}} \caption{Evolution of the elemental abundances in
the Solar Neighborhood as measured in  a sample of F and G stars.
The evolution of $[\mathrm{A}/\mathrm{Fe}]$ vs.
$[\mathrm{Fe}/\mathrm{H}]$ is shown  for six elements of interest,
that is C, O, Mg, Si, S and Ca. For the sake of comparison we plot
the results by \citet{Zhukovska08} for the SNa yields by
\citet{Woosley95} where Mg and Fe have been corrected to get a
better agreement with the observations (the thick dotted line), and
the yields by \citet{Nomoto06}, the thin dotted line. The meaning of
the symbols for the observational data is the same as in Fig.
\ref{EvolutionAbundancesAH}.} \label{EvolutionAbundancesAFe}
\end{figure*}

In Fig. \ref{MgCORR} we show the evolution of the Mg abundance in
the SoNe once corrected for the under-abundance problem. The same
models already presented in Fig. \ref{EvolutionAbundancesAH} are
shown, but for the correction of the Magnesium abundance. With this
correction not only we have a better pattern of depletions (See Fig.
\ref{DepletionFITB}), but also a better reproduction of the local
data. \\ \indent To enforce the argument we examine the evolution of
the most important elements involved in the dust formation process
looking  at  the $[\mathrm{A}/\mathrm{Fe}]$ vs.
$[\mathrm{Fe}/\mathrm{H}]$ diagrams relative to the SoNe. The most
striking points of disagreement with the observational data are  the
under-abundance of Mg and the evolution of Sulphur. For the sake of
comparison, we also show the models by  \citet{Zhukovska08},
calculated after revising the yields by  \citet{Woosley95} and
\citet{Nomoto06}. In both cases the yields of Sulphur  do not lead
to a good fit of the observations. The yields by  \citet{Nomoto06}
give indeed the worst model.  Our  yields produce similar results in
the sense that they  tend to keep the abundance of Sulphur too high.
Fortunately,  this has no effect on the overall production of dust
as Sulphur drives its own accretion efficiency (See Sect.
\ref{Equations:Sulfur}) and does not affect  other channels of dust
production. Apart from these minor uncertainties,  the observational
data and theoretical abundances fairly agree and consequently  dust
formation stands on a realistic description of the evolution of the
elemental abundances in the Solar Neighborhood.

\section{Discussion and conclusions}\label{Discus_Concl}

In the first paper of this series of three \citep{Piovan11a}, we
presented and discussed the prescriptions currently in use to
describe the type and amounts of dust injected into the ISM by AGB
stars and SNa explosions.  The  condensation coefficients of the
dust have tested and analyzed referring to a suitable  chemical
model for the MW Disk and the Solar Neighbourhood in particular.
This reference model and its physical ingredients are described in
great detail in this paper.  In particular, we focused on the
mechanism of dust accretion in the ISM,  dust destruction by various
processes, the main parameters of the chemical model, namely IMF,
star formation efficiency and infall timescale. The main conclusions
can be summarized as follows:

\begin{itemize}
\item The CO molecules influence the formation of C-based dust, thus
introducing into the estimates an unavoidable uncertainty. The
higher the amount of carbon embedded into CO, the slower is the
accretion process and the smaller the amount of carbonaceous grains
that are  formed.

\item During most of the MW evolutionary history,  the main process
enriching the ISM in dust is the accretion in the cold regions. Only
in the very early stages,  SN{\ae} dominate, and the duration of
this phase tends to shorten in regions of  high star formation
 and fast enrichment in metals (innermost regions of the MW). The
opposite for the regions of low star formation (the outskirts of the
MW), where the accretion in the ISM becomes significant much later
so that for many Gyr SN{\ae} govern the total dust budget. In this
case, AGB stars play an important  role, because there is enough
time for them to significantly contribute to the  dust budget,
without being overwhelmed by the SN{\ae} (earlier phases) or the ISM
accretion (later phases). The time interval during which SN{\ae} are
the main dust producer can slightly change at varying the upper mass
limit of  stars undergoing  the AGB phase.

Our conclusions for the MW Disk could be extended to  galaxies
characterized by  continuous star formation on the notion that the
outer regions of the MW Disk might correspond to low star forming
galaxies and the  inner ones to high star forming objects. Very high
values of SFR \citep[See for example] [for detailed simulations of
starburst galaxies and QSOs]{Gall11a,Gall11b}, which are not reached
in the MW, not even in the early phases of the evolution in the
inner regions, somehow elude this simple scheme. In any case,
star-dust dominates the mild SF environments for a long  period of
time. This time scale tends to decrease in environments with IMFs
skewed toward the mass interval in which  the star-dust/metals
factories (AGB stars and SN{\ae}) are important, thus helping the
onset of the accretion phase in the ISM.

\item In the high SFR/high
metallicity regions AGB stars mainly produce silicates, whereas in
the low SFR/low metallicity ones, carbon stars can contribute
significantly to the C-based dust.

\item The IMF plays a fundamental role because it controls the
relative amounts of low, intermediate and massive stars (see the
entries of Table \ref{IMFfractions}). We found that IMFs skewed
toward massive stars are not suitable to reproduce the properties of
dust in the  MW, the SoNe in particular, \citep[see
also][]{Dwek98,Zhukovska08,Calura08}, whereas it seems that for
star-burst galaxies \citep{Gall11a}, ellipticals \citep{Pipino11}
and QSOs \citep{Gall11b,Valiante11} IMFs more biased toward higher
mass stars are required, because otherwise it would be more
difficult  to reproduce the amount of dust observed already on site
in very high-z objects. The alternative is to introduce very high
condensation efficiencies in SN{\ae}. This possibility is somehow
supported by the recent FIR/sub-mm observations on the SN 1987A by
\citet{Matsuura11}. In such a case normal  IMFs in high-z galaxies
and QSOs cannot be excluded. The use of different IMFs for different
galaxies in order to reproduce dust properties is a subject of vivid
debate, and tightly related to the wider question about the
universality of the IMF. The debate seems to favour an IMF sensitive
to the initial conditions of star formation \citep[see for instance
the recent] [\,to mention a
few]{Elmegreen09,Bastian10,Myers11,Gunawardhana11,Kroupa11}. This
possibility was suggested long ago by \citet{Chiosi1998} to solve
the apparent contradiction between the spectral and chemical
properties of early type galaxies. Similar conclusions are reached
by \citet{Valiante11}, trying to match the observed properties of
the QSOs SDSS J1148+5251, where a top-heavy IMF allow a more
coherent match between different observations. Finally, when the SFR
is high (inner regions), the effect of the IMF on the dust budget
tends to disappear at the current time, because the accretion in the
ISM becomes dominant early on  in the evolution, while if the SFR is
low we can see the effect of different IMFs spreading until the
current time.

\item Lower efficiencies $\nu$ of the SF, in the range adopted for the SoNe, correspond to
a slower onset of the accretion in the ISM and a lower final dust budget.

\item We tested different descriptions of the process of dust
accretion in the ISM. A simple approach adopts an average time-scale
of accretion to estimate the \textit{total} dust budget at least in
the early stages and for normal star forming environments. More
complicated descriptions are require to  follow the evolution of the
abundance of single elements in dust or low star forming regions.

\item The range of depletions observed for most of the elements, that is
C, N, O, Mg, Si and S, is nicely  reproduced by  reasonable
combinations of the parameters. IMFs with the classical slope in the
high mass range,  more easily fit  the data for depletion. The only
exceptions are: (i) iron, for which probably a more complex
mechanism of accretion in the ISM than the simple one on cold
regions is  required \citep{Zhukovska08}; (ii)  calcium whose
extreme depletion can not be reproduced by our model.

\item The ratio between the abundances of Mg and Si is crucial in the
formation of silicates and it is important to tune it according to
the ratio expected from models.  We had to  correct  the slightly
under-abundance of the Mg in our gaseous yields, to obtain  a better
agreement with the observed depletion for the refractory elements
involved into the silicates formation.
\end{itemize}

 To conclude, the classical chemical models nicely reproduce the
observed depletion and properties of the SoNe. The theoretical
ingredients, like condensation coefficients and accretion models,
behave in a satisfactory way. What can be improved? Clearly the
major weak point is the one-phase description of the ISM. At least a
two-phases ISM is required. Finally, work is in progress to
introduce a multi-phase description of the ISM with dust in
N-Body-TSPH simulations of galaxies.\\

\textit{Acknowledgements}. L. Piovan acknowledges A. Weiss and the
Max Planck Institut F\"ur AstroPhysik (Garching - Germany) for the
very warm and friendly hospitality and for providing unlimited
computational support during the visits as EARA fellow when a
significant part of this study has been carried out.
 The authors are also deeply grateful to  S. Zhukovska and H. P. Gail
for many explanations and clarifications about their model of dust
accretion, T. Nozawa and H. Umeda for many fruitful discussions
about SNa dust yields. This work has been financed by the University
of Padua with the dedicated fellowship "Numerical Simulations of
galaxies (dynamical, chemical and spectrophotometric models),
strategies of parallelization in dynamical lagrangian approach,
communication cell-to-cell into hierarchical tree codes, algorithms
and optimization techniques" as  part of the AACSE Strategic
Research Project.

\begin{scriptsize}
\bibliographystyle{apj}                          
\bibliography{mnemonic,PiovanII}    
\end{scriptsize}

\end{document}